\documentclass[11pt,english,twoside]{article}

\usepackage[T1]{fontenc}
\usepackage[latin1]{inputenc}
\usepackage[english]{babel}
\usepackage{lmodern}
\usepackage{a4wide}
\usepackage{amssymb, amsmath, amsthm}
\usepackage{slashed}
\usepackage{float}
\usepackage{graphicx}
\usepackage[dvips]{epsfig}
\usepackage{psfrag}
\usepackage{lscape}
\usepackage[all]{xy}
\usepackage{hyperref}
\usepackage{enumerate}
\usepackage{dsfont}
\usepackage{color}
\usepackage{mathabx}
\usepackage{mathtools}
\usepackage{framed}

\newcommand{\beq}{\begin{equation}}
\newcommand{\eeq}{\end{equation}}
\def\bea#1\eea{\begin{align}#1\end{align}}
\newcommand{\nn}{\nonumber}
\newcommand{\w}{\wedge}
\newcommand{\id}{\mathds{1}}
\newcommand{\ov}{\overline}

\def\del {\partial}
\def\d {{\rm d}}
\def\R {\mathcal{R}}
\def\L {\mathcal{L}}
\def\tL {\tilde{\mathcal{L}}}
\def\hhh {\mathcal{H}}

\def\tg {\tilde{g}}
\def\b {\beta}
\def\tp {\tilde{\phi}}
\def\te {\tilde{e}}

\def\Ga{\Gamma}
\def\p {\phi}

\def\cR {\widecheck{\cal R}}
\def\cG {\widecheck{\Gamma}}
\def\cN {\widecheck{\nabla}}
\def\g {\gamma}
\def\G {\Gamma}
\def\o {\omega}
\def\O {\Omega}

\def\hO {\hat{\O}}
\def\eps {\epsilon}

\def\La {\Lambda}
\def\N {\nabla}

\def\T {\mathcal {T}}

\def\eee {\mathcal{E}}
\def\teee {\tilde{\mathcal{E}}}
\def\mmm {\mathcal{M}}
\def\reee {\mathring{\mathcal{E}}}

\def\D {\mathcal{D}}
\def\w {\wedge}
\def\NS {N\!S}
\def\KK {K\!K}

\makeatletter
\@addtoreset{equation}{section}
\makeatother

\begin{document}

\begin{titlepage}

\begin{center}

\rightline{\small AEI-2014-004}

\rightline{\small MPP-2014-34}

\vskip 3cm

{\fontsize{16.1}{21}\selectfont \noindent\textbf{$\NS$-branes, source corrected Bianchi identities, \\ \vskip 0.25cm and more on backgrounds with non-geometric fluxes} }

\vskip 2.1cm

\textbf{David Andriot${}^{a,b}$, Andr\'e Betz${}^{c}$}

\vskip 0.6cm

\begin{enumerate}[$^a$]
\item \textit{Max-Planck-Institut f\"ur Gravitationsphysik, Albert-Einstein-Institut,\\Am M\"uhlenberg 1, 14467 Potsdam-Golm, Germany}
\vskip 0.1cm
\item \textit{Institut f\"ur Mathematik, Humboldt-Universit\"at zu Berlin, IRIS-Adlershof,\\Zum Gro\ss en Windkanal 6, 12489 Berlin, Germany}
\vskip 0.1cm
\item \textit{Max-Planck-Institut f\"ur Physik,\\F\"ohringer Ring 6, 80805 M\"unchen, Germany}
\end{enumerate}

\vskip 0.2cm

{\small \texttt{david.andriot@aei.mpg.de}, \texttt{abetz@mpp.mpg.de}}

\end{center}

\vskip 2.0cm

\begin{center}
{\bf Abstract}
\end{center}

\noindent In the first half of the paper, we study in details $\NS$-branes, including the $\NS5$-brane, the Kaluza-Klein monopole and the exotic $5_2^2$- or $Q$-brane, together with Bianchi identities for NSNS (non)-geometric fluxes. Four-dimensional Bianchi identities are generalized to ten dimensions with non-constant fluxes, and get corrected by a source term in presence of an $\NS$-brane. The latter allows them to reduce to the expected Poisson equation. Without sources, our Bianchi identities are also recovered by squaring a nilpotent $Spin(D,D) \times \mathbb{R}^+$ Dirac operator. Generalized Geometry allows us in addition to express the equations of motion explicitly in terms of fluxes. In the second half, we perform a general analysis of ten-dimensional geometric backgrounds with non-geometric fluxes, in the context of $\beta$-supergravity. We determine a well-defined class of such vacua, that are non-geometric in standard supergravity: they involve $\beta$-transforms, a manifest symmetry of $\beta$-supergravity with isometries. We show as well that these vacua belong to a geometric T-duality orbit.

\vfill

\end{titlepage}

\tableofcontents

\addtocontents{toc}{\protect\enlargethispage{2\baselineskip}}

\newpage

\section{Introduction and main results}

In the last few years, there has been a renewed interest in the topic of non-geometry and non-geometric fluxes (for reviews see \cite{Wecht:2007wu, Andriot:2011uh, Andriot:2013txa}). The non-geometric backgrounds of string theory exhibit unusual behaviors, leading to new possibilities and opening still fairly unexplored directions. Their study has been conducted from various angles, including world-sheet and CFT approaches, target space constructions such as Double Field Theory (DFT) and its U-duality extensions (for reviews see \cite{Aldazabal:2013sca, Berman:2013eva, Hohm:2013bwa}), ten-dimensional supergravities and Generalized Geometry, and four-dimensional supergravities. We take in this paper the last two points of view, and study the Bianchi identities for NSNS fluxes, the related $\NS$-branes, and properties of further ten-dimensional backgrounds with non-geometric fluxes.

Some four-dimensional gauged supergravities have as gaugings or components of the embedding tensor the so-called non-geometric fluxes \cite{Shelton:2005cf, Dabholkar:2002sy, Dabholkar:2005ve}. In the NSNS sector, those are given by $Q_c{}^{ab}$ and $R^{abc}$. These $Q$- and $R$-fluxes give rise to specific terms in the four-dimensional potential that are of phenomenological interest. They were shown in various examples to help in stabilising moduli \cite{Shelton:2006fd, Micu:2007rd, Palti:2007pm} or in obtaining de Sitter vacua \cite{deCarlos:2009qm, Danielsson:2012by, Blaback:2013ht, Damian:2013dq, Damian:2013dwa, Catino:2013syn}. Then, it is natural to ask whether such configurations with non-zero $Q$- and $R$-fluxes can be obtained as backgrounds of string theory. To answer this question, we follow here the approach of flux compactifications, that considers dimensional reductions from ten- to four-dimensional supergravity on an internal compact manifold $\mmm$. Traditionally, four-dimensional vacua with $Q$- and $R$-fluxes are then rather believed to uplift to non-geometric backgrounds, where $\mmm$ can be a non-geometry. In these backgrounds, stringy symmetries such as T-duality are used instead of diffeomorphisms or gauge transformations \cite{Hellerman:2002ax, Dabholkar:2002sy, Flournoy:2004vn} (a more precise definition is given in section \ref{sec:GNG}). This results mostly in non-standard spaces for $\mmm$, on which the compactification procedure cannot be applied. The relation between these four- and ten-dimensional perspectives looks thus not well established.

Progress on these aspects have been made recently thanks to local reformulations of standard supergravity into new ten-dimensional theories, in \cite{Andriot:2011uh, Andriot:2012wx, Andriot:2012an, Andriot:2013xca} and \cite{Blumenhagen:2012nk, Blumenhagen:2012nt, Blumenhagen:2013aia}. This is achieved in the NSNS sector, with the standard Lagrangian $\L_{{\rm NSNS}}$ \eqref{LNSNS}, by redefining the metric $g_{mn}$, $b$-field $b_{mn}$ and dilaton $\p$ into a new set of fields $\tg_{mn},\ \b^{mn},\ \tp$, where $\b$ is an antisymmetric bivector. As a consequence, the standard $H$-flux is traded for two new fluxes, identified as the ten-dimensional $Q$- and $R$-fluxes. Their definition depends on the theory, and we follow here $\b$-supergravity \cite{Andriot:2013xca}, where
\beq
Q_{c}{}^{ab} = \del_c \b^{ab} - 2 \b^{d[a} f^{b]}{}_{cd}\ ,\quad R^{abc} = 3 \b^{d[a}\N_d \b^{bc]} \ , \label{fluxesintro}
\eeq
as in \cite{Grana:2008yw, Aldazabal:2011nj, Andriot:2012an}.\footnote{Throughout the paper, $a \dots l$ denote tangent space flat indices and $m \dots z$ curved space indices. The structure constant or geometric flux $f^{a}{}_{bc}$ is defined in \eqref{fabc} and we refer to appendix \ref{ap:conv} for more conventions.} The Lagrangian of the NSNS sector of $\b$-supergravity is given by
\bea
\tL_{\b}  = e^{-2d} \ \bigg( & \R(\tg) +4(\del \tp)^2 + 4 (\b^{ab}\del_b \tp - \T^a)^2 - \frac{1}{2} \eta_{ab} R^{acd} f^b{}_{cd} -\frac{1}{12} \eta_{ad} \eta_{be} \eta_{cf} R^{abc} R^{def} \label{Lbetaintro} \\
&\ \ + 2 \eta_{ab} \b^{ad}\del_d Q_c{}^{bc} - \eta_{cd} Q_a{}^{ac} Q_b{}^{bd} - \frac{1}{2} \eta_{cd} Q_a{}^{bc} Q_b{}^{ad} - \frac{1}{4} \eta^{ad} \eta_{be} \eta_{cf} Q_a{}^{bc} Q_d{}^{ef} \bigg) \ , \nn
\eea
as detailed in section \ref{sec:betasugra}. It looks very similar to the four-dimensional scalar potential of gauged supergravities with $Q$- and $R$-fluxes. So $\b$-supergravity appears to be a good candidate to uplift four-dimensional gauged supergravities, as argued in \cite{Andriot:2013xca}. A dimensional reduction on a concrete background can only be performed though, if at least, the metric $\tg$ describes a standard manifold. Fortunately, this reformulation of standard supergravity not only provides ten-dimensional non-geometric fluxes, but it also transforms in some examples a non-geometry given by $g$ into a standard geometry described by $\tg$. The information on the former non-geometry gets encoded in the new non-geometric fluxes. This reformulation allows eventually to relate these backgrounds properly to the four-dimensional description.\\

Using $\b$-supergravity, one can now study backgrounds with non-geometric fluxes directly in ten dimensions; this is the main purpose of this paper. In a first half, we focus on Bianchi identities (BI) for the NSNS fluxes, and how they are corrected on specific backgrounds corresponding to $\NS$-branes. The corrections show that these branes actually source those fluxes. In a second half, we make a generic study of (the NSNS sector of) geometric backgrounds of $\b$-supergravity, and try to determine whether those lead to new physics.

While the BI bring constraints to be satisfied by the vacua, the equations of motion should be verified in the first place. Those were derived in \cite{Andriot:2013xca} in curved indices. We rewrite them here in flat indices, so that $f^a{}_{bc}$ and $Q_c{}^{ab}$ appear: this simplifies the study of solutions. We use two methods for this rewriting: a direct reformulation, and a Generalized Geometry approach, following \cite{Coimbra:2011nw}. The non-trivial result is the $\b$ equation of motion given in \eqref{beomflat2}.

\subsubsection*{Bianchi identities and $\NS$-branes}

We study in section \ref{sec:BI} a particular type of backgrounds: the $\NS$-branes. The $\NS5$-brane is a codimension $4$ brane and a known vacuum of standard supergravity. Smearing it along one direction and T-dualising leads to the Kaluza-Klein (KK) monopole, that can be viewed as a codimension $3$ brane. The latter is a solution of general relativity, and as such, it is a vacuum of both standard supergravity and $\b$-supergravity. Smearing it once and T-dualising again leads finally to the $5^2_2$-brane \cite{deBoer:2010ud, deBoer:2012ma} (the former two were denoted there $5^0_2$ and $5^1_2$), that we prefer to call here the $Q$-brane \cite{Hassler:2013wsa}. This brane is codimension $2$. It appears in terms of standard supergravity as a non-geometric background \cite{deBoer:2010ud, deBoer:2012ma}, but a geometric description is restored in $\b$-supergravity \cite{Hassler:2013wsa, Geissbuhler:2013uka}. We verify in appendix \ref{ap:Qbranevac} that it satisfies the $\b$-supergravity equations of motion. More details on these branes, their smearing and T-duality relations, are given in section \ref{sec:Tbranes}.

The BI of supergravity fluxes can get corrected in presence of a brane: the latter provides a source term. The resulting equation usually boils down to the Poisson equation on a warp factor. Let us recall the standard case of the $H$-flux with an $\NS5$-brane, before presenting our extensions to the other NSNS fluxes and branes. The BI for the $H$-flux is given by the four-form $\d H$. In terms of its coefficient in flat indices, the BI, in presence of an $\NS5$-brane, is written
\beq
\NS5 {\rm -brane:}\qquad \del_{[a} H_{bcd]} - \frac{3}{2} f^e{}_{[ab} H_{cd]e} = \frac{C_H}{4}\ \epsilon_{4\bot abcd} \ \delta^{(4)} (r_4) \ .\label{BINS5intro}
\eeq
The right-hand side (RHS) localises the brane in its four transverse directions (as indicated by the ${}_{\bot}$) at the radius $r_4=0$. The factor $C_H$ will be specified in the paper, and conventions on the $\epsilon_4$ are given in appendix \ref{ap:conv}. With the fluxes of the $\NS5$-brane background, \eqref{BINS5intro} becomes the Poisson equation on the warp factor $f_H$ (with a normalisation constant $c_H$)
\beq
\Delta_4 f_H = c_H\ \delta^{(4)} (r_4) \ ,\label{PoissonNS5intro}
\eeq
as we will verify explicitly. Another BI that the background should satisfy is given below by equation \eqref{delf}. This condition is obtained either by considering $\d^2 =0$ in flat indices (more precisely $\d(\d e^a)=0$), or from the Jacobi identity of the Lie bracket on Cartan one-forms \cite{Blumenhagen:2012pc}, or the first BI of the Riemann tensor. This BI \eqref{delf} on the geometric flux is automatically satisfied when expressing $f$ in terms of vielbeins. The BI for the $H$-flux without source behaves similarly: $\d H$ vanishes when replacing $H$ by $\d b$. This property holds if the fields have no singularity. A source is responsible for a singular point, hence the RHS in \eqref{BINS5intro} and \eqref{PoissonNS5intro}. These two equations still vanish locally at any point away from the source. We will recover the same behaviour in what follows. Finally, the two BI without a source verify another important property: they are recovered by setting to zero the square of the "derivative" $\d - H\w$ acting on a form $A$. We can as well introduce a dilaton factor, and write
\beq
{\rm For} \ \D A= 2 e^{\p} (\d - H\w) (e^{-\p} A)\ ,\qquad \D^2=0\ \Leftrightarrow\ \d^2=0 \ {\rm and}\ \d H=0 \ . \label{Dintro}
\eeq

For constant $H_{abc}$ and $f^{a}{}_{bc}$, their BI without source can also be obtained from the Jacobi identities of some algebra. This algebra can be extended to the gauging algebra of four-dimensional gauged supergravity: it then includes all NSNS fluxes
\bea
& [Z_a, Z_b]= H_{abc} X^c + f^c{}_{ab} Z_c \label{algebra}\\
& [Z_a, X^b]= -f^b{}_{ac} X^c + Q_a{}^{bc} Z_c \nn\\
& [X^a, X^b]= Q_c{}^{ab} X^c - R^{abc} Z_c \nn \ .
\eea
The Jacobi identities of \eqref{algebra}, given in \eqref{BIstw1} - \eqref{BIstw5}, were thus proposed as the BI for constant NSNS fluxes (without source) \cite{Shelton:2005cf}. For a vanishing $H$-flux, we propose here a ten-dimensional generalization of those, for non-constant fluxes
\bea
\del_{[b}f^{a}{}_{cd]} - f^{a}{}_{e[b}f^{e}{}_{cd]} = 0&\ ,\label{delf}\\
\del_{[c}Q_{d]}{}^{ab}-\b^{e[a}\del_e f^{b]}{}_{cd} - \frac{1}{2}Q_{e}{}^{ab}f^{e}{}_{cd} + 2Q_{[c}{}^{e[a}f^{b]}{}_{d]e} = 0&\ ,\label{delQ}\\
\del_d R^{abc} -3\b^{e[a}\del_e Q_{d}{}^{bc]} + 3R^{e[ab}f^{c]}{}_{de} - 3Q_{d}{}^{e[a}Q_{e}{}^{bc]} = 0&\ ,\label{delR}\\
\b^{e[a}\del_e R^{bcd]} + \frac{3}{2}R^{e[ab}Q_{e}{}^{cd]} = 0&\ .\label{bdelR}
\eea
It is worth stressing that for $H=0$ and constant fluxes, our BI boil down to those of \cite{Shelton:2005cf}. Such a generalization was already obtained in \cite{Blumenhagen:2012pc} from Jacobi identities of Lie brackets, and at the level of DFT in \cite{Geissbuhler:2013uka}. We show in appendix \ref{ap:BIlit} that those match the simpler expressions given by our BI \eqref{delf} - \eqref{bdelR}. These equations are meaningful in $\b$-supergravity, where fluxes can be expressed in terms of vielbeins and $\b$. Interestingly, using these explicit local expressions, the four BI are then automatically satisfied, exactly as above for $\d H$. This is actually how these four conditions were discovered in \cite{Andriot:2013xca} (see appendix C.3). These BI are therefore natural candidates to have non-zero RHS in the presence of $\NS$-branes. We propose indeed the following BI for the geometric flux $f$ in presence of a $\KK$-monopole (see also \cite{Villadoro:2007yq})
\beq
\KK {\rm -monopole:}\qquad \del_{[b}f^{a}{}_{cd]} - f^{a}{}_{e[b}f^{e}{}_{cd]}= \frac{C_K}{3}\ \epsilon_{3\bot bcd}\ \epsilon_{1|| e}\ \eta^{ea} \ \delta^{(3)} (r_3) \ , \label{BIsourcedKKm}
\eeq
where $\epsilon_{1|| e}$ is non-zero and equal to one for $e$ being the direction along the brane, and the factor $C_K$ will be specified in the paper. All other BI should as well be satisfied with a vanishing RHS. In presence of a $Q$-brane, we propose the following BI for the $Q$-flux
\bea
& Q {\rm -brane:} \label{BIsourcedQbrane}\\
& \quad \del_{[c}Q_{d]}{}^{ab}-\b^{e[a}\del_e f^{b]}{}_{cd}-\frac{1}{2}Q_{e}{}^{ab}f^{e}{}_{cd}+2Q_{[c}{}^{e[a}f^{b]}{}_{d]e} = \frac{C_Q}{2}\ \epsilon_{2\bot cd}\ \epsilon_{2|| ef}\ \eta^{ea} \eta^{fb}\ \delta^{(2)} (r_2) \ ,\nn
\eea
and all other BI should again be satisfied with a vanishing RHS. We will verify that these sourced BI boil down to Poisson equations on warp factors once evaluated on the brane solutions
\beq
\KK {\rm -monopole:}\quad \Delta_3 f_K = c_K\ \delta^{(3)} (r_3) \ ,\qquad Q {\rm -brane:}\quad \Delta_2 f_Q = c_Q\ \delta^{(2)} (r_2) \ .
\eeq

Similarly to \eqref{Dintro}, a "derivative" $\D_{\sharp}$ was built for constant fluxes \cite{Shelton:2006fd, Ihl:2007ah}, such that $\D_{\sharp}^2=0$ would be equivalent to the (sourceless) BI of the NSNS fluxes \eqref{BIstw1} - \eqref{BIstw5}, i.e. the Jacobi identities of the algebra \eqref{algebra}, together with a further scalar condition \cite{Ihl:2007ah} given in \eqref{Dsharpsquare}. Here we generalize this idea for non-constant fluxes and $H=0$: we introduce a $\D$ such that
\beq
\D^2=0\ \Leftrightarrow\ \mbox{BI}\ \eqref{delf} - \eqref{bdelR}\ + \ \mbox{scalar condition} \ . \label{Dsquareresultintro}
\eeq
As explained in section \ref{sec:BIsourceless}, $\D$ is the Dirac operator associated to the $Spin(D,D) \times \mathbb{R}^+$ covariant derivative $D_A$ that can be built from the Generalized Geometry approach
\beq
\D\Psi=\Ga^{A}D_{A}\Psi=\Big(\Ga^{A}\del_{A}+\frac{1}{4}\hat \Omega_{ABC}\Ga^{ABC}+\frac{1}{2}\hat\Omega_{D}{}^{D}{}_{C}\Ga^{C}\Big)\Psi \ ,\label{Dgenintro}
\eeq
where the $\Ga^A$ satisfy the $Spin(D,D)$ Clifford algebra, and we represent them with forms and contractions using a Clifford map. Similarly, $\Psi$ is a spinor and can be viewed as a polyform. Using the connection coefficients computed in \cite{Andriot:2013xca}, we recover \eqref{Dintro} for standard supergravity with $b$-field, and get for $\b$-supergravity
\beq
\D A= 2 e^{\tp} (\N_a \cdot \te^a \w - \cN^a \cdot \iota_a + \T \vee + R \vee ) (e^{-\tp} A)\ . \label{Dbetaintro}
\eeq
We recall that $\N_a \cdot \te^a \w=\d$, and understand the dot as acting only on the coefficient of the form $A$; we denote by $\iota_a$ or $\vee$ the contractions (see appendix \ref{ap:conv}), and $\cN^a$ is a covariant derivative containing the $Q$-flux \eqref{defof}. This $\D$ of \eqref{Dbetaintro} verifies \eqref{Dsquareresultintro}. We also show that $\D_{\sharp}$ corresponds to the second term in the RHS of \eqref{Dgenintro}, clarifying how our $\D$ generalizes the constant flux situation. An explicit expression for $\D$ in terms of fluxes is given in \eqref{Dtotal}, while tensorial formulations of the BI are discussed around \eqref{Riemanndelf} and \eqref{tensorialformBI}.

\subsubsection*{Geometric vacua of $\b$-supergravity}

In section \ref{sec:geomvac}, we study vacua of $\b$-supergravity more generally. $\b$-supergravity is of particular interest with respect to standard supergravity when its solutions are geometric. As explained above, such backgrounds can provide a ten-dimensional uplift to some four-dimensional solutions of gauged supergravities with non-geometric fluxes. In addition, a geometric vacuum of $\b$-supergravity is non-geometric when expressed in standard supergravity, at least in the examples considered so far. A geometric, target space, description of a non-geometric string background is therefore restored. Those are the two main achievements of $\b$-supergravity. So the first question we study is to determine the conditions for a geometric vacuum of $\b$-supergravity. Two examples (or at least their NSNS sector) are helpful: the $Q$-brane mentioned previously, and the toroidal example studied in details in \cite{Andriot:2011uh, Andriot:2012vb, Andriot:2013xca}. For both, their standard supergravity description is non-geometric, but also T-dual to a geometric one. From a four-dimensional point of view, such backgrounds are then said to be on a geometric (T-duality) orbit. All theories on an orbit are the same, up to a redefinition of the four-dimensional fields. So the theory obtained from the toroidal example does not describe new physics, with respect to the one from the T-dual configuration that is geometric in standard supergravity. The second question is then whether geometric vacua of $\b$-supergravity ever lead to new physics. To address these questions, we pursue the following reasoning:
\begin{enumerate}
\item Consider a field configuration defined on a set of patches of a space. To form a valid vacuum of a theory, these fields should at least glue from one patch to the other with symmetries of that theory. This allows to describe the configuration on all patches with only one theory (here one Lagrangian) \cite{Blumenhagen:2013aia}.

\item A symmetry leaves a Lagrangian invariant up to a total derivative, and the two Lagrangians $\L_{{\rm NSNS}}$ and $\tL_{\b}$ only differ by a total derivative (see section \ref{sec:betasugra}), so they share the same symmetries. These are diffeomorphisms and $b$-field gauge transformations, where the latter can be translated in terms of the fields of $\b$-supergravity \cite{Andriot:2013xca}. Field configurations gluing with such symmetries are geometric for standard supergravity.\footnote{Definitions of geometric and non-geometric field configurations are given in section \ref{sec:GNG}.} They may as well, under some restrictions, be geometric in terms of $\b$-supergravity, but there is no need for such a description, since standard supergravity already gives a proper one \cite{Blumenhagen:2013aia, Andriot:2013xca}.

\item Getting an interesting geometric background of $\b$-supergravity therefore requires other symmetries. This can be achieved by considering a modification, e.g. a restriction, of the theory, that would lead to a symmetry enhancement \cite{Andriot:2013xca}. Here, the restriction to be made is to consider the presence of $N$ isometries. This provides a further symmetry to $\L_{{\rm NSNS}}$ and $\tL_{\b}$, that is T-duality. We prove this in appendix \ref{ap:proofs}.

\item One of the T-duality transformations that brings some novelty is the $\b$-transform. Expressing it in $\b$-supergravity is simple: it results in a constant shift of $\b$ along isometries. The Lagrangian $\tL_{\b}$ is manifestly invariant under such a transformation, as we show in details. In particular, the $Q$- and the $R$-flux are invariant under this symmetry. Field configurations gluing with $\b$-transforms and diffeomorphisms are thus geometric for $\b$-supergravity, and in most cases non-geometric for standard supergravity: this defines a class of interesting geometric vacua of $\b$-supergravity. The two examples mentioned above are of this type.

\item We however show that such vacua (or at least their NSNS sector) are always T-dual to geometric ones for standard supergravity, i.e. they are on a geometric orbit. So they do not give new physics. The converse point of view remains interesting: we know precisely when geometric backgrounds of standard supergravity have non-geometric T-duals that can be described geometrically by $\b$-supergravity. The latter then provides an uplift to some non-geometric points on the four-dimensional orbit. We still list various possibilities beyond the setting just mentioned, that could circumvent the result, and maybe lead to new physics.

\end{enumerate}

The paper is structured as follows. $\b$-supergravity is reviewed in section \ref{sec:betasugra}, with conventions in appendix \ref{ap:conv}. Equations of motion are rewritten in flat indices in section \ref{sec:eom} and appendix \ref{ap:eom}. We then turn to the sourceless BI in section \ref{sec:BIsourceless} and appendix \ref{ap:BI}, where we review the literature and construct the $Spin(D,D) \times \mathbb{R}^+$ covariant derivative and Dirac operator $\D$. We study $\NS$-branes in section \ref{sec:Tbranes} and appendix \ref{ap:Qbrane} by showing their smearing and T-duality relations, the source corrections to BI and the derivation of Poisson equations. Finally, we detail in section \ref{sec:sym} and appendix \ref{ap:proofs} the symmetries of $\L_{{\rm NSNS}}$ and $\tL_{\b}$, including the T-duality for $N$ isometries. We study how using them leads to geometric or non-geometric vacua in section \ref{sec:nongeoglobal}. T-duals of some geometric vacua of $\b$-supergravity are analyzed in section \ref{sec:geombcgdorbit}. An outlook is eventually provided in section \ref{sec:Ccl}.

\section{$\b$-supergravity and its equations of motion}

We gave in the Introduction several motivations to consider $\b$-supergravity, a ten-dimensional theory that contains non-geometric $Q$- and $R$-fluxes. In this section, we briefly review this theory by providing the technical material needed in the rest of the paper. We mostly follow \cite{Andriot:2013xca}. Then, we turn to the rewriting of its equations of motion in flat indices.

\subsection{Technical review of $\b$-supergravity}\label{sec:betasugra}

A local reformulation of the NSNS sector of standard supergravity was proposed in  \cite{Andriot:2011uh, Andriot:2012wx, Andriot:2012an}. It is based on a field redefinition transforming the standard NSNS fields into a new metric $\tg_{mn}$, an antisymmetric bivector $\b^{mn}$ and a new dilaton $\tp$
\beq
\begin{drcases} \tg^{-1}=(g+b)^{-1} g (g-b)^{-1}\\ \b= - (g+b)^{-1} b (g-b)^{-1} \end{drcases} \Leftrightarrow (g+b)^{-1}=(\tg^{-1}+\b) \ , \quad e^{-2 \tp} \sqrt{|\tg|} = e^{-2 \p} \sqrt{|g|} \equiv e^{-2d} \ ,\label{fieldredef}
\eeq
where we introduce the quantity $d$. This field redefinition was read-off from a reparametrization of the generalized metric $\hhh$, that usually depends on $g$ and $b$. This is equivalent to choosing another generalized vielbein $\teee$ instead of the usual $\eee$, where $\teee$ depends on the new fields
\bea
& \eee= \begin{pmatrix} e & 0 \\ e^{-T} b & e^{-T} \end{pmatrix} \ , \ \teee= \begin{pmatrix} \te & \te \b \\ 0 & \te^{-T} \end{pmatrix} \ , \ \mathbb{I}= \begin{pmatrix} \eta_D & 0 \\ 0 & \eta_D^{-1} \end{pmatrix} \ , \label{genvielb}\\
& \hhh= \begin{pmatrix} g-b g^{-1} b & -b g^{-1} \\ g^{-1} b & g^{-1} \end{pmatrix} = \eee^T \ \mathbb{I} \ \eee = \teee^T \ \mathbb{I} \ \teee = \begin{pmatrix} \tg & \tg \b \\ - \b \tg &  \tg^{-1}-\b \tg \b \end{pmatrix} \ , \label{fieldredefH}
\eea
where $\hhh$ is a $2D \times 2D$ matrix for a $D$-dimensional space-time, $\eta_D$ denotes the flat metric, and the vielbeins $e$ and $\te$ are associated to the metrics $g=e^T \eta_D e$ and $\tg=\te^T \eta_D \te$. This reparametrization was inspired from earlier Generalized Complex Geometry papers \cite{Grange:2006es, Grange:2007bp, Grana:2008yw}. The field redefinition is then an $O(D-1,1) \times O(1, D-1)$ transformation \cite{Andriot:2013xca}.

The standard NSNS Lagrangian, where $H_{mnp}= 3 \del_{[m}b_{np]}$, is given by
\beq
\L_{{\rm NSNS}}= e^{-2\p} \sqrt{|g|} \left(\R(g) + 4(\del \p)^2 - \frac{1}{2} H^2 \right) \ , \label{LNSNS}
\eeq
with conventions in appendix \ref{ap:conv}. Building on the above, the field redefinition \eqref{fieldredef} performed on $\L_{{\rm NSNS}}$ lead in \cite{Andriot:2013xca} to the Lagrangian $\tL_{\b}$ of the NSNS sector of $\b$-supergravity
\beq
\L_{{\rm NSNS}} (g,b,\p) =  \tL_{\b}(\tg,\b,\tp) + \del(\dots) \ ,
\eeq
up to a total derivative $\del(\dots)$ detailed in section \ref{sec:symgen}. In curved indices, $\tL_{\b}$ is given by
\beq
\tL_{\b} = e^{-2\tp} \sqrt{|\tg|}\ \bigg( \R(\tg) +4(\del \tp)^2 + 4 (\b^{mp}\del_p \tp - \T^m)^2 + \cR(\tg) -\frac{1}{2} R^2 \bigg) \ ,\label{Lb}
\eeq
\bea
\mbox{with}\ \ & \cR=\tg_{mn} \cR^{mn} \ , \ \cR^{mn} = -\b^{pq}\del_q \cG_p^{mn} + \b^{mq}\del_q \cG_p^{pn} + \cG_p^{mn} \cG_q^{qp} - \cG_p^{qm} \cG_q^{pn} \ ,\label{cR}\\
& \cG_p^{mn} = \frac{1}{2}\tg_{pq}\left(-\b^{mr}\del_r \tg^{nq} - \b^{nr}\del_r \tg^{mq} +  \b^{qr}\del_r \tg^{mn} \right) + \tg_{pq} \tg^{r(m} \del_r \b^{n)q}-\frac{1}{2} \del_p \b^{mn}  \ ,\label{cG}\\
& \T^n \equiv \cG_p^{pn} = \del_p \b^{np} - \frac{1}{2} \b^{nm} \tg_{pq} \del_m \tg^{pq} = \frac{1}{\sqrt{|\tg|}} \del_p \left(\b^{np} \sqrt{|\tg|} \right) = \N_p \b^{np} \ , \label{tracecG} \\
& R^{mnp}\equiv 3 \b^{q[m}\del_q \b^{np]} = 3 \b^{q[m}\N_q \b^{np]} \ , \label{fluxes}
\eea
and conventions in appendix \ref{ap:conv}. Note that $R^{mnp}$, $\T^m$ and $\cR^{mn}$ are tensors. This last "Ricci tensor" is related to a new covariant derivative $\cN^m$ built from $\b^{mn} \del_n$ and the connection $\cG^{mn}_p$
\beq
\cN^m V^p = -\b^{mn} \del_n V^p - \cG^{mp}_n V^n \ , \ \cN^m V_p = -\b^{mn} \del_n V_p + \cG^{mn}_p V_n \ . \label{defcN}
\eeq
That derivative plays a crucial role, as we will see. Another useful tensor is $\cG_{\!\!(t)}{}_p^{mn}$
\beq
\cG^{mn}_p= \cG_{\!\!(t)}{}_p^{mn} + \b^{ms} \G^n_{ps} \ , \qquad \cG_{\!\!(t)}{}_p^{mn}= \frac{1}{2} \tg_{pq} \left( \tg^{rm} \N_r \b^{nq} + \tg^{rn} \N_r \b^{mq} - \tg^{qr} \N_r \b^{mn} \right) \ . \label{cG_t}
\eeq
It allows to relate the covariant derivatives $\N$ and $\cN$, and then to rewrite the $R$-flux
\bea
& \cN^m V^p = -\b^{mn} \N_n V^p - \cG_{\!\!(t)}{}^{mp}_n V^n \ , \ \cN^m V_p = -\b^{mn} \N_n V_p + \cG_{\!\!(t)}{}^{mn}_p V_n \ ,\label{relnacN}\\
& R^{mnp} = 3\ \b^{q[m}\N_q \b^{np]} = \frac{3}{2}\ \cN^{[m} \b^{np]} \ . \label{RfluxcN}
\eea
Imposing the condition $\b^{mn} \del_n \cdot =0$ (as well as $\del_p \b^{np}=0$), where the dot stands for any field, reduces $\tL_{\b}$ to the Lagrangian obtained in \cite{Andriot:2011uh}. One gets $R^{mnp}=0$, $\T^m=0$, and $\cR$ results only in a $(\del \b)^2$. This subcase is useful in some examples, like the $Q$-brane: see appendices \ref{ap:relbetadel} and \ref{ap:Qbranevac}.

We now turn to flat indices: this reveals the $Q$-flux given in \eqref{fluxesintro}, since it is not a tensor. It rather plays an analogous role in $\cN$ as $f$ does in $\N$ with Levi-Civita connection \cite{Andriot:2013xca}
\bea
\!\!\!\!\!\!\!\!\! \te^a{}_m \te^n{}_b \N_n V^m = \N_b V^a \equiv \del_b V^a + \o^a_{bc} V^c \Leftrightarrow \ & \o^a_{bc} = \frac{1}{2} \left(f^{a}{}_{bc} + \eta^{ad} \eta_{ce} f^{e}{}_{db} + \eta^{ad} \eta_{be} f^{e}{}_{dc} \right)  \label{defof}\\
\!\!\!\!\!\!\!\!\! \te^m{}_a \te^b{}_n \cN^n V_m = \cN^b V_a \equiv -\b^{bd}\del_d V_a - {\o_Q}^{bc}_a V_c \Leftrightarrow \ & {\o_Q}^{bc}_a = \frac{1}{2} \left(Q_{a}{}^{bc} + \eta_{ad} \eta^{ce} Q_{e}{}^{db} + \eta_{ad} \eta^{be} Q_{e}{}^{dc} \right)  \ ,\nn
\eea
where we introduced $\o_Q$, (the opposite of) the spin connection associated to $\cG$. This $\o_Q$ enjoys similar properties as those of \eqref{prop}
\beq
\eta_{dc} {\o_Q}_a^{bc} = - \eta_{ac} {\o_Q}_d^{bc} \ , \ Q_{a}{}^{bc} = 2 {\o_Q}_a^{[bc]} \ , \ {\o_Q}_a^{ad} = Q_a{}^{ad}  \ ,\ \eta_{bc} {\o_Q}_a^{bc} = \eta_{ad} Q_b{}^{db} \ . \ \label{troQ}
\eeq
From it, we can define a quantity $\R_Q$ analogous to the standard Ricci scalar $\R(\tg)$
\bea
& \R(\tg)=2 \eta^{bc} \del_a \o^a_{bc} + \eta^{bc} \o^a_{ad} \o^d_{bc} - \eta^{bc} \o^a_{db} \o^d_{ac} \label{Ricflat}\\
& \qquad \qquad \qquad \qquad \qquad = 2 \eta^{ab} \del_a f^c{}_{bc} - \eta^{cd} f^a{}_{ac} f^b{}_{bd} - \frac{1}{4} \left( 2 \eta^{cd} f^a{}_{bc} f^b{}_{ad} + \eta_{ad} \eta^{be} \eta^{cg} f^a{}_{bc} f^d{}_{eg} \right) \ ,\nn \\
& \R_Q \equiv 2 \eta_{bc} \b^{ad} \del_d {\o_Q}_a^{bc} + \eta_{bc} {\o_Q}_a^{ad} {\o_Q}_d^{bc} - \eta_{bc} {\o_Q}_a^{db} {\o_Q}_d^{ac} \label{defRQ} \\
& \qquad \qquad \qquad \qquad = 2 \eta_{ab} \b^{ad}\del_d Q_c{}^{bc} - \eta_{cd} Q_a{}^{ac} Q_b{}^{bd} - \frac{1}{4} \left( 2 \eta_{cd} Q_a{}^{bc} Q_b{}^{ad} + \eta^{ad} \eta_{be} \eta_{cg} Q_a{}^{bc} Q_d{}^{eg} \right) \ ,\nn
\eea
and $\R_Q$ is related to $\cR$ as follows
\beq
\cR= \R_Q - \frac{1}{2} R^{acd} f^b{}_{cd} \eta_{ab}\ . \label{cRRqRf}
\eeq
The Lagrangian $\tL_{\b}$ \eqref{Lb} can then be rewritten as in \eqref{Lbetaintro}, where the four terms in $Q$ match $\R_Q$. Finally, let us give a few useful expressions, such as $\cR^{ab}$ in \eqref{cRtensor}, and
\bea
& 2\ \R_{cd} = \del_a f^a{}_{cd} + 2 \eta^{ab} \del_a f^g{}_{b(c} \eta_{d)g} - 2 \del_c f^b{}_{bd} \label{Ricciflat} \\
&\phantom{2\ \R_{cd} } + f^a{}_{ab} \left(f^b{}_{cd} + 2 \eta^{bg} f^h{}_{g(c} \eta_{d)h} \right) - f^b{}_{ac} f^a{}_{bd} - \eta^{bg} \eta_{ah} f^h{}_{gc} f^a{}_{bd} - \frac{1}{2} \eta^{ah}\eta^{bj}\eta_{ci}\eta_{dg} f^i{}_{ja} f^g{}_{hb} \ ,\nn\\
& R^{abc}= 3 \b^{d[a} \del_d \b^{bc]} - 3 \b^{d[a} f^b{}_{de} \b^{c]e} = 3 \b^{d[a} Q_d{}^{bc]} + 3 \b^{d[a} f^b{}_{de} \b^{c]e} \ ,\ \T^a = -Q_b{}^{ba}+\frac{1}{2}\beta^{cd}f^{a}{}_{cd} \ .\nn
\eea

We rederived in \cite{Andriot:2013xca} the Lagrangian $\tL_{\b}$ \eqref{Lbetaintro} and most of the structures just presented (in particular $\cN$ and $\o_Q$) from the Generalized Geometry formalism, building on \cite{Coimbra:2011nw}. Choosing the generalized vielbein $\teee$ in \eqref{genvielb} plays a crucial role for this purpose. We recall some results of this derivation in section \ref{sec:eom}, and use them in section \ref{sec:Spinder} to compute the $Spin(D,D) \times \mathbb{R}^+$ covariant derivative. In addition, $\beta$-supergravity can be derived from DFT \cite{Andriot:2012wx, Andriot:2012an, Aldazabal:2011nj, Geissbuhler:2013uka}.

Finally, the equations of motion for the NSNS sector of $\b$-supergravity were derived in \cite{Andriot:2013xca}
\bea
 & \! \frac{1}{4} \left(\R(\tg) + \cR(\tg) -\frac{1}{2} R^2 \right) = (\del \tp)^2 - \N^2 \tp + (\b^{mr}\del_r \tp - \T^m)^2 + \tg_{mn}\cN^m(\b^{nr}\del_r \tp - \T^n) \label{dileom}\\
 & \!\!\!\!\!\!\!\!\!\!\!\!\! \R_{pq} - \tg_{m(p} \tg_{q)n} \cR^{mn} + \frac{1}{4} \tg_{pm} \tg_{qn} \tg_{rs} \tg_{uv} R^{mru} R^{nsv} = - 2 \N_p \del_q \tp  - 2 \tg_{m(p} \tg_{q)n} \cN^m(\b^{nr} \del_r \tp - \T^n) \label{Einstein}\\
&\! \frac{1}{2} \tg_{ms} \tg_{ru} \tg_{np}  \left( e^{2\tp} \cN^m (e^{-2\tp}  R^{sun}) - 2 \T^m R^{sun}\right) \label{beom} \\
& \!\!\!\!\!\!\!\!\! = \frac{1}{2} \tg_{np} \tg_{rq} \tg^{sm} e^{2\tp} \N_m (e^{-2\tp} \N_s \b^{nq} ) + 2 \tg_{n[p} \R_{r]s} \b^{ns} - e^{-2\tp} \N_q ( e^{2\tp} \tg_{n[p} \N_{r]} \b^{nq}) + 4 \tg_{n[p} \N_{r]} (\b^{nq} \del_q \tp )  \nn \ .
\eea
Those are given in curved indices. We now turn to their rewriting in flat indices.

\subsection{Equations of motion in flat indices and Generalized Geometry formalism}\label{sec:eom}

The equations of motion for the NSNS sector of $\b$-supergravity, derived from $\tL_{\b}$ \eqref{Lb} in \cite{Andriot:2013xca}, have just been given: the one for the dilaton \eqref{dileom}, the Einstein equation \eqref{Einstein}, and the $\b$ equation of motion \eqref{beom}. They are in curved indices; in this section, we rewrite them with flat indices: this allows to make the fluxes $f^a{}_{bc}$ and $Q_c{}^{ab}$ appear, since those are not tensors. Having an explicit dependence on the fluxes is more convenient when looking for solutions. It will indeed be the case in appendix \ref{ap:Qbranevac} when verifying that the $Q$-brane is a vacuum of $\b$-supergravity. To perform this rewriting, we follow two methods: first, a direct approach is detailed in appendix \ref{ap:eomdirect}, and secondly we use the Generalized Geometry formalism, building on \cite{Coimbra:2011nw} and the results of \cite{Andriot:2013xca}. This second method is presented below.

Since all terms in the above equations are tensors, going to flat indices is only a multiplication by vielbeins. The difficulty is rather to make the fluxes appear explicitly. For the dilaton and Einstein equations, this essentially amounts to give the expressions of the Ricci scalars and tensors in terms of the fluxes: those can be found in \eqref{Ricflat}, \eqref{defRQ}, \eqref{cRRqRf} for the scalars, and \eqref{Ricciflat}, \eqref{cRtensor} for the tensors. The equation of motion for $\b$ requires more work. Both methods lead to the following result for this equation
\bea
& -\frac{1}{2} \eta_{ab} \eta_{cd} \eta_{ef} \cN^a R^{bdf} + Q_a{}^{gf} f^{a}{}_{g[c} \eta_{e]f} + \frac{1}{2} f^{f}{}_{ha} Q_{[c}{}^{ha} \eta_{e]f} -\frac{1}{2} Q_{a}{}^{ag} f^{i}{}_{ec} \eta_{gi}  \label{beomflat2} \\
& + \frac{1}{2} \eta_{ef} \eta_{cd} \eta^{gk} Q_g{}^{fd} f^{a}{}_{ak} + \eta_{gi} \eta^{ab} Q_a{}^{dg} f^{i}{}_{b[e} \eta_{c]d} \nn\\
= &\ - \frac{1}{2} \eta_{gi} \b^{ga} \del_a f^{i}{}_{ce} - \b^{df} \del_{d} f^{a}{}_{a[c} \eta_{e]f} + \eta_{f[e} \del_{c]} Q_{a}{}^{af} - \frac{1}{2} \eta_{ef} \eta_{cd} \eta^{ab} \del_a Q_b{}^{fd} + 2 \eta_{f[e} \N_{c]} \T^f \nn\\
& + \eta_{ab} \eta_{cd} \eta_{ef} R^{bdf} \left(\b^{ag} \del_g \tp - \T^a \right) + \eta^{ab} \eta_{cd} \eta_{ef} \N_b \b^{fd}\ \del_a \tp + 4 \b^{ab} \eta_{a[c} \N_{e]} \del_b \tp + 2 \eta_{a[c} \N_{e]} \b^{ab}\ \del_b \tp \ . \nn
\eea
Although it looks at first complicated, many terms would drop out upon reasonable assumptions: we argued in \cite{Andriot:2013xca} in favor of an ansatz with $\forall b\ ,\ f^a{}_{ab}=0 \ , \ Q_a{}^{ab}=0 \ , \ \T^b=0  \ ,\ \del_b \tp = 0 $ that would make several terms vanish, e.g. the last row. Finally, let us mention that a complete use of \eqref{beomflat2} would require to expand $\cN^a R^{bdf}$, but the procedure should be straightforward. The resulting terms would not mix with the others, given the number of $\b$.

\subsubsection*{Derivation using the Generalized Geometry formalism}

We presented in \cite{Andriot:2013xca} a useful formulation of $\b$-supergravity based on the formalism of Generalized Geometry, established in \cite{Coimbra:2011nw} for standard type II supergravities. This formulation clarified the origin of the various structures appearing in $\b$-supergravity, including the fluxes, the covariant derivative $\cN^a$, and $\T^a$. It also lead us to reobtain the Lagrangian $\tL_{\b}$ \eqref{Lbetaintro}. Using these tools, we derive here the three equations of motion in flat indices. This amounts to compute generalized quantities analogous to a Ricci scalar and a Ricci tensor.

The starting point of Generalized Geometry is to consider a generalized bundle with structure group $O(D,D) \times \mathbb{R}^+$. Various objects, covariant with respect to this structure group, can then be constructed. The crucial one is the generalized (flat) covariant derivative
\beq
D_A V^B = \del_A V^B + \hO_A{}^{B}{}_{C} V^C\ , \label{genconn}
\eeq
that acts on a generalized vector component $V^B$. To reproduce $\b$-supergravity, we chose a generalized frame related to the generalized vielbein $\teee$ given in \eqref{genvielb}; standard supergravity is rather obtained from $\eee$. Then, using metric compatibility, a constraint on the generalized torsion and some further fixing, we showed in great details in \cite{Andriot:2013xca} how to determine the generalized connection coefficients $\hO_A{}^{B}{}_{C}$ (as well as $\del_A $). Those are essentially given in terms of fluxes. This is analogous to the standard case of the spin connection for Levi-Civita connection. We then restricted the structure group to $O(D-1,1)\times O(1,D-1)$, leading to covariant derivatives with respect to that subgroup. Going to the spinorial version $Spin(D-1,1)\times Spin(1,D-1)$, we obtained as well derivatives on spinors, in particular
\bea
\gamma^a D_a \epsilon^+ & = \left( \gamma^a \N_a - \gamma^a \eta_{ad} \cN^d  + \frac{1}{24} \eta_{ad} \eta_{be} \eta_{cf} R^{def} \gamma^{abc} - \frac{1}{2} \gamma^c \Lambda_c \right) \epsilon^+\ , \label{gDaeps} \\
D_{\ov{a}} \epsilon^+ & = \left( \N_{\ov{a}} + \ov{\eta_{ad}} \cN^{\ov{d}} - \frac{1}{8} \ov{\eta_{ad}} \eta_{be} \eta_{cf} R^{\ov{d}ef} \gamma^{bc} \right) \epsilon^+ \ ,\label{Dabeps} \\
D_{\ov{a}} w^{\ov{a}} & = \N_{\ov{a}} w^{\ov{a}} + \ov{\eta_{ad}} \cN^{\ov{d}} w^{\ov{a}} - \Lambda_{\ov{a}} w^{\ov{a}} \ , \label{traceDwb}\\
D_a w^{\ov{b}} & = \nabla_a w^{\ov{b}} - \eta_{ad} \cN^d w^{\ov{b}} - \frac{1}{2} \eta_{ad} \ov{\eta_{cf}} R^{d\ov{bf}} w^{\ov{c}} \ ,
\eea
where in \eqref{gDaeps} and \eqref{Dabeps}, $\N$ and $\cN$ are the spinorial derivatives naturally defined from \eqref{defof}. Conventions for $\gamma$-matrices are given in appendix \ref{ap:conv}, and the unbarred-barred notation refers to the two orthogonal groups. This notation disappears when choosing aligned vielbeins \cite{Andriot:2013xca}. These derivatives can be rewritten as in \cite{Andriot:2013xca} using only the following quantities
\bea
X_{abc}&= \frac{1}{4} \eta_{be} \left(\omega^e_{ac} - \eta_{ad} {\omega_{Q}}_c^{de} + \frac{1}{6} \eta_{ad} \eta_{cf} R^{def} \right)\ , \\
X_a &= \frac{1}{2} \left( \o^d_{da} + \eta_{ac} {\o_Q}_d^{dc} - \Lambda_a  \right)\ , \\
Y_{\ov{a}bc} & = \frac{1}{4} \eta_{be} \left(\omega^e_{\ov{a}c} + \ov{\eta_{ad}} {\omega_{Q}}_c^{\ov{d}e} - \frac{1}{2} \ov{\eta_{ad}} \eta_{cf} R^{\ov{d}ef} \right)\ , \\
Z_{\ov{a}} & = \o^{\ov{d}}_{\ov{da}} - \ov{\eta_{ac}} {\o_Q}_{\ov{d}}^{\ov{dc}} - \Lambda_{\ov{a}} \ ,\\
&\!\!\!\!\!\!\!\!\!\!\!\! \begin{cases} \Lambda_a = \lambda_a + \eta_{ad} \xi^d \\ \Lambda_{\ov{a}} = \lambda_{\ov{a}} - \ov{\eta_{ad}} \xi^{\ov{d}} \end{cases} \!\! , \quad \lambda_a=2 \del_a \tp \ , \ \xi^{a} =  2 ( \b^{ad}\del_d\tp - \T^a) \ .
\eea

From those derivatives, we calculated in \cite{Andriot:2013xca} the scalar $S$, defined in \cite{Coimbra:2011nw} as
\beq
-\frac{1}{4}S\eps^+= \left( \g^a D_a\g^b D_b- \ov{\eta^{ab}} D_{\ov{a}}D_{\ov{b}} \right)\eps^+ \ .\label{defS}
\eeq
This quantity is related to the Lagrangian, and we reproduced from it $\tL_{\b}$ \eqref{Lbetaintro}. We obtained
\bea
S  = &\ \R(\tg) + \R_Q - \frac{1}{2} R^{acd} f^b{}_{cd} \eta_{ab} -\frac{1}{2} R^2 \label{Sfinalsec}\\
& -4(\del \tp)^2 + 4\N^2 \tp - 4 (\b^{ab}\del_b \tp - \T^a)^2 -4\eta_{ab}\cN^a (\b^{bc}\del_c \tp - \T^b) \  \nn \ .
\eea
In addition, it was shown in \cite{Coimbra:2011nw} to encode the dilaton equation of motion for standard supergravity, by considering $S=0$. Here, we get the analogous result: $S=0$ reproduces the dilaton equation of motion \eqref{dileom} in flat indices.

To derive the two other equations of motion, we calculate the generalized Ricci tensor
\beq
\frac{1}{2} R_{a\ov{b}}\g^a\eps^+=[\g^a D_a,D_{\ov{b}}]\eps^+\ ,\label{defR_ab}
\eeq
that depends on the above derivatives.\footnote{Analogous quantities to $S$ and $R_{a\ov{b}}$ were considered before in \cite{Siegel:1993xq, Siegel:1993th, Hohm:2010pp, Kwak:2010ew, Hohm:2010xe, Jeon:2010rw, Jeon:2011cn}; their relations to the Lagrangian and the equations of motion were as well studied. The DFT quantities were shown in \cite{Hohm:2011nu} to match those of \eqref{defS} and \eqref{defR_ab} for standard supergravity.} For standard supergravity, it was shown in \cite{Coimbra:2011nw} that setting the symmetric part to zero, $R_{(ab)}=0$, corresponds to the Einstein equation, while the antisymmetric part $R_{[ab]}=0$ yields the equation of motion for the $b$-field. In analogy here we should obtain the equations of motion for $\tg$ and $\b$ taking respectively the symmetric or antisymmetric part of $R_{ab}$. Using the quantities defined above, \eqref{defR_ab} becomes
\bea
\frac{1}{2} R_{a\ov{b}}\g^a\eps^+=&\left( \gamma^a \del_a + \gamma^a \eta_{ad} \b^{dc} \del_c + X_{acd} \gamma^{acd} + X_a \gamma^a \right)\left( \del_{\ov{b}} - \ov{\eta_{bg}} \b^{\ov{ge}} \del_{\ov{e}} + Y_{\ov{b}gh} \gamma^{gh} \right) \eps^+ \label{GenRicci1} \\
&-\g^a\o_{a\ov{b}}^{\ov{c}}\left( \del_{\ov{c}} - \ov{\eta_{cg}} \b^{\ov{ge}} \del_{\ov{e}} + Y_{\ov{c}gh} \gamma^{gh} \right) \eps^+ + \g^a\eta_{ad}{\o_Q}^{d\ov{c}}_{\ov{b}}\left( \del_{\ov{c}} - \ov{\eta_{cg}} \b^{\ov{ge}} \del_{\ov{e}} + Y_{\ov{c}gh} \gamma^{gh} \right) \eps^+ \nn \\
&-\frac{1}{2}\g^a \eta_{ad}\ov{\eta_{bf}}R^{d\ov{fc}}\left( \del_{\ov{c}} - \ov{\eta_{cg}} \b^{\ov{ge}} \del_{\ov{e}} + Y_{\ov{c}gh} \gamma^{gh} \right) \eps^+ \nn \\
&-\left( \del_{\ov{b}} - \ov{\eta_{bg}} \b^{\ov{ge}} \del_{\ov{e}} + Y_{\ov{b}gh} \gamma^{gh} \right)\left( \gamma^a \del_a + \gamma^a \eta_{ad} \b^{dc} \del_c + X_{acd} \gamma^{acd} + X_a \gamma^a \right)\eps^+ \ .\nn
\eea
We leave the computational details of the above expression to appendix \ref{ap:GGeom}, and give here the result. After aligning the vielbeins, and considering only the first order in $\gamma$-matrices, $\frac{1}{2}R_{ab}\g^a$ gives
\bea
&\Big(\frac{1}{2} \R_{ba}-\frac{1}{2}  \eta_{e(a}\eta_{b)g} \cR^{ge}+\frac{1}{8} \eta_{ae}\eta_{bg}\eta_{if}\eta_{cd} R^{igc}  R^{dfe} \label{Rab}\\
&\ +\N_{b} \N_a\tp -  \eta_{e(a}\eta_{b)g} \cN^{g}(\cN^{e}\tp)-\eta_{e(a}\eta_{b)g} \cN^{g} \T^e \nn \\
&\ +\frac{1}{4} \eta_{ae}\eta_{bg}\eta^{df}\del_d  Q_{f}{}^{eg}-\frac{1}{2} \eta_{e[a}\del_{b]} Q_d{}^{de} -\frac{1}{4}  \b^{gc} \del_c  f^{e}{}_{ab}\eta_{ge}+\frac{1}{2}   \b^{gc} \del_{c} f^d{}_{d[a}\eta_{b]g} \nn \\
&\ +\frac{1}{4}\eta_{bg}\eta_{ae}\eta^{ch}f^d{}_{dc}Q_h{}^{eg}- \frac{1}{4}\eta_{ch} Q_d{}^{dc}f^{h}{}_{ab}\nn \\
&\ +\frac{1}{4}f^{g}{}_{cd}Q_{[a}{}^{dc}\eta_{b]g}+\frac{1}{2}\eta_{e[a}f^h{}_{b]d}Q_{i}{}^{ec}\eta_{ch} \eta^{di}+\frac{1}{2}\eta_{e[a} f^{h}{}_{b]c} Q_{h}{}^{ec} \nn  \\
&\ -\eta_{e[a}\N_{b]}(\cN^{e}\tp)-\eta_{e[a}\N_{b]}\T^e+  \eta_{g[b} \cN^{g} \N_{a]}\tp \nn \\
&\ - \frac{1}{2}\eta_{ae}\eta_{bg}\eta_{fc} R^{gfe}\T^c+\frac{1}{4} \eta_{ae}\eta_{bg} \eta_{df} e^{2\tp}\cN^{d}(e^{-2\tp} R^{gfe})\Big) \g^{a}\nn\ .
\eea
The first order in $\gamma^a$ will be enough to recover the equations of motion derived above, i.e. the higher orders in $\gamma^a$ should vanish, as they did for $S$ \cite{Andriot:2013xca}.

As explained above, setting $R_{ab}=0$ and therefore the expression \eqref{Rab} to vanish, we should obtain the equations of motion for $\tg$ and $\b$. More precisely, setting the symmetric part of \eqref{Rab} to vanish gives
\bea
&\frac{1}{2} \R_{ba}-\frac{1}{2}  \eta_{e(a}\eta_{b)g} \cR^{ge}+\frac{1}{8} \eta_{ae}\eta_{bg}\eta_{if}\eta_{cd} R^{igc}  R^{dfe} \label{Einsteinflat}\\
+& \N_{b} \N_a\tp -  \eta_{e(a}\eta_{b)g} \cN^{g}(\cN^{e}\tp)-\eta_{e(a}\eta_{b)g} \cN^{g} \T^e=0 \nn\ ,
\eea
that matches the Einstein equation \eqref{Einstein}. Similarly, the antisymmetric part of \eqref{Rab} gives
\bea
&\frac{1}{4} \eta_{ae}\eta_{bg}\eta^{df}\del_d  Q_{f}{}^{eg}-\frac{1}{2} \eta_{e[a}\del_{b]} Q_d{}^{de}-\frac{1}{4}  \b^{gc} \del_c  f^{e}{}_{ab}\eta_{ge}+\frac{1}{2}   \b^{gc} \del_{c} f^d{}_{d[a}\eta_{b]g} \label{beomflat3}\\
+&\frac{1}{4}\eta_{bg}\eta_{ae}\eta^{ch}f^d{}_{dc}Q_h{}^{eg}-\frac{1}{4}\eta_{ch} Q_d{}^{dc}f^{h}{}_{ab} \nn \\
+&\frac{1}{4}f^{g}{}_{cd}Q_{[a}{}^{dc}\eta_{b]g}+\frac{1}{2}\eta_{e[a}f^h{}_{b]d}Q_{i}{}^{ec}\eta_{ch} \eta^{di}+\frac{1}{2}\eta_{e[a} f^{h}{}_{b]c} Q_{h}{}^{ec} \nn  \\
-&\eta_{e[a}\N_{b]}(\cN^{e}\tp)-\eta_{e[a}\N_{b]}\T^e+  \eta_{g[b} \cN^{g} \N_{a]}\tp \nn \\
-& \frac{1}{2}\eta_{ae}\eta_{bg}\eta_{fc} R^{gfe}\T^c+\frac{1}{4} \eta_{ae}\eta_{bg} \eta_{df} e^{2\tp}\cN^{d}(e^{-2\tp} R^{gfe})=0 \nn \ .
\eea
This last result matches \eqref{beomflat2}, the equation of motion for $\b$ in flat indices.\footnote{To verify this, one should multiply the equation \eqref{beomflat3} by $2$ and match its indices $(a,b)$ with those $(e,c)$ of \eqref{beomflat2}. In addition, one can use \eqref{relnacN} and \eqref{cG_t} on the term in $ \cN \N \tp$.}

\section{Bianchi identities and $\NS$-branes}\label{sec:BI}

\subsection{NSNS Bianchi identities without sources}\label{sec:BIsourceless}

In this section, we first review the appearance of NSNS Bianchi identities (BI) through the literature. As mentioned in the Introduction, the BI in the NSNS sector have been treated in different ways. We recall approaches based on algebras with various brackets, that eventually lead to the BI using their Jacobi identities. The BI have also been derived from a nilpotency condition on generalizations of the standard exterior derivative, where including the geometric and non-geometric fluxes plays an important role. We will then make use of these ideas, and rederive the BI \eqref{delf} - \eqref{bdelR} by considering the square of a $Spin(D,D)\times \mathbb R^{+}$ derivative.

\subsubsection{Sourceless NSNS Bianchi identities through the literature}\label{sec:BIsourcelesslit}

In the Introduction, we gave our BI for the NSNS fluxes in the absence of source \eqref{delf} - \eqref{bdelR}. We repeat them here for convenience
\bea
\del_{[b}f^{a}{}_{cd]} - f^{a}{}_{e[b}f^{e}{}_{cd]} = 0&\ ,\label{delf2}\\
\del_{[c}Q_{d]}{}^{ab}-\b^{e[a}\del_e f^{b]}{}_{cd} - \frac{1}{2}Q_{e}{}^{ab}f^{e}{}_{cd} + 2Q_{[c}{}^{e[a}f^{b]}{}_{d]e} = 0&\ ,\label{delQ2}\\
\del_d R^{abc} -3\b^{e[a}\del_e Q_{d}{}^{bc]} + 3R^{e[ab}f^{c]}{}_{de} - 3Q_{d}{}^{e[a}Q_{e}{}^{bc]} = 0&\ ,\label{delR2}\\
\b^{e[a}\del_e R^{bcd]} + \frac{3}{2}R^{e[ab}Q_{e}{}^{cd]} = 0&\ .\label{bdelR2}
\eea
Let us first make a few remarks on them. As mentioned in the Introduction, the conditions \eqref{delf2} - \eqref{bdelR2} are actually identities: they hold automatically if one uses the definitions of the fluxes, and this is how we obtained them in the first place in \cite{Andriot:2013xca} (see appendix C.3). Moreover, \eqref{delf2} corresponds to the first BI of the Riemann tensor, as can be seen in the following equalities derived from the torsionless Cartan equations
\beq
\frac{1}{2}\R^{a}{}_{[bcd]}=\del_{[c}\o^{a}_{db]} - \frac{1}{2} \o^{a}_{e[b}f^{e}{}_{cd]}+\o^{e}_{[cd}\o^{a}_{b]e}= \frac{1}{2}\left( \del_{[c}f^{a}{}_{db]}+ f^{e}{}_{[cd}f^{a}{}_{b]e}\right) \ ,\label{Riemanndelf}
\eeq
or using $\d (\d \te^a)$. Similarly, \eqref{delR2} should correspond to the BI for the Riemann tensor associated to $\cR$, given in (3.44) or (3.47) of \cite{Andriot:2012an}. Finally, \eqref{bdelR2} can be derived from $\cN^{[m} R^{npq]}=0$ obtained in \cite{Andriot:2012wx, Andriot:2012an}. The case of \eqref{delQ2} is discussed around \eqref{tensorialformBI}. Let us now review the appearance of NSNS BI in the literature and draw a connection to the above relations.

\subsubsection*{Algebraic interpretation}

This approach is based on having an algebra where the geometric and non-geometric fluxes appear as structure constants; the NSNS BI are then obtained by considering the Jacobi identities of the algebra. This idea first appeared for standard geometric backgrounds: the algebra was that of the gaugings of four-dimensional gauged supergravity, and the generators $Z$ and $X$ were understood as descending from ten-dimensional ones, for diffeomorphisms and $b$-field gauge transformation respectively \cite{Kaloper:1999yr, Derendinger:2004jn, Dall'Agata:2005ff, Hull:2005hk}. For T-duality covariance in four dimensions, this algebra was extended towards the famous one \eqref{algebra} to include non-geometric fluxes \cite{Shelton:2005cf, Dabholkar:2005ve}.\footnote{Our conventions differ by a minus sign on the $R$-flux with those of \cite{Shelton:2005cf}.} A further extension was considered in \cite{Aldazabal:2008zza} to include other sectors of supergravities. As mentioned already, the Jacobi identities of the algebra \eqref{algebra} generate the following set of NSNS BI \cite{Shelton:2005cf}
\bea
f^e{}_{[ab} H_{cd]e} = 0 & \label{BIstw1}\\
H_{e[bc}Q_{d]}{}^{ae} + f^{a}{}_{e[b}f^{e}{}_{cd]} = 0&\label{BIstw2}\\
\frac{1}{2} H_{ecd} R^{abe} - \frac{1}{2}Q_{e}{}^{ab}f^{e}{}_{cd} + 2Q_{[c}{}^{e[a}f^{b]}{}_{d]e} = 0&\label{BIstw3}\\
R^{e[ab}f^{c]}{}_{de} - Q_{d}{}^{e[a}Q_{e}{}^{bc]} = 0&\label{BIstw4}\\
R^{e[ab}Q_{e}{}^{cd]} = 0&\label{BIstw5}
\eea
Setting the $H$-flux to vanish, one can see that these BI exactly match our relations \eqref{delf2} - \eqref{bdelR2} for constant fluxes. Our BI can thus be thought of as a generalization when fluxes are not constant.\footnote{It was argued in \cite{Shelton:2005cf} that the BI \eqref{BIstw1}-\eqref{BIstw5} could be obtained one from the other by applying T-duality in four dimensions as described there. It would be interesting to study the behaviour of our \eqref{delf2} - \eqref{bdelR2} under such a transformation.}

Such a generalization has already been obtained in \cite{Blumenhagen:2012pc}.\footnote{Relations similar to our \eqref{delf2} - \eqref{bdelR2} were also obtained in \cite{Blumenhagen:2012ma}, although they do not match exactly, as the $Q$-flux defined there is different, and there is no geometric flux turned on.} There, a quasi-Poisson structure given by $\b$ is considered. Applying in ten dimensions the Lie bracket on the generators $Z_a=\del_a$, $X^a=\b^{ab}\del_b$, the algebra \eqref{algebra} for $H=0$ is precisely reproduced, where the definition of the fluxes there match ours (up to a sign on $R$). A further deformation allows to include an $H$-flux. The Jacobi identities of that algebra then provide NSNS BI for non-constant fluxes. These identities are given for $H=0$ in \eqref{BIBlum1} - \eqref{BIBlum4}, and we verify in appendix \ref{ap:BIlit} that they match with our \eqref{delf2} - \eqref{bdelR2}. This explains in another way why our BI hold automatically: they correspond to ten-dimensional identities derived from Lie brackets.

Finally, other approaches made use of different brackets to obtain similar results. The algebra \eqref{algebra}, at least for $H=0$, was derived from a Generalized Complex Geometry perspective \cite{Grana:2008yw} by considering the Courant bracket acting on generalized $O(D,D)$ frames. The $R$-flux there however does not match our definition. The algebra \eqref{algebra} is obtained again with the Courant bracket, acting this time on standard frames and co-frames (flat vectors and one-forms) in \cite{Blumenhagen:2012pc}; similar results appear in \cite{Chatzistavrakidis:2013wra} with an emphasis on the related Dirac structures. The corresponding Jacobiators derived in \cite{Blumenhagen:2012pc} contain some terms encoding the aforementioned BI \eqref{BIBlum1}-\eqref{BIBlum4}. A Double Field Theory (DFT) extension of these ideas can be found in \cite{Geissbuhler:2013uka}, where the C-bracket \cite{Siegel:1993th, Hull:2009zb} is used: this DFT generalization of the Courant bracket reduces to the latter upon the strong constraint $\tilde{\del}=0$. Acting this way on generalized vielbeins, the algebra \eqref{algebra} is reproduced in an $O(D,D)$ covariant manner. The related Jacobi identity would be given by two terms, one of which is proportional to a quantity $\mathcal{Z}_{ABCD}$ that can be decomposed and reduced into the various BI \eqref{delf2} - \eqref{bdelR2}, as detailed in appendix \ref{ap:BIlit}. Another generalization of the Courant bracket, called the Roytenberg bracket, was also used in \cite{Halmagyi:2009te} to write the algebra \eqref{algebra}. In Exceptional Field Theory, a generalized Lie derivative is introduced \cite{Aldazabal:2013via} and its closure conditions, that can be thought of as related to Jacobi identities of a bracket, are shown to generate BI, including \eqref{delf2}. Finally, in the CFT approach of \cite{Condeescu:2013yma}, the algebra \eqref{algebra} is directly reproduced from actions of (asymmetric) orbifolds.

\subsubsection*{Nilpotent derivative}

Besides the algebraic approach to derive the BI by evaluating Jacobi identities, there is a second proposal using a generalization of the standard exterior derivative. Imposing a nilpotency condition on this derivative is equivalent to a set of constraints that turn out to be the BI. The first simple illustration of that idea is given in the Introduction, particularly in \eqref{Dintro}, with the square of the derivative $\d- H\wedge$ on a $p$-form $A$. In \cite{Shelton:2006fd}, a generalization of $\d- H\wedge$ that includes all NSNS geometric and non-geometric fluxes was proposed. It is given here in our conventions by
\bea
& \D_{{\rm stw}} A= (-H\wedge -f\cdot -Q\cdot + R  \vee) A \ ,\\
& f \cdot = \frac{1}{2!} {f}^{a}{}_{bc}  \ \te^b \wedge \te^c \wedge \iota_a \ ,\ Q \cdot = \frac{1}{2!} Q_c{}^{ab} \  \te^c \wedge \iota_a \ \iota_b \ ,\ R\vee = \frac{1}{3!} R^{abc} \ \iota_a \ \iota_b \ \iota_c \ ,\nn
\eea
where $\iota_a$ and $\vee$ denote contractions on forms, and we refer to appendix \ref{ap:conv} for more conventions. More precisely, this derivative was given without the numerical coefficients that we add here, and was rather specified on the component of the form $A$, i.e. without the contractions. This corresponds to a four-dimensional perspective, where fluxes and $A$ only appear through constant components after being integrated over an internal space. This explains the absence of a derivative on the component of $A$. It was then claimed that the nilpotency condition $\D_{{\rm stw}}^2=0$ would reproduce the NSNS BI for constant fluxes \eqref{BIstw1} - \eqref{BIstw5}. This claim was made more precise in \cite{Ihl:2007ah} where the previous derivative was completed by two more terms as
\bea
\D_{\sharp} A= \Big(&-\frac{1}{3!} H_{abc} \te^a \wedge \te^b \wedge \te^c \wedge - \frac{1}{2!} {f}^{a}{}_{bc}  \ \te^b \wedge \te^c \wedge \iota_a - \frac{1}{2!} Q_c{}^{ab} \  \te^c \wedge \iota_a \ \iota_b + \frac{1}{3!} R^{abc} \ \iota_a \ \iota_b \ \iota_c  \nn\\
& - \frac{1}{2} {f}^{a}{}_{ab}  \ \te^b \wedge + \frac{1}{2} Q_a{}^{ab} \ \iota_b \Big) A  \ .\label{Dc1}
\eea
More precisely, we again rewrite a formula that was given on form components, namely (B.3) of \cite{Ihl:2007ah}, using here forms and contractions; also, our conventions differ by a minus sign on the $H$-flux. The two new terms given by the traces of $f$ and $Q$ will play an important role, together with dilaton terms, when we define later on the $Spin(D,D)\times \mathbb R^{+}$ derivative. They were already important in \cite{Ihl:2007ah}, where an explicit computation of the nilpotency condition for the derivative \eqref{Dc1} lead to
\beq
\D_{\sharp}^2=0\ \Leftrightarrow\ \mbox{BI}\ \eqref{BIstw1} - \eqref{BIstw5}\ \ \mbox{and}\ \ \frac{1}{3} H_{abc}R^{abc} + \frac{1}{2} f^a{}_{ab} Q_a{}^{ab} = 0 \ .\label{Dsharpsquare}
\eeq
The nilpotency condition reproduces the NSNS BI (with constant fluxes) together with an extra scalar constraint that includes the traces of $f$ and $Q$. Note that particular indices contractions of the BI also appear in this computation; the same will happen for our derivative in section \ref{sec:Spinder}.\\

As mentioned already in \cite{Shelton:2006fd} (see also \cite{Villadoro:2007tb}), the derivative $\d- H\wedge$ enters the BI of the RR fluxes for type II supergravities, given by $(\d- H\wedge) F=0$ in the sourceless case. Here, $F$ is the polyform given by the sum of the RR fluxes (we set $F_0=0$ for simplicity); one has $F=(\d -H\w)C$ for a polyform gauge potential $C$. The polyforms $F$ and $C$ can actually be interpreted as an $O(D,D)$ spinor: this was pointed out in \cite{Hull:1994ys, Brace:1998xz, Fukuma:1999jt, Hassan:1999bv, Hassan:1999mm}, and it could be guessed from the SUSY conditions of \cite{Grana:2005sn}. This idea lead in \cite{Hohm:2011dv, Geissbuhler:2013uka} to define at the level of DFT $\mathcal F=\mathcal D\mathcal C$, where $\mathcal D=\G^{A}D_A$ denotes the Dirac operator associated to a $Spin(D,D)\times \mathbb R^{+}$ covariant derivative $D_A$, and $\Gamma^A$ are $Spin(D,D)$ Clifford matrices. A related derivative appeared already in \cite{Grana:2008yw, Jeon:2012kd, Jeon:2012hp}.

So this spinorial derivative is somehow natural to consider, and we will do so in section \ref{sec:Spinder} at the level of standard supergravity and $\b$-supergravity, using its generic Generalized Geometry definition; the one of \cite{Geissbuhler:2013uka} is then the DFT extension. The non-trivial point we make in this paper is that {\it the vanishing square of this spinorial derivative should give the NSNS BI}, in analogy to $\d- H\wedge$. In other words, as we will show using the Clifford map on the $\Gamma$-matrices, this $Spin(D,D)\times \mathbb R^{+}$ derivative reproduces and generalizes the above $\D_{\sharp}$. Although this idea is not explicitly mentioned in \cite{Geissbuhler:2013uka}, $\D^2$ is already computed there in (4.13) at a generic level, and it gives a hint on the results to be derived. Indeed, this square depends on various quantities among which $\mathcal Z_{ABCD}$ and $\mathcal Z$. We show in appendix \ref{ap:BIlit} that the former reduces to our BI \eqref{delf2} - \eqref{bdelR2} while the latter contains the scalar quantity appearing \eqref{Dsharpsquare}. So a nilpotency condition of this spinorial derivative does look relevant; we now turn to it.

\subsubsection{The $Spin(D,D)\times \mathbb R^{+}$ covariant derivative}\label{sec:Spinder}

We have just motivated the introduction of the $Spin(D,D)\times \mathbb R^{+}$ covariant derivative, that we consider here at the level of the Generalized Geometry formalism. We first construct it generically, as well as the corresponding Dirac operator, and further express it for different generalized frames: the one with a $b$-field for standard supergravity, and the one with a $\b$ for $\b$-supergravity. To do so, we use conventions and results of \cite{Andriot:2013xca}, especially the value of connection coefficients. We verify in a second part that the nilpotency condition on this spinorial derivative for $\b$-supergravity exactly reproduces our BI \eqref{delf2} - \eqref{bdelR2}, together with the scalar condition mentioned in \eqref{Dsharpsquare}. We also clarify the relation to the above $\D_{\sharp}$ of \cite{Ihl:2007ah}.

We start with the $O(D,D)\times \mathbb R^{+}$ generalized covariant derivative of \eqref{genconn}. From it, the corresponding spinorial derivative $D_A$ (with generalized flat index) can be written down,\footnote{In \eqref{spinderiv}, the index ${}_B$ of the generalized connection coefficient has been lowered with the $O(D,D)$ metric.} as well as the Dirac operator $\D$ on a spinor $\Psi \in \Ga(S^{\pm}_{(1/2)})$ \cite{Coimbra:2011nw}
\beq
\mathcal D \Psi=\Ga^{A}D_{A}\Psi=\Ga^{A}\Big(\del_{A}+\frac{1}{4}\O_{ABC}\Ga^{BC}-\frac{1}{2}\La_{A}\Big)\Psi\ . \label{spinderiv}
\eeq
The $\Ga$-matrices satisfy the Clifford algebra
\beq
\{\Ga^{A},\Ga^{B}\}=2\eta^{AB}, \quad \eta=\frac{1}{2}\begin{pmatrix}0&1\\1&0\end{pmatrix}, \quad \eta^{-1}=2 \begin{pmatrix}0&1\\1&0\end{pmatrix} \label{Cliffalg}\ .
\eeq
Here $\eta$ of coefficients $\eta_{AB}$ denotes the $O(D,D)$ metric. A particular representation of this algebra is given by the Clifford map
\beq
\Ga^{A}=\begin{cases} \Ga^a=2 \te^a\ ,\\ \Ga_a=2 \iota_{a}\ ,\end{cases} \text{with}\quad \{\te^{a},\te^{b}\}=0\ ,\quad \{\te^{a},\iota_{b}\}=\delta^{a}_{b}\ , \quad \{\iota_{a},\iota_{b}\}=0 \ . \label{Cliffrepr}
\eeq
We will use it to express the Dirac operator with fluxes, forms and contractions, in a generalization of the standard exterior derivative acting on a $p$-form $A$. The spinor $\Psi$ should then be understood as polyform \cite{Grana:2005sn}. For now, we can simplify \eqref{spinderiv} using the identity $\Ga^{A}\Ga^{BC}=\Ga^{ABC}+\eta^{AB}\Ga^{C}-\eta^{AC}\Ga^{B}$ that relates antisymmetrized products of $\Ga$-matrices. Using the compatibility condition, we get
\bea
\D\Psi=\Ga^{A}D_{A}\Psi=&\Big(\Ga^{A}\del_{A}+\frac{1}{4}\Omega_{ABC}\Ga^{ABC}+\frac{1}{2}(\Omega_{D}{}^{D}{}_{C}-\La_{C})\Ga^{C}\Big)\Psi \label{Diracopgen}\\
=&\Big(\Ga^{A}\del_{A}+\frac{1}{4}\hat \Omega_{ABC}\Ga^{ABC}+\frac{1}{2}\hat\Omega_{D}{}^{D}{}_{C}\Ga^{C}\Big)\Psi\nn\\
\equiv &\Big(\D_1 + \D_2 + \D_3\Big) \Psi\nn\ .
\eea
Let us point out that $\D_3$ denotes the trace part due to the extension of the $O(D,D)$ by the conformal factor $\mathbb{R}^{+}$, that usually combines the determinant of the metric and the dilaton.

We now determine these three terms for different choices of generalized frames. Following \cite{Andriot:2013xca}, such a choice can fix $\del_A$, $\Omega_{A}{}^{B}{}_{C}$  and $\La_{A}$. For the $\G$-matrices, we use the Clifford map \eqref{Cliffrepr}: forms and contractions act on the one-forms in $A$ while a derivative $\del_a \cdot$ only acts on the (flat indices) component of $A$. Details on the computation of $\D_2$ are given in appendix \ref{ap:spinder}.

\subsubsection*{Standard supergravity}

Using the generalized frames with $b$-field, we obtain
\bea
\D_1 &= 2 \del_a \cdot e^a\! \w \\
\D_2 &= - f^{c}{}_{ab} e^{a}\!\w e^{b}\!\w \iota_c -f^{d}{}_{dc}e^c\!\w  - \frac{1}{3}H_{abc} e^a\!\w e^b\!\w e^c\!\w \\
\D_3 &= {f}^{a}{}_{ab}  \ e^b \wedge - 2 \del_a\p\ e^a \wedge \ ,
\eea
that sums up to $\D$ given by
\bea
\frac{1}{2}\D A &= \left( \del_a \cdot e^a\! \w - \frac{1}{2} f^{c}{}_{ab} e^{a}\!\w e^{b}\!\w \iota_c - \frac{1}{6}H_{abc} e^a\!\w e^b\!\w e^c\!\w - \del_a\p\ e^a \wedge \right) A \\
& = e^{\p} \left(\d - H \w \right) (e^{-\p} A) \label{Dbstandard} \ .
\eea

\subsubsection*{$\b$-supergravity}

Using the generalized frames with $\b$, we obtain
\bea
\D_1 &= 2\del_a\cdot \te^a\!\w + 2\b^{ab}\del_b\cdot\iota_{a}\\
\D_2 &= - f^{c}{}_{ab} \te^{a}\!\w \te^{b}\!\w \iota_c -f^{d}{}_{dc}\te^c\!\w -Q_{a}{}^{bc}\te^a\!\w \iota_b\,\iota_c +Q_{d}{}^{dc}\iota_c + \frac{1}{3}R^{abc} \iota_a\,\iota_b\,\iota_c\\
\D_3 &= {f}^{a}{}_{ab}  \ \te^b \wedge - 2 \del_a\tp\ \te^a \wedge  + Q_a{}^{ab} \ \iota_b - 2 (\b^{ab}\del_b \tp -\T^a) \  \iota_a \ .
\eea
Adding up these various pieces, we find
\bea
\D  = & 2\del_a\cdot \te^a\!\w + 2\b^{ab}\del_b\cdot\iota_{a}- f^{c}{}_{ab}\, \te^{a}\!\w \te^{b}\!\w \iota_c -2\del_a\tp\, \te^{a}\w \label{Dtotal}\\
&-Q_{a}{}^{bc}\,\te^a\!\w \iota_b\,\iota_c +2Q_{d}{}^{dc}\,\iota_c - 2(\b^{ab}\del_b \tp -\T^a)\, \iota_a + \frac{1}{3} R^{abc}\,\iota_a\,\iota_b\,\iota_c \nn \ ,
\eea
where the second row could be further simplified using the definition of $\T^a$. We can rewrite this result differently, using the following relations for a $2$-form $A$ (easily extendable to higher forms)
\bea
& \frac{1}{2} \iota_a \cN^a (A_{bd}) \te^b \w \te^d = ( - \b^{ac}\del_c A_{ad} + Q_a{}^{ac} A_{dc} - \frac{1}{2} Q_d{}^{ac} A_{ac} ) \te^d \ ,\\
& Q_{a}{}^{bc}\,\te^a\!\w \iota_b\,\iota_c (\frac{1}{2} A_{ef} \te^e \w \te^f) = - Q_a{}^{ef} A_{ef} \te^a \ ,\ Q_{c}{}^{ca} \iota_a (\frac{1}{2} A_{bd} \te^b \w \te^d)= Q_c{}^{ca} A_{ad} \te^d \ .
\eea
These relations are derived using the definitions and properties of $\cN$, $Q$, and conventions of appendix \ref{ap:conv}. From them, we deduce, as given in \eqref{Dbetaintro}
\beq
\frac{1}{2} \D A= e^{\tp} (\N_a \cdot \te^a \w - \cN^a \cdot \iota_a + \T \vee + R \vee ) (e^{-\tp} A) \ ,\label{Dbetasection}
\eeq
where $\N_a \cdot \te^a \w =\d$, as we act on forms. The second term gives an interesting counterpart to the exterior derivative.\\

The resulting $\D$ for standard supergravity is a known spinorial derivative \cite{Grana:2005sn}, and its square gives the standard NSNS BI as mentioned in \eqref{Dintro}. We are now going to show the analogous result for the $\b$-supergravity derivative and our BI \eqref{delf2} - \eqref{bdelR2}. A first hint is given by the comparison to the above derivative $\D_{\sharp}$ of \cite{Ihl:2007ah} given in \eqref{Dc1}. For constant forms and fluxes, we recognise that in both cases ($\b$ or $b$ vanishes), one has
\beq
\D_{\sharp} = \frac{1}{2}\D_2\ .
\eeq
The natural completion of $\D_{\sharp}$ in the case of non-constant fluxes would have been by derivatives, as given by $\D_1$. Interestingly, we will see that this is not enough to recover the BI: the additional traces and dilaton terms of $\D_3$ are also needed. So we now turn to the study of the nilpotency condition for the above derivative $\D$ of \eqref{Dtotal}
\bea
\D^2 A=0\ . \label{nilpocond}
\eea
We compute in appendix \ref{ap:spinder} this condition in details. It produces the following set of seven equations
\bea
- \frac{1}{2}\del_{[a} f^{d}{}_{bc]}+ \frac{1}{2} f^{d}{}_{g[a}f^{g}{}_{bc]}&=0 \label{BIdelf}\\
-\frac{1}{2}Q_{d}{}^{da}f^{g}{}_{ga}&=0\label{fullcontr}\\
-\frac{3}{2}\b^{de}\del_{[e} f^{b}{}_{da]}+\frac{3}{2}\b^{de}f^{b}{}_{h[a}f^{h}{}_{ed]}&=0\label{betadelf}\\
-\frac{1}{2}(\del_{[a} Q_{c]}{}^{de}-\b^{g[d}\del_g f^{e]}{}_{ac})+\frac{1}{4}(-4f^{[d}{}_{g[a}Q_{c]}{}^{e]g}+f^{g}{}_{ac}Q_{g}{}^{de})&=0 \label{BIdelQ}\\
-\frac{1}{2}\b^{dc}\del_c Q_{d}{}^{ab}-\frac{1}{2}\b^{cd}\b^{g[a}\del_g f^{b]}{}_{cd} -\b^{dc}Q_{c}{}^{g[a}f^{b]}{}_{dg}+\frac{1}{4}\b^{dc}Q_{g}{}^{ab}f^{g}{}_{cd}&=0\label{betadelQ}\\
 \frac{1}{6} (\del_a R^{bcd}-3\b^{e[b}\del_e Q_{a}{}^{cd]})- \frac{1}{2} ( -R^{g[bc}f^{d}{}_{a]g}+Q_{a}{}^{g[d}Q_{g}{}^{bc]} )&=0 \label{BIdelR}\\
-\frac{1}{6}\b^{g[a}\del_g R^{bcd]}-\frac{1}{4}Q_{g}{}^{[ab} R^{cd]g}&=0\ . \label{BIbdelR}
\eea
It is remarkable that the dilaton terms completely cancel out. All of the above equations are not independent. \eqref{betadelf} is a contraction of \eqref{BIdelf} by $\b$, and similarly \eqref{betadelQ} is a contraction of \eqref{BIdelQ}. We are then left with a set of five independent identities. These are exactly the four Bianchi identities listed before: \eqref{BIdelf} matches \eqref{delf2}, \eqref{BIdelQ} matches \eqref{delQ2}, \eqref{BIdelR} matches \eqref{delR2}, \eqref{BIbdelR} matches \eqref{bdelR2}. So the square of this spinorial derivative \eqref{Dtotal} precisely produces the BI. In addition we find the scalar condition derived in \cite{Ihl:2007ah}, and given in \eqref{Dsharpsquare}, from the fully contracted terms \eqref{fullcontr}.

Given this result, and the expression of $\D$ given in \eqref{Dbetasection}, we deduce on a two-form $A$
\beq
\Big\{ \N_a \cdot \te^a \w\ , \cN^b \cdot \iota_b - \T \vee \Big\} A = -\frac{1}{2} \left(3\ \b^{eb}S^{a}_{ebc} \ A_{ad} + S^{ab}_{cd}\ A_{ab} \right)\te^c \w \te^d \ ,\label{tensorialformBI}
\eeq
where the quantities $S$ are defined in section \ref{sec:TdBI} and correspond to the LHS of the BI \eqref{delf2} and \eqref{delQ2}. This gives a tensorial form to \eqref{delQ2}, since such a form for \eqref{delf2} was already mentioned around \eqref{Riemanndelf}. The cases of \eqref{delR2} and \eqref{bdelR2} were discussed below the latter.

\subsection{T-dual $\NS$-branes sourcing the Bianchi identities}\label{sec:Tbranes}

As presented in the Introduction, the Bianchi identity (BI) for the $H$-flux gets modified with a source term on its right-hand side (RHS) in the presence of an $\NS 5$-brane. We show in this section that the BI \eqref{delf2}-\eqref{bdelR2} just studied get corrected similarly if other $\NS$-branes are present, namely for a Kaluza-Klein ($\KK$) monopole or a $Q$-brane. These are vacua of standard supergravity and $\b$-supergravity. Up to smearing, they are T-dual to the $\NS 5$-brane. We first present these solutions following the literature. We then focus on the smearing procedure that allows T-dualities along isometry directions. This clarifies how the different warp factors can be the appropriate Green functions in the Poisson equations of each brane. We finally verify how the branes are related by T-duality. We further show that the above BI on the brane vacua boil down to the Poisson equations, allowing the emergence of the source term. This study establishes $\b$-supergravity as a nice framework to describe $Q$-branes.

\subsubsection{$\NS$-branes solutions}\label{sec:NSbranesol}

We present here the various $\NS$-branes, starting with the $\NS 5$-brane that sources the $H$-flux. The $\NS 5$-brane solution was first given in the limit of zero size instanton in \cite{Strominger:1990et}, and presented in a broader context in \cite{Callan:1991dj} as corresponding to the case where the gauge field vanishes. More generalizations and references can be found in \cite{Nolle:2012hf, Gemmer:2012pp}. Smearing and T-dualising it along one direction leads to the $\KK$-monopole, which was first discovered as a solution to pure five-dimensional general relativity (see \cite{Gross:1983hb, Sorkin:1983ns}, and \cite{Villadoro:2007yq} for more references); it sources the geometric flux. A further smearing and T-duality along another direction leads to a new brane known as the $5_{2}^{2}$-brane \cite{deBoer:2010ud, deBoer:2012ma} or $Q$-brane \cite{Hassler:2013wsa}. It is one of the exotic branes \cite{deBoer:2010ud, Bergshoeff:2011se, deBoer:2012ma, Hassler:2013wsa, Kimura:2013fda, Kimura:2013zva, Kimura:2013khz}: those recently received much attention, as being related to standard branes by different U-dualities. $Q$-branes are non-geometric vacua of standard supergravity, but become geometric in $\b$-supergravity \cite{Hassler:2013wsa, Geissbuhler:2013uka} and then source the $Q$-flux.

\subsubsection*{$\NS 5$-brane}

The $\NS 5$-brane is physically a codimension $4$ object, i.e. it is located in four dimensions that are singled out as we will see below; it is the magnetic counterpart of the fundamental string. The original solution takes the following form\footnote{We have a factor of $2$ difference for the $H$-flux with respect to the conventions of \cite{Callan:1991dj}. Note that the warp factor given here is not considered in \cite{Jensen:2011jna, Geissbuhler:2013uka}, as only the $\KK$-monopole and T-duals are used there. In particular, only the smeared warp factor of the $\NS 5$-brane is present there.}
\bea
& \d s^2 = \d s_6^2 + f_{H}\ \d \hat{s}_4^2 \ ,\ H_{mnp}=-\sqrt{|g_4|} \epsilon_{4 mnpq} g^{qr} \del_r \ln f_{H} \ ,\ e^{2\p}= f_{H} \label{NS5sol}\\
& {\rm where}\ \d \hat{s}_4^2= \sum_{m=1 \dots 4} (\d x^m)^2 \ ,\ r_4^2=\sum_{m=1 \dots 4} (x^m)^2\ ,\ f_{H}=e^{2\p_H} + \frac{q}{r_4^2}\ ,\nn
\eea
and $\d s^2_6$ is the Minkowski metric. $\d \hat{s}_4^2$ is the flat Euclidian metric, and gives the transverse directions. The warp factor $f_{H}$ depends on the radius $r_4$ and on two constants, the value at $\infty$ of the dilaton $\p_H$, and $q$ that is related to the tension of the brane. The $H$-flux is proportional to the volume form coefficient of the transverse four-dimensional space $\sqrt{|g_4|} \epsilon_{4 mnpq}$ (see appendix \ref{ap:conv} for conventions). Given the transverse metric, we can simplify the expression for the $H$-flux towards
\beq
H_{mnp}=- \epsilon_{4 mnpq} \delta^{qr} \del_r f_{H} \ .\label{HNS5simple}
\eeq

\subsubsection*{Kaluza-Klein monopole}

The $\KK$-monopole is considered here as a codimension $3$ brane. This solution is given by
\bea
& \d s^2 = \d s_6^2 + f_K\ \d \hat{s}_3^2 +  f_K^{-1} (\d x + a \d y )^2 \ ,\ H_{mnp}=0\ ,\ e^{2\p}= 1 \label{KKsol}\\
& {\rm where}\ \d \hat{s}_3^2= \d \rho^2+ \rho^2 \d \varphi^2+ \rho^2 \sin^2 \varphi\ \d y^2 \ ,\ f_{K}=e^{2\p_K} - \frac{q_K}{\rho} \ .\nn
\eea
The metric $\d s^2_6$ is still that of Minkowski, and the metric $\d \hat{s}_3^2$ is the flat space one. But we prefer here to use spherical coordinates $\{ \rho, \varphi, y \}$ for the three transverse directions. The radius $\rho$ will sometimes be denoted $r_3$ below. The warp factor $f_K$ depends on two constants, $\p_K$ denoted this way for convenience, and $q_K$ that we will relate to the above $q$ in section \ref{sec:smearing}.\footnote{A warp factor for the $\KK$-monopole depending on $x$ was considered in \cite{Gauntlett:1992nn, Gregory:1997te}, and related to world-sheet instantons corrections \cite{Harvey:2005ab} (see also \cite{Kimura:2013zva}). One can verify that it matches ours far away from the brane
\beq
f_K(\rho,x)=\frac{1}{g^{2}}+\frac{1}{2\rho}\frac{\sinh \rho}{\cosh \rho -\cos x}\ .
\eeq} Finally, the important quantity in the solution is $a$. It is like a connection one-form coefficient and is a priori not gauge invariant. Away from the singularity, one has
\beq
a(\varphi)= q_K \cos \varphi \ {\rm for}\ \rho >0 \ .
\eeq
We will complete it towards
\beq
a(\rho, \varphi)= \cos \varphi\ \rho^2 \del_\rho f_K \ ,
\eeq
for reasons to be detailed in section \ref{sec:TdBI}. From this we will deduce the corresponding (geometric) flux; the latter will be a better defined quantity to consider. It will be given by
\beq
f^x{}_{\varphi y}= f_K^{-\frac{3}{2}} \del_{\rho} f_K \ .
\eeq

\subsubsection*{$Q$-brane}

The $Q$-brane is a codimension $2$ brane. This solution is better described in terms of $\beta$-supergravity as
\bea
& \d \tilde{s}^2 = \d s_6^2 + f_Q\ \d \hat{s}_2^2 +  f_Q^{-1} (\d x^2 + \d y^2 ) \ ,\ {\rm only}\ \b^{xy}=- \b^{yx} \neq 0 \ ,\ e^{2\tp}= f_Q^{-1} \label{Qbrsol}\\
& {\rm where}\ \d \hat{s}_2^2= \d \rho^2+ \rho^2 \d \varphi^2 \ ,\ f_Q=e^{-2\tp_Q} - q_Q \ln \rho \nn \ .
\eea
Its expression in terms of standard supergravity is given below in \eqref{bQ-brane}. The metric $\d s_6^2$ is again Minkowski, and $\d \hat{s}_2^2$ is the flat metric, given this time using polar coordinates $\{ \rho, \varphi \}$ for the transverse directions. The radius $\rho$ will sometimes be denoted $r_2$ below. The warp factor $f_Q$ depends on two constant, $\tp_Q$ denoted this way for convenience, and $q_Q$ that we will relate to $q$ in section \ref{sec:smearing}. $\tp_Q$ may contain a cutoff when $\rho \rightarrow \infty$, as mentioned in \cite{deBoer:2010ud, deBoer:2012ma}; we will rediscuss this point in section \ref{sec:smearing}. Finally, as for the $\KK$-monopole and $a$, the field $\b$ is here not a well-defined quantity. Still, we will consider (in curved indices)
\beq
\b^{xy}= - \varphi\  \rho \del_{\rho} f_Q\ \Rightarrow \ \b^{xy}= q_Q\ \varphi  \ {\rm for} \rho >0 \ .
\eeq
The $Q$-flux is a better defined quantity. It will be given by (in flat indices)\footnote{As usual, the three fluxes are the same in flat indices, up to a sign on the structure constant. For the $H$-flux, one can choose coordinates that isolate the coordinate $r_4$. The corresponding metric element would still only be given by a warp factor, so one would get
\beq
H_{mnp}=-\sqrt{|g_3|} \epsilon_{4 mnp (r_4)} f_H^{-\frac{3}{2}} \del_{r_4} f_{H} \ .
\eeq
The remaining volume factor is then removed when going to flat indices (see the conventions on $\epsilon$ in the appendix \ref{ap:conv}). So the three fluxes are the same in flat indices, although one needs to take the same warp factor. This only happens when there is smearing, i.e. in the case of T-duality, as we will show below. It is definitely in that case that we expect the equality of the fluxes, as given in the T-duality chain of \cite{Shelton:2005cf}.}
\beq
Q_{\varphi}{}^{x y}= - f_Q^{-\frac{3}{2}} \del_{\rho} f_Q \ .
\eeq
We verify explicitly in appendix \ref{ap:Qbranevac} that the $Q$-brane is a solution to the equations of motion of $\b$-supergravity. In \cite{Hassler:2013wsa}, using a different method, this result is somehow obtained away from the singularity.

\subsubsection{Smearing warp factors and Poisson equations}\label{sec:smearing}

The brane solutions that we have just presented are related by smearing and T-dualising along transverse directions. We focus here on the different warp factors, and show how smearing relates one warp factor to the other. This explains how each of those can satisfy the appropriate Poisson equation. To get familiar with these ideas, we start with the well-known case of $p$-branes solutions, before turning to $\NS$-branes.

\subsubsection*{Warm-up: $D_p$-branes}
A $p$-brane is a type II supergravity background that provides an effective description of a $D_p$-brane in some regime. This solution contains in particular a dilaton that depends on the warp factor $Z_p(r)$, and the metric is given by
\beq
\d s^2 = Z_p^{-\frac{1}{2}} \d s_{||}^2 + Z_p^{\frac{1}{2}} \d s_{\bot}^2 \ ,
\eeq
where $\d s_{||}^2$ is the Minkowski space-time along the brane, $\d s_{\bot}^2$ the flat Euclidian space transverse to the brane, $r$ the Euclidian radius for the latter, and
\beq
Z_p(r)=1+\frac{q_p}{r^{7-p}} \ ,\ \mbox{for}\ p\leq 6 \ ,\label{Zp}
\eeq
with $q_p$ a constant related to the tension of the brane. The Ramond-Ramond (RR) flux $F$ of this background verifies typically a BI of the form
\beq
\d F = Q\ \delta(x_{\bot}) \ . \label{BIgen}
\eeq
The flux is sourced by the brane, localised by the $\delta$ in its $9-p$ transverse directions,\footnote{\label{foot:currents}The $(9-p)$-form $\delta(x_{\bot})$ of \eqref{BIgen} can also be viewed as a current, and defined through
\beq
\int_{||} A_{p+1} = \int_{10} A_{p+1} \wedge \delta(x_{\bot})
\eeq
for any $(p+1)$-form $A_{p+1}$ (see for instance \cite{Mourad:1997uc, Koerber:2006hh}).} and carrying a charge $Q$. Using for instance the transverse Hodge star $*_{\bot}$, one can extract the forms to leave only coefficients, in particular the density $\delta^{(9-p)}(x_{\bot})$. The Bianchi identity then typically boils down to the scalar equation (up to a proper normalisation)
\beq
\Delta_{9-p} Z_p = \delta^{(9-p)}(r) \ ,
\eeq
where $\Delta_{9-p}$ is the Laplacian of the unwarped metric $\d s_{\bot}^2$. The appearance of the latter can be understood for $F=*\d C$ with $C$ the dual potential.\footnote{The BI and resulting scalar equation are sometimes more complicated, depending on what exactly is $F$. For example, an additional constant next to the $\delta$ can be obtained, see for instance \cite{Schulz:2004ub}.} This scalar equation is a Poisson equation; solving it means finding the Green function for the Laplacian given some boundary conditions. The solutions to this problem are known: for two dimensions, one has $\ln r$, and for $d_{\bot}\geq 3$, one has $\frac{1}{r^{d_{\bot}-2}}$. For $d_{\bot}=3$, this is the well-known electrostatic potential. The radial dependence in the transverse space directions $d_{\bot}=9-p$ coincides precisely with that of $Z_p$ \eqref{Zp} as expected.

We now consider T-dualities on these branes. T-dualising along a transverse direction is known to extend a $D_p$-brane to a $D_{p+1}$-brane. Can this be seen on the above solutions? The standard "radius inversion" of T-duality inverts a warp factor in the metric, so the correct powers of warp factor are obtained by applying the Buscher rules. However, the warp factor itself should also be changed from $Z_p$ to $Z_{p+1}$, as well as the radius of the transverse directions, from $r_{9-p}$ to $r_{9-(p+1)}$. This is rather obtained from the smearing required by T-duality, as explained in \cite{Schulz:2004ub}: a transverse direction of coordinate $x$, along which we want to T-dualise, is a priori not an isometry, since $Z_p$ depends on $x$. To allow the T-duality, we first make it an isometry by smearing, that amounts to averaging in this direction
\beq
Z_{p+1} (r_{9-(p+1)}) \sim \int \d x\ Z_p (r_{9-p}) \ , \quad r_{9-p}^2= x^2 + r_{9-(p+1)}^2 \ .\label{relwarpfactors}
\eeq
The smeared $p$-brane is then T-dual to the $(p+1)$-brane. Interestingly, the Poisson equations are also consistent under this procedure
\bea
& \Delta_{9-p} Z_p =\left((\del_x)^2 + \Delta_{9-(p+1)} \right) Z_p=\delta^{(9-p)}(r_{9-p}) \label{intBI1}\\
& \Rightarrow  \int \d x\ \left((\del_x)^2 + \Delta_{9-(p+1)} \right) Z_p =  \int \d x\ \delta^{(9-p)}(r_{9-p}) \\
&\Leftrightarrow \left( 0 + \Delta_{9-(p+1)} \int \d x\ \right) Z_p =  \delta^{(9-(p+1))}(r_{9-(p+1)}) \\
&\Leftrightarrow \ \Delta_{9-(p+1)} Z_{p+1} =  \delta^{(9-(p+1))}(r_{9-(p+1)}) \ . \label{intBI5}
\eea
In the last but one line, we use conditions on the warp factor and its derivatives that will be verified in the examples below. In this derivation, we actually only need the warp factor without its pure constant part, since only its derivatives are involved. So that is what we meant in \eqref{relwarpfactors}, and what will be used in the following.

The $\NS$-branes share many features with the $p$-brane solutions. They both have warp factors that determine the transverse directions. The constants in the warp factors are related to the tension of the brane, although they scale differently in $e^{\p_0}=g_s$. Finally, these warp factors take analogous forms, corresponding to the various Green functions in different (co)dimensions. As we will see, the $\NS$-branes satisfy as well Poisson equations. They actually follow the same logic as the $D_p$-branes: up to smearing, they are T-dual. Their (co)dimension, metric and warp factors given above match all the criteria just discussed for that to hold. We will verify explicitly the T-duality relations and derive the Poisson equations from the Bianchi identities in section \ref{sec:TdBI}. Before doing so, let us first relate their different warp factors by smearing as just explained for the $p$-branes.

\subsubsection*{$\NS 5$-brane}

The Bianchi identity for the $H$-flux of the $\NS5$-brane is given by $\d H$, proportional to ($\propto$)
\beq
\del_{[m} H_{npq]}=- \del_{[m} \epsilon_{4 npq]r} \delta^{rs} \del_s f_{H}\ \propto\ \epsilon_{4 mnpq} \delta^{rs}  \del_{r} \del_s f_{H} \ ,\label{BIHcurved}
\eeq
where we used the expression of the $H$-flux \eqref{HNS5simple}. One therefore gets that
\beq
\d H \ \propto\ \hat{{\rm vol}}_4 \ \Delta_4 f_H \ ,\quad \Delta_4=  \sum_{m=1 \dots 4} (\del_m)^2 \ ,
\eeq
with the four-dimensional volume form $\hat{{\rm vol}}_4$. The Bianchi identity in presence of a source is given by $\d H\ \propto\ \hat{{\rm vol}}_4\ \delta^{(4)} (r_4)$, so the warp factor has to solve the Poisson equation
\beq
\Delta_4 f_H = c_H \ \delta^{(4)} (r_4)\ ,\label{PoissonfH}
\eeq
with a constant $c_H$. In other words, $f_H / c_H$ should be a Green function for the four-dimensional Laplacian $\Delta_4$. A known Green function for this problem is $\frac{1}{r_4^2}$, so $f_H$ given in \eqref{NS5sol} certainly solves the Poisson equation. A crosscheck of this result is that away from the singularity $r_4=0$, the Poisson equation boils down to the Laplace equation, meaning
\beq
\Delta_4 f_H = 0 \ {\rm for}\ r_4>0 \ .
\eeq
One can verify that this holds for $f_H$ of \eqref{NS5sol}.

\subsubsection*{Kaluza-Klein monopole}

We turn to the $\KK$-monopole. We follow the procedure explained above, by smearing the $\NS 5$-brane along one direction $x$. First, we introduce the new three-dimensional radius $r^2_3=r^2_4 - x^2$. Then, we smear the warp factor without its constant $f_H-e^{2\p_H}$ to get the new one $f_K$ up to its constant $e^{2\p_K}$, as follows
\beq
f_K (r_3) - e^{2\p_K}= \int_{-\infty}^{+\infty} \d x\ (f_H (r_4) - e^{2\p_H}) =  \left[\frac{q}{r_3}\arctan\left(\frac{x}{r_3} \right)\right]_{-\infty}^{+\infty} = \frac{q \pi}{r_3} \ .
\eeq
This new warp factor matches the one given in \eqref{KKsol} with $q_K=-\pi q$. In addition, it is a known solution to the three-dimensional Poisson equation
\beq
\Delta_3 f_K = c_K\ \delta^{(3)} (r_3) \ , \label{PoissonfK}
\eeq
the well-known electrostatic potential. One can straightforwardly verify that
\beq
\Delta_3 f_K = 0 \ {\rm for}\ r_3>0 \ .
\eeq
This result was expected from the discussion around \eqref{intBI1} - \eqref{intBI5}. One condition for this procedure to work is that the derivative of the warp factor vanishes on the boundary. Here this holds, as $\del_m f_H = -\frac{2q\ x^m}{r_4^4} \sim_{\infty} -\frac{2q}{(x^m)^3}$. The same will be true for the further warp factors (the power of $x^m$ in the denominator decreases by one at each step).

\subsubsection*{$Q$-brane}

We should now obtain the warp factor $f_Q$ of the $Q$-brane by smearing the previous one along a further direction $y$. We introduce the two-dimensional radius $r_2^2= r_3^2 - y^2$, and the boundary constant $e^{-2\tp_Q}$. We introduce further $\epsilon$ that will be sent to $\infty$, and the function ${\rm arsinh} x= \ln (x + \sqrt{(x^2 +1)})$. Then
\bea
f_Q (r_2) - e^{-2\tp_Q}= \int_{-\epsilon}^{+\epsilon} \d y\ (f_K (r_3) - e^{2\p_K}) & = q\pi \left[\ln \left(\frac{y + \sqrt{(y^2 + r_2^2)} }{r_2}\right)\right]_{-\epsilon}^{+\epsilon} \\
& = q\pi \left[\ln \left(y + \sqrt{(y^2 + r_2^2)} \right)\right]_{-\epsilon}^{+\epsilon} \ .\nn
\eea
The function ${\rm arsinh} x$ is odd, from which we get the property
\beq
\ln \left(-y + \sqrt{(y^2 + r_2^2)} \right)= - \ln \left(y + \sqrt{(y^2 + r_2^2)} \right) + 2 \ln r_2 \ .\nn
\eeq
We deduce
\beq
f_Q (r_2) - e^{-2\tp_Q}= 2 q\pi \ln \left(\epsilon + \sqrt{(\epsilon^2 + r_2^2)} \right) - 2 q\pi \ln r_2 \ .\nn
\eeq
This diverges when taking the limit $\epsilon \rightarrow \infty$. We therefore need a cutoff, as argued in \cite{deBoer:2012ma}, to remove this divergence.\footnote{It would be interesting to study whether the divergence is related to the non-geometry, and thus whether the field redefinition could avoid it, by for instance including volume factors in the integral relation \eqref{relwarpfactors}.} Up to a redefinition of the constant $\tp_Q$ to absorb it, one obtains
\beq
f_Q (r_2) = e^{-2\tp_Q} - 2 q\pi \ln r_2 \ .
\eeq
This warp factor matches the solution \eqref{Qbrsol} with $q_Q= 2 \pi q$. In addition, it is a known solution to the two-dimensional Poisson equation
\beq
\Delta_2 f_Q = c_Q\ \delta^{(2)} (r_2) \ .\label{PoissonfQ}
\eeq
One can straightforwardly verify that
\beq
\Delta_2 f_Q = 0 \ {\rm for}\ r_2>0 \ .
\eeq

\subsubsection*{$R$-brane ?}

It is tempting to go one step further: we smear along the direction $z$ to get the warp factor $f_R$ of a hypothetical $R$-brane, with constant $e^{2\tp_R}$. We introduce the one-dimensional radius that depends on the left-over coordinate $w$: $r_1^2= r_2^2 - z^2 = w^2$. We introduce again an $\epsilon$ that will be sent to $\infty$. Then
\bea
f_R (r_1) - e^{2\tp_R}= \int_{-\epsilon}^{+\epsilon} \d z\ (& f_Q (r_2) - e^{-2\tp_Q}) =  - q\pi \int_{-\epsilon}^{+\epsilon} \d z\ \ln (z^2 + r_1^2)\\
& = - q\pi \left[ z \ln (z^2 + r_1^2) \right]_{-\epsilon}^{+\epsilon} + q\pi \int_{-\epsilon}^{+\epsilon} \d z\ z \frac{2z}{z^2 + r_1^2}\\
& = - 2q\pi \epsilon \ln (\epsilon^2 + r_1^2) + 2q\pi \int_{-\epsilon}^{+\epsilon} \d z\ \left( 1 - \frac{r_1^2}{z^2 + r_1^2} \right)\\
& = - 2q\pi \left( \epsilon \ln (\epsilon^2 + r_1^2) - 2 \epsilon \right)- 2q\pi r_1 \left[ \arctan\left( \frac{z}{r_1} \right) \right]_{-\epsilon}^{+\epsilon} \ .
\eea
As for the $Q$-brane, the first term diverges. We consider again a cutoff and absorb it in a redefinition of the constant. We are then left with the second term, that gives for $\epsilon \rightarrow \infty$
\beq
f_R (r_1) = e^{2\tp_R} - 2q\pi^2 r_1 = e^{2\tp_R} - 2q\pi^2 |w| \ .\label{fR}
\eeq
The absolute value is known to be a solution of the one-dimensional Poisson equation
\beq
\Delta_1 f_R = c_R\ \delta^{(1)} (r_1) \ ,
\eeq
and one can again verify that away from the singularity,
\beq
\Delta_1 f_R = 0 \ {\rm for}\ r_1>0 \ .
\eeq

Although smearing the warp factor seems to work and to yield a consistent result, performing a T-duality along $z$ is more challenging. It would require to smear as well the $b$-field or the $\b$, for which there is no clear procedure. Maybe one could rather consider a direct T-duality transformation of the flux, as proposed in \cite{Plauschinn:2013wta}, since the flux is a better defined quantity that does not depend on $z$. We hope to come back to this possible $R$-brane solution in a future work. Note that it should be different than the one proposed in \cite{Hassler:2013wsa}, that rather involves a dual coordinate.

\subsubsection{Smeared branes, T-duality and sourced Bianchi identities}\label{sec:TdBI}

We have just shown how the warp factors of the different branes are related by smearing, and how this allowed them to solve the various Poisson equations. We have now all the tools necessary to T-dualise the (smeared) $\NS$-branes into one another, and then verify that the Bianchi identities \eqref{delf2} - \eqref{bdelR2} for their fluxes lead to the Poisson equations. We start with the $Q$-brane, as it involves most of the ingredients needed for the others.

\subsubsection*{$Q$-brane}

We are going to obtain the $Q$-brane by T-dualising the $\NS 5$-brane along two directions. To do so, we should first smear the latter. This amounts to consider the smeared warp factor $f_Q$ of \eqref{Qbrsol} instead of the standard $f_H$ of \eqref{NS5sol}, and to use cylindrical coordinates: $\rho=r_2$ and $\varphi$ for polar coordinates, and $x,y$ cartesian for the two smeared directions. Those coordinates are the most appropriate, not only because of the two-dimensional radius in $f_Q$, but also for T-duality. Unless one uses a procedure as the one of \cite{Plauschinn:2013wta}, T-duality requires to have a $b$-field. Given the expression of the $H$-flux in \eqref{HNS5simple} and the relation $H_{mnp}= 3 \del_{[m} b_{np]}$, it is much simpler to obtain a $b$-field that respects the isometries using those coordinates. So starting with \eqref{NS5sol}, the (twice) smeared $\NS 5$-brane is given by
\bea
& \d s^2 = \d s_6^2 + f\ \d \hat{s}_4^2 \ ,\ H_{mnp}=- \rho\ \epsilon_{4 mnp \rho} \del_{\rho} f \ ,\ e^{2\p}= f\\
& {\rm where}\ \d \hat{s}_4^2= \d \rho^2+ \rho^2 \d \varphi^2+\d x^2+\d y^2 \ ,\ f=f_Q \ ,
\eea
in curved cylindrical indices. Fixing $\epsilon_{4 \rho \varphi xy}=+1$ (see conventions in appendix \ref{ap:conv}), one computes away from the singularity the only non-trivial component of the $H$-flux
\beq
H_{\varphi x y}= q_Q \ {\rm for}\ \rho>0 \ ,
\eeq
in curved indices. We then choose the following gauge for the $b$-field
\beq
b_{xy}=-b_{yx}= q_Q\ \varphi \ {\rm for}\ \rho>0 \ ,
\eeq
so that it respects the isometries. To include the singularity, it is tempting to define
\beq
b_{mn}= \epsilon_{4 \rho \varphi mn}\ a(\rho,\varphi)\ , \ {\rm with}\ a= - \varphi\ \rho \del_{\rho} f\ ,\label{bsing}
\eeq
that gives the correct expression when acting with $\del_{\varphi}$. But it leads to undesired $H$-flux components at the singularity when acting with $\del_{\rho}$. This same ambiguity will appear below for the $\KK$-monopole and the $Q$-brane. So it is important to keep it in mind: we consider this completed but ambiguous $b$-field, and the trick to get the good fluxes is to set $\del_{\rho} a=0$.

We now T-dualise along $x$. Applying the Buscher rules,\footnote{In \cite{Andriot:2012vb} are given Buscher rules in terms of $g$ and $b$ that are equivalent to the transformation \eqref{Td}. We use those, with a minus sign difference on the $b$-field, due to conventions.} we get no $b$-field and
\bea
& \d s^2 = \d s_6^2 + f\ \d \hat{s}_3^2 +  f^{-1} (\d x + a \d y )^2 \ ,\ H_{mnp}=0\ ,\ e^{2\p}= 1\\
& {\rm where}\ \d \hat{s}_3^2= \d \rho^2+ \rho^2 \d \varphi^2+ \d y^2 \ .
\eea
This corresponds to the $\KK$-monopole \eqref{KKsol} smeared along $y$, as can be seen from the warp factor and the coordinates. The smeared $a$ present here can only be understood through this T-duality procedure though. Finally, we T-dualise along $y$ and get
\bea
& \d s^2 = \d s_6^2 + f\ \d \hat{s}_2^2 +  f^{-1} (1+\frac{a^2}{f^2})^{-1} (\d x^2 + \d y^2 ) \ ,\ b_{xy}=- b_{yx} = -a f^{-2} (1+\frac{a^2}{f^2})^{-1} \ ,\label{bQ-brane}\\
& e^{2\p}= f^{-1} (1+\frac{a^2}{f^2})^{-1}\ , \ {\rm where}\ \d \hat{s}_2^2= \d \rho^2+ \rho^2 \d \varphi^2 \ ,\nn
\eea
which has been argued in \cite{deBoer:2010ud} to be non-geometric. Using the field redefinition \eqref{fieldredef}, we get precisely the $Q$-brane solution \eqref{Qbrsol}
\bea
& \d \tilde{s}^2 = \d s_6^2 + f\ \d \hat{s}_2^2 +  f^{-1} (\d x^2 + \d y^2 ) \ ,\ \b^{xy}=- \b^{yx} = a \ ,\ e^{2\tp}= f^{-1} \\
& {\rm where}\ \d \hat{s}_2^2= \d \rho^2+ \rho^2 \d \varphi^2 \ .
\eea
Going around the singularity (i.e. moving along $\varphi$ at $\rho>0$), $\b$ gets shifted by a constant along the isometry directions: the gluing is then done by a $\b$-transform, and this solution is part of the class studied in sections \ref{sec:nongeoglobal} and \ref{sec:geombcgdorbit}. The T-dual background given by the smeared $\NS 5$-brane also has a linear $b$-field. As described in those sections, such a situation leads typically to a non-geometry, as in \eqref{bQ-brane}.

Let us now determine the fluxes of this solution. The vielbein is given by
\beq
\te=\begin{pmatrix} f^{\frac{1}{2}} & & & \\ & f^{\frac{1}{2}} \rho & & \\ & & f^{-\frac{1}{2}} & \\ & & & f^{-\frac{1}{2}} \end{pmatrix} \ ,
\eeq
from which we deduce the non-zero structure constants or geometric flux \eqref{fabc}
\beq
f^{\varphi}{}_{\rho \varphi} = - \frac{1}{2} f^{-\frac{3}{2}} \del_{\rho} f - f^{-\frac{1}{2}} \rho^{-1} \ ,\ f^x{}_{\rho x}=f^y{}_{\rho y}= \frac{1}{2} f^{-\frac{3}{2}} \del_{\rho} f \ ,\ f^{a}{}_{bc}=-f^{a}{}_{cb}\ ,
\eeq
where with some abuse of notation we denote on the LHS the flat indices with the corresponding curved space coordinate, and on the RHS the derivative has a curved index. We now compute the $Q$-flux. It is worth noticing that the $Q$-brane solution verifies the condition $\b^{mn} \del_n \cdot =0$, as pointed out in \cite{Hassler:2013wsa}; this holds even at the singularity. Then, one has
\beq
\xymatrix{Q_{c}{}^{ab} \equiv \del_c \b^{ab} - 2 \b^{d[a} f^{b]}{}_{cd}\ \ar@{=}[rr]^{\quad \quad \b^{mn} \del_n \cdot =0} & & \ \te^p{}_c \te^a{}_m \te^b{}_n \del_p \b^{mn}} \ ,\label{Qsimplif}
\eeq
as can be seen from \eqref{Qfluxeee}, while $R^{abc}=0$. Recalling the ambiguity of the $b$-field and $a$ in the $\NS 5$-brane discussed around \eqref{bsing}, one gets the only non-trivial component of the $Q$-flux
\beq
Q_{\varphi}{}^{xy}=  - f^{-\frac{3}{2}} \del_{\rho} f \ ,
\eeq
where we mean again flat indices, and the derivative has a curved index. This result matches precisely the smeared $\NS 5$ $H$-flux in flat indices, which confirms the validity of our procedure.

Finally, we turn to the BI. Given the fluxes just determined and using some antisymmetry arguments, one can see that \eqref{delf2}, \eqref{delR2} and \eqref{bdelR2} are satisfied. Let us rather focus on \eqref{delQ2}, and the quantity
\beq
S^{ab}_{cd}= \del_{[c}Q_{d]}{}^{ab}-\b^{e[a}\del_e f^{b]}{}_{cd}-\frac{1}{2}Q_{e}{}^{ab}f^{e}{}_{cd}+2Q_{[c}{}^{e[a}f^{b]}{}_{d]e}\ .
\eeq
The second term vanishes here. We fix $(c,d,a,b)$ to be $(\rho, \varphi, x,y)$: this is the only non-trivial choice, up to antisymmetries. One gets
\bea
S^{xy}_{\rho \varphi} & = \frac{1}{2}f^{-\frac{1}{2}} \del_{\rho}Q_{\varphi}{}^{xy}- \frac{1}{2} Q_{\varphi}{}^{xy} (f^{\varphi}{}_{\rho \varphi} - f^{y}{}_{\rho y} - f^{x}{}_{\rho x} ) \label{computationSQbr} \\
& = - \frac{1}{2} f^{-2} \left( \del_{\rho}^2 f + \rho^{-1} \del_{\rho} f \right) = - \frac{1}{2} f^{-2} \Delta_2 f \ ,
\eea
where $\Delta_2$ is the two-dimensional Laplacian obtained here in polar coordinates, since $f$ does not depend on $\varphi$. As argued in \eqref{PoissonfQ}, $f$ is here the Green function for $\Delta_2$ up to a constant $c_Q$. So we propose the following correction of the BI \eqref{delQ2} due to the source
\beq
S^{ab}_{cd} = - \frac{c_Q}{2} f^{-2}\ \epsilon_{2\bot cd}\ \epsilon_{2|| ef}\ \eta^{ea} \eta^{fb}\ \delta^{(2)} (\rho)  \ ,
\eeq
where we took into account the constraints on the indices. This results in the BI \eqref{BIsourcedQbrane}, and we have just shown that the $Q$-brane solves it.

Let us mention that a BI with a $Q$-brane source term was proposed in \cite{Chatzistavrakidis:2013jqa}. We comment on it in appendix \ref{ap:Hannover} and conclude on a mismatch with our proposal \eqref{BIsourcedQbrane}.

\subsubsection*{$\KK$-monopole}

We follow a similar procedure to show that the $\KK$-monopole is obtained by T-dualising the $\NS5$-brane along one direction. We first smear the $\NS5$-brane along $x$. Doing so amounts to choose the smeared warp factor $f_K$ of \eqref{KKsol} instead of $f_H$, and to use the better suited spherical coordinates $\rho=r_3,\ \varphi,\ y$. Then the (once) smeared $\NS 5$-brane is given by
\bea
& \d s^2 = \d s_6^2 + f\ \d \hat{s}_4^2 \ ,\ H_{mnp}=- \rho^2 \sin \varphi\ \epsilon_{4 mnp \rho}  \del_{\rho} f \ ,\ e^{2\p}= f\\
& {\rm where}\ \d \hat{s}_4^2= \d \rho^2+ \rho^2 \d \varphi^2+ \rho^2 \sin^2 \varphi\ \d y^2 + \d x^2 \ ,\ f=f_K .\nn
\eea
Similarly to the discussion for the $Q$-brane, we introduce (in curved indices)
\bea
& b_{mn}= \epsilon_{4 \rho \varphi mn}\ a(\rho,\varphi)\ , \ {\rm with}\ a= \cos \varphi\ \rho^2 \del_{\rho} f\ ,\\
&  \ b_{xy}= q_K \cos \varphi\ , \ H_{\varphi xy}=- q_K \sin \varphi\ ,\ {\rm for}\ \rho>0 \ .
\eea
We can then perform the T-duality along $x$. It is formally the same as above, giving
\bea
& \d s^2 = \d s_6^2 + f\ \d \hat{s}_3^2 +  f^{-1} (\d x + a \d y )^2 \ ,\ H_{mnp}=0\ ,\ e^{2\p}= 1\\
& {\rm where}\ \d \hat{s}_3^2= \d \rho^2+ \rho^2 \d \varphi^2+ \rho^2 \sin^2 \varphi\ \d y^2 \ ,
\eea
where now $f$ and $a$ are precisely those of the $\KK$-monopole \eqref{KKsol}, that is thus recovered.

To proceed further, we consider the following vielbein and its inverse (in the basis $(\rho, \varphi, y, x)$)
\beq
\te=\begin{pmatrix} f^{\frac{1}{2}} & & & \\ & f^{\frac{1}{2}} \rho & & \\ & & f^{\frac{1}{2}} \rho \sin \varphi & \\ & & f^{-\frac{1}{2}} a & f^{-\frac{1}{2}} \end{pmatrix} \ ,\ \te^{-1}=\begin{pmatrix} f^{-\frac{1}{2}} & & & \\ & f^{-\frac{1}{2}} \rho^{-1} & & \\ & & f^{-\frac{1}{2}} \rho^{-1} \sin^{-1} \varphi & \\ & & - f^{-\frac{1}{2}} a \rho^{-1} \sin^{-1} \varphi & f^{\frac{1}{2}} \end{pmatrix} \ ,\nn
\eeq
from which we compute the following non-trivial structure constants \eqref{fabc}
\bea
& f^{\varphi}{}_{\rho \varphi}= f^{y}{}_{\rho y}= \rho\ \del_{\rho} ( f^{-\frac{1}{2}} \rho^{-1} ) \ ,\ f^{x}{}_{\rho x}= f^{-1} \del_{\rho}  f^{\frac{1}{2}}\ ,\ f^{y}{}_{\varphi y}= f^{-\frac{1}{2}} \rho^{-1} \sin \varphi\ \del_{\varphi} ( \sin^{-1} \varphi ) \ ,\nn \\
& f^{x}{}_{\varphi y}= - f^{-\frac{3}{2}} \rho^{-2} \sin^{-1} \varphi\ \del_{\varphi} a = f^{-\frac{3}{2}} \del_{\rho} f\ ,\quad f^{a}{}_{bc}=-f^{a}{}_{cb} \ .
\eea
As above, we mean flat indices on the LHS, and the derivatives carry curved indices on the RHS. Due to the ambiguity of the $b$-field of the $\NS5$-brane and of $a$ discussed around \eqref{bsing}, we do not consider an $f^{x}{}_{\rho y}$, that would have been non-zero at the singularity. This way, for all T-dual branes, the important component of the flux has the (flat) indices $(\varphi, x, y)$ and is due to the potential, being here $a$. The value of these components even matches, up to a sign. The other $f$ present here are mostly artefacts of the metric and do not play the same role. Finally, the absence of $b$-field for the $\KK$-monopole makes the other type of fluxes vanish.

We finally turn to BI: \eqref{delQ2} - \eqref{bdelR2} are trivially satisfied, while \eqref{delf2} involves the quantity
\beq
S^a_{bcd} = \del_{[b}f^{a}{}_{cd]} - f^{a}{}_{e[b}f^{e}{}_{cd]} \ .
\eeq
By antisymmetry, $S^{\varphi}_{bcd} =0$. In addition, one can verify
\beq
S^y_{\rho \varphi y}= \frac{1}{3} \left( f^{-\frac{1}{2}} \del_{\rho} f^y{}_{\varphi y} + f^y{}_{y \varphi} f^{\varphi}{}_{\rho \varphi} \right) = 0 \ .
\eeq
Therefore, the only non-zero $S^a_{bcd}$ is given by
\bea
S^x_{\rho \varphi y} &= \frac{1}{3} \left( f^{-\frac{1}{2}} \del_{\rho} f^x{}_{\varphi y} - f^x{}_{\varphi y} \left( f^x{}_{x \rho} + f^y{}_{\rho y}+ f^{\varphi}{}_{\rho \varphi} \right) \right)\\
&= -\frac{1}{3} \sin^{-1} \varphi f^{-2} \rho^{-2} \del_{\rho} \del_{\varphi} a \\
&= \frac{1}{3} f^{-2} \left( \del_{\rho}^2 f + \frac{2}{\rho} \del_{\rho} f \right) = \frac{1}{3} f^{-2} \Delta_3 f \ ,
\eea
where $\Delta_3$ is the three-dimensional Laplacian, here in spherical coordinates, since $f$ only depends on $\rho$. We mentioned that $f$ is the Green function for $\Delta_3$ up to a constant $c_K$ \eqref{PoissonfK}. So we propose the following correction of the BI \eqref{delf2} due to the source
\beq
S^a_{bcd}= \frac{c_K}{3} f^{-2} \ \epsilon_{3\bot bcd}\ \epsilon_{1|| e}\ \eta^{ea} \ \delta^{(3)} (\rho)  \ ,
\eeq
where the constraints on the indices were taken into account, and $\epsilon_{1|| e}$ is only non-zero and equal to one if $e$ is the direction along the brane. This results in the BI \eqref{BIsourcedKKm}, and we have just shown that the $\KK$-monopole solves it.

\subsubsection*{$\NS 5$-brane}

For completeness, let us come back to the BI of the $H$-flux for the $\NS5$-brane. We showed below \eqref{BIHcurved} how this BI in curved indices would lead to the Poisson equation. Going to flat indices amounts to multiplying by vielbeins since $\d H$ is a tensor. One gets the quantity
\beq
S_{abcd}=\te^m{}_a \te^n{}_b \te^p{}_c \te^q{}_d \del_{[m} H_{npq]} = \del_{[a} H_{bcd]} - \frac{3}{2} f^e{}_{[ab} H_{cd]e}\ .
\eeq
In cartesian coordinates, the vielbeins are just given by $f^{\frac{1}{2}}$. So from \eqref{BIHcurved}, \eqref{PoissonfH}, and the above, we propose the following contribution of the source
\beq
S_{abcd} = - \frac{c_H}{4} f^{-2} \ \epsilon_{4\bot abcd} \ \delta^{(4)} (r_4)\ ,
\eeq
where only the numerical factor should be verified, and the convention for $\epsilon_4$ is in appendix \ref{ap:conv}. This results in the BI \eqref{BINS5intro}, and we have shown that the $\NS5$-brane solves it.\footnote{For the three branes, we obtained a factor $f^{-2}$ next to the $\delta$ in the source contributions to the BI. It would be better to have a generic formula that reproduces this factor, for instance with volumes or vielbeins, but we did not find any.}

\section{Geometric vacua of $\b$-supergravity}\label{sec:geomvac}

In this section, we study the conditions for a vacuum of $\b$-supergravity to be geometric, while its formulation in standard supergravity would be non-geometric. As explained in the Introduction, such backgrounds are those for which $\b$-supergravity description is truly useful. In the context of compactification, those backgrounds allow a dimensional reduction to four-dimensional gauged supergravities with non-geometric fluxes; the latter would not have a ten-dimensional uplift otherwise. Whether a background is of this type is related to the symmetries used to glue its fields from one patch to another, as mentioned in the Introduction. We mostly follow the reasoning presented there, and clarify on the way several concepts such as geometry and non-geometry, that is a theory dependent notion. We end this section by studying the properties of some of these backgrounds, namely those that use $\b$-transforms, determining in particular whether they eventually lead to new four-dimensional physics.

\subsection{Symmetries of the NSNS sector}\label{sec:sym}

We consider a field configuration in a theory (possibly a vacuum), in a target space picture, as given by a set of fields defined locally on several patches of the space, and gluing from one to the other by some transformations. In order for this field configuration to be described by a single theory, as it should be to have a good description of the physics, or in other words, in order to use only one Lagrangian over the whole space, the gluing transformations should be symmetries of that theory \cite{Blumenhagen:2013aia}. It is therefore important to first identify these symmetries, as we now turn to. In section \ref{sec:nongeoglobal}, we will then look at what type of background the symmetries lead to when used as gluing transformation.

\subsubsection{General case}\label{sec:symgen}

We will be mostly interested in the NSNS sector of standard supergravity given by the Lagrangian $\L_{{\rm NSNS}}$ \eqref{LNSNS} and the NSNS sector of $\b$-supergravity given by the Lagrangian $\tL_{\b}$ \eqref{Lb}. Up to the field redefinition, they differ as explained in section \ref{sec:betasugra} by a total derivative. In \cite{Andriot:2013xca}, we had
\bea
& \L_{{\rm NSNS}} - \del_m\left(e^{-2d}\big(\tg^{mn}\tg^{pq}\del_n\tg_{pq} - g^{mn} g^{pq} \del_n g_{pq} + \del_n(\tg^{mn}-g^{mn})\big)\right) \label{relLag}\\
= & \ \tL_{\b} + \del_m \left(\frac{e^{-2d}}{|\tg|} \del_n \big(\tg_{pq} \b^{pm} \b^{qn} |\tg| \big) - 4 e^{-2d} \b^{pm} \tg_{pq} \T^q \right) \ .\nn
\eea
The total derivative can be simplified by noticing as in \cite{Andriot:2011uh, Andriot:2012an} that $\tg^{mn}-g^{mn} = -\tg_{pq} \b^{pm} \b^{qn}$. Using in addition that $\del_n \ln |\tg| = \tg^{pq}\del_n\tg_{pq}$, one obtains
\beq
\L_{{\rm NSNS}} + \del_m\left(e^{-2d}\big(g^{mn} \del_n \ln \frac{|g|}{|\tg|} + 4 \b^{pm} \tg_{pq} \T^q )\right) = \tL_{\b}  \ .
\eeq
The field redefinition also gives that $|g|=|\tg|^{-1} |\tg^{-1} + \b|^{-2}$, from which we get
\beq
\L_{{\rm NSNS}} + \del_m\left(e^{-2d}\big(-2g^{mn} \del_n \ln |\id + \tg \b| + 4 \b^{pm} \tg_{pq} \T^q )\right) = \tL_{\b}  \ .\label{totderfinal}
\eeq
The fact they differ only by a total derivative has two crucial consequences: first the equations of motion are then the same, up to the field redefinition, so a vacuum of one theory is then, at least locally, a vacuum of the other theory. Secondly, a symmetry of a theory usually leaves its Lagrangian invariant up to a total derivative (the case of supersymmetry for instance), so here, a symmetry of one theory will be a symmetry of the other one.

The symmetries of both theories are well known and were studied in details in \cite{Andriot:2013xca}. The Lagrangians are invariant under diffeomorphisms: this is manifest in their expressions \eqref{LNSNS} and \eqref{Lb}. In addition, $\L_{{\rm NSNS}}$ is invariant under the $b$-field gauge transformation. This can be translated as a transformation on the $\b$-supergravity fields, and was called a $\b$ gauge transformation \cite{Andriot:2013xca}. $\tL_{\b}$ is then invariant under it up to a total derivative.

A field configuration that uses diffeomorphisms or $b$-field gauge transformations to glue is certainly geometric in standard supergravity (see the definition in section \ref{sec:GNG}). As we will see, it may or may not be geometric in terms of $\b$-supergravity, but in any case, such a description is not really necessary, as standard supergravity is then appropriate \cite{Blumenhagen:2013aia, Andriot:2013xca}. Therefore, it would be interesting for $\b$-supergravity to have more symmetries at hand. To reach such a situation, we necessarily have to modify the theories in some manner: we will consider a further constraint, or restriction, or subcase, that will generate an enhancement of symmetries, as suggested in \cite{Andriot:2013xca}. Let us motivate the restriction to be considered by a new symmetry that appears manifestly in $\tL_{\b}$.

\subsubsection{A new symmetry of $\b$-supergravity}\label{sec:newsym}

We present here a new symmetry of $\b$-supergravity (under some conditions), that we will later relate to the $\b$-transforms of T-duality. The Lagrangian $\tL_{\b}$, given in curved indices in \eqref{Lb}, only contains $\b$ through either $\del_m \b^{pq}$ or $\b^{pr} \del_r \cdot$, where the dot stands for any of the three fields or their derivatives. Therefore, the following holds
\begin{framed}
\beq
\begin{array}{r|}
\b^{pq} \rightarrow \b^{pq} + \varpi^{pq} \\
{\rm with}\ \forall\ m,p,q,\ \varpi^{pr} \del_r \cdot = 0 \ ,\ \del_m \varpi^{pq} = 0
\end{array}\ \ \mbox{is a symmetry of}\ \tL_{\b} \ . \label{newsym}
\eeq
\end{framed}
\noindent In others words, a constant shift of $\b$ by (an antisymmetric) $\varpi^{pq}$ satisfying $\varpi^{pr} \del_r \cdot = 0$ leaves $\tL_{\b}$ invariant. Can the two requirements on $\varpi$ in \eqref{newsym} be relaxed to a more general one, which would, for instance, not require $\varpi$ to be constant? It does not seem possible,\footnote{\label{foot:gaugebeta}It is tempting to consider the conditions
\bea
\forall\ m,p,q,\ \ \varpi^{pr} \del_r \tg^{mq} +  \tg^{pr} \del_r \varpi^{mq} = 0 & \ ,\label{gcond}\\
\varpi^{pr} \del_r \b^{mq} +  \b^{pr} \del_r \varpi^{mq} = 0 & \ . \label{bcond}
\eea
\eqref{gcond} implies the invariance of $\cG^{mn}_p$ under the shift, and so of $\T^n=\cG^{pn}_p$. In addition, \eqref{bcond} makes the linear terms in $\varpi$ in the variation of the $R$-flux vanish. One could then hope for a more general symmetry. However, using the (anti)symmetry of $m,q$ in \eqref{gcond}, one obtains that this condition and \eqref{bcond} are actually equivalent to the two of \eqref{newsym}, at least for $\tg$ and $\b$ instead of the dot.} and the relation we will establish to T-duality suggests that there is no such generalization. So we stick to this form \eqref{newsym} of the symmetry. It is now important to understand the two conditions on $\varpi$ in \eqref{newsym}, i.e. how can this symmetry be concretely realised. To that end, let us consider the following equivalence, given a field configuration and an integer $N>1$
\begin{framed}
\beq
\!\!\!\!\!\! \begin{array}{l|}
\mbox{$\exists$ $N$ isometries generated by $N$ independent}\\
\mbox{constant Killing vectors $V_{\iota},\ \iota \in \{1 \dots N\}$} \ .
\end{array}\ \Leftrightarrow \ \begin{array}{|l}
\mbox{Any constant $\varpi^{pq}$, that is only non-zero}\\
\mbox{along a specific $N\times N$ (diagonal) block,}\\
\mbox{satisfies $\varpi^{pr} \del_r \cdot =0$} \ .
\end{array}  \label{condnewsym}
\eeq
\end{framed}
\noindent We provide a rigorous proof of this equivalence in appendix \ref{ap:proofs}. As shown in that proof, the left-hand side of \eqref{condnewsym} can be translated as \eqref{proofcond2}, i.e. as the independence of the fields (and their derivatives, by commutation) on $N$ coordinates. In addition, the right-hand side of \eqref{condnewsym} gives conditions on the $\varpi$ that are precisely those needed to realise the symmetry \eqref{newsym}, up to the restriction of having a non-zero block. So this equivalence can be translated in particular into the implication\footnote{The reverse can only be formulated with the $\varpi^{pr} \del_r \cdot =0$ condition, because it is not clearly the same as the constant shift being a symmetry.}
\begin{framed}
\beq
\!\!\!\!\!\! \begin{array}{l|}
\mbox{The fields are independent}\\
\mbox{of $N$ coordinates} \ .
\end{array}\ \Rightarrow \ \begin{array}{|l}
\mbox{The shift $\b^{pq} \rightarrow \b^{pq} + \varpi^{pq}$, for any constant $\varpi^{pq}$}\\
\mbox{that is non-zero only along the $N\times N$ block,}\\
\mbox{is a symmetry of $\tL_{\b}$} \ .
\end{array}  \label{condnewsym2}
\eeq
\end{framed}
The symmetry can thus be realised provided the fields are independent of $N$ ($>1$) coordinates; the allowed shifts are then those along these isometry directions, and constant. The new symmetry \eqref{newsym} is therefore tied to having isometries: it is not a symmetry of general $\b$-supergravity, but requires to focus on the subcase (in particular, on the set of backgrounds) that have isometries. In this sense, it is reminiscent of T-duality for string theory; we will see that the two are actually related.

As this symmetry of $\tL_{\b}$ is only present in a subcase, one may wonder under what conditions it can also be a symmetry of $\L_{{\rm NSNS}}$. The field redefinition relating only the fields among themselves, the independence on the coordinates of one set of fields translates in that of the other set. So the conditions for the symmetry to be realised is the same on both sides: given the discussion made below \eqref{totderfinal}, we deduce that in this subcase, this symmetry of $\tL_{\b}$ is also a symmetry of $\L_{{\rm NSNS}}$, up to a total derivative. We can actually be more precise on this last point: in the total derivative \eqref{totderfinal}, $\b$ appears again through $\del_m \b^{pq}$ and $\b^{pr} \del_r \cdot$, but also through a determinant. The variation of this determinant does not seem to vanish, so $\L_{{\rm NSNS}}$ would be invariant under \eqref{newsym} only up to a non-vanishing total derivative. The same may happen reverse wise with constant $b$-shifts, although one should rewrite the total derivative in terms of $g$ and $b$ to verify this.

\subsubsection{Elements of the T-duality symmetry}\label{sec:Td}

We now turn to T-duality. When the target-space fields are independent of $N$ coordinates in a $D$-dimensional space-time, the bosonic string sigma-model gets an additional symmetry, that is T-duality (see the reviews \cite{Giveon:1994fu, Thompson:2010sr, Maharana:2013uvy} and references therein). This symmetry translates in the NSNS sector into the action of a constant $O(N,N)$ group on the fields. Therefore, if the latter are independent of $N$ coordinates, the target-space theory, namely $\L_{{\rm NSNS}}$, should inherit this symmetry: $\L_{{\rm NSNS}}$ is then invariant under the $O(N,N)$ transformation (up to a total derivative).\footnote{Its regime of validity as an effective theory might however be changed accordingly to the transformation.} This invariance is not often mentioned, as one usually considers a full supergravity, for instance type IIA/B, that also contains a RR sector. The latter is on the contrary not always preserved by T-duality, so T-duality is generically not a symmetry of type II supergravities, but only a transformation. Here, we only focus on the NSNS sector, and we recall in appendix \ref{ap:proofs} two approaches to show the invariance of $\L_{{\rm NSNS}}$ under this transformation, up to a total derivative. The first one is the work by Maharana and Schwarz \cite{Maharana:1992my} that considers a compactification along the isometries, and the second one is the relation between $\L_{{\rm NSNS}}$ and the Double Field Theory Lagrangian, which is invariant under the bigger group $O(D,D)$. We conclude that this $O(N,N)$ transformation is a symmetry of $\L_{{\rm NSNS}}$ (up to a total derivative) when the fields are independent of $N$ coordinates. As discussed above, the same then holds for $\tL_{\b}$ and its fields.

Let us now present in more details the action of the T-duality group $O(N,N)$. Its action on the fields is better characterised by considering the $2D \times 2D$ matrix $\hhh$, the generalized metric that depends on the metric $g$ and $b$-field, and the quantity $d$ related to the dilaton, that we introduced in section \ref{sec:betasugra}. In addition, one should consider $O(D,D)$ elements $O$ in their fundamental representation: they preserve the $2D \times 2D$ matrix
\beq
\eta= \frac{1}{2} \begin{pmatrix} 0 & \id \\ \id & 0 \end{pmatrix} \ ,\ O^T \eta O = \eta \ .
\eeq
The T-duality transformations then consist in taking a trivial embedding of $O(N,N)$ into $O(D,D)$, and acting with the corresponding elements on $\hhh$; the transformed dilaton is defined so that $d$ remains invariant
\bea
& \begin{pmatrix} a & c \\ f & h \end{pmatrix} \in O(N,N)\ ,\ O=\left(\begin{array}{cc|cc} a &  & c & \\  & \id_{D-N} & & 0_{D-N} \\ \hline f &  & h &  \\  & 0_{D-N} & & \id_{D-N} \end{array}\right) \in O(D,D) \ , \label{ONNODD}\\
& \hhh'=O^T \hhh O \ ,\ e^{-2d} = e^{-2 \p} \sqrt{|g|} = e^{-2 \p'} \sqrt{|g'|} \ . \label{Td}
\eea
Only the components along the $N$ directions are then transformed. A particular example is the Buscher transformation \cite{Buscher:1987sk, Buscher:1987qj} along all $N$ directions given by $a=h=0_{N}$, $c=f=\id_{N}$.

Let us now present the content of this $O(N,N)$ group. For string theory, any element of $O(N,N,\mathbb{Z})$ can be generated by the following three types of elements \cite{Giveon:1994fu, Thompson:2010sr}:
\begin{itemize}
\item the $GL(N,\mathbb{Z})$ subgroup: for $a\in GL(N,\mathbb{Z})$, one considers\footnote{This subgroup can be further decomposed into generators, see e.g. \cite{Giveon:1994fu} and references therein.}
\beq
O_a= \begin{pmatrix} a & 0_N \\ 0_N & a^{-T} \end{pmatrix} \in O(N,N,\mathbb{Z}) \ .
\eeq

\item the $b$-transforms: for $\varpi$ an $N\times N$ antisymmetric integer matrix, one considers
\beq
O_{\varpi}= \begin{pmatrix} \id_N & 0_N \\ \varpi & \id_N \end{pmatrix} \in O(N,N,\mathbb{Z}) \ .
\eeq

\item the Buscher transformations \cite{Buscher:1987sk, Buscher:1987qj}: for $c_i$ the $N\times N$ matrix with only one non-zero entry, equal to $1$ and placed in the $(i,i)$ position, one considers
\beq
O_{t_i}= \begin{pmatrix} \id_N -c_i & c_i \\ c_i & \id_N-c_i \end{pmatrix} \in O(N,N,\mathbb{Z}) \ .
\eeq
\end{itemize}
Let us introduce yet another set of elements
\begin{itemize}
\item the $\b$-transforms: for an integer $N\times N$ antisymmetric matrix $\varpi$, one considers
\beq
\begin{pmatrix} \id_N & \varpi \\ 0_N & \id_N \end{pmatrix} = \begin{pmatrix} 0_N & \id_N \\ \id_N & 0_N \end{pmatrix} \begin{pmatrix} \id_N & 0_N \\ \varpi & \id_N \end{pmatrix} \begin{pmatrix} 0_N & \id_N \\ \id_N & 0_N \end{pmatrix} = O_{t}^T O_{\varpi} O_t \ ,
\eeq
where we denote by $O_t$ the Buscher transformation along all $N$ directions
\beq
O_t = O_{t_1} \dots O_{t_N} = \begin{pmatrix} 0_N & \id_N \\ \id_N & 0_N \end{pmatrix} \ .
\eeq
\end{itemize}
At the level of supergravity, the stringy T-duality group just discussed is extended to $O(N,N,\mathbb{R})$. We then consider the natural extensions of the above elements towards the $GL(N,\mathbb{R})$ subgroup, the real $b$- and $\b$-transforms, where $a$ and $\varpi$ are now real. Those three sets form three independent subgroups of $SO(N,N,\mathbb{R})$ (they only contain elements that have a determinant equal to $1$). So they do not generate the whole $O(N,N,\mathbb{R})$, in particular no combination can reproduce an $O_{t_i}$ as $\det O_{t_i} = -1$. There might even be some elements of $O(N,N,\mathbb{R})$ that are not generated by a simple extension from $O(N,N,\mathbb{Z})$. Nevertheless, we will mainly focus in the following on these three subgroups of $SO(N,N,\mathbb{R})$, but we can keep in mind the possibility of further T-duality transformations.

We now look at the action of these three subgroups on the NSNS fields. We explained above that when fields are independent of $N$ coordinates, the $O(N,N)$ T-duality group is a symmetry of the Lagrangians (up to a total derivative). So each of these three transformations should then correspond to a symmetry. The action of the three subgroups of interest can be read from \eqref{ONNODD} and \eqref{Td}, but also from the corresponding action on a generalized vielbein $\reee$ (up to Lorentz transformations)
\beq
\reee'= \reee O \ .
\eeq
By considering respectively $\eee$ and $\teee$ of \eqref{genvielb}, one gets simple expressions for the $b$-transforms, resp. $\b$-transforms: they just consist in shifting the $b$-field, resp. $\b$
\bea
& \mbox{$b$-transform:}\quad e'=e,\ b'=b+\begin{pmatrix} \varpi & \\ & 0_{D-N} \end{pmatrix} \ , \label{bshifts}\\
& \mbox{$\b$-transform:}\quad \te'=\te,\ \b'=\b+\begin{pmatrix} \varpi & \\ & 0_{D-N} \end{pmatrix} \ , \label{betashifts}
\eea
along the $N$ directions. In addition, we read the $GL(N,\mathbb{R})$ action on either set of fields as
\bea
O_a:\ & e'=e \begin{pmatrix} a & \\ & \id_{D-N} \end{pmatrix},\ b'=\begin{pmatrix} a & \\ & \id_{D-N} \end{pmatrix}^{T} b \begin{pmatrix} a & \\ & \id_{D-N} \end{pmatrix} \label{Oabbeta} \ ,\\
 & \te'=\te \begin{pmatrix} a & \\ & \id_{D-N} \end{pmatrix},\ \b'=\begin{pmatrix} a & \\ & \id_{D-N} \end{pmatrix}^{-1} \b \begin{pmatrix} a & \\ & \id_{D-N} \end{pmatrix}^{-T} \ .\nn
\eea
Let us now identify the corresponding symmetries. The $b$-transforms \eqref{bshifts} are an obvious symmetry of $\L_{{\rm NSNS}}$: first, constant shifts of $b$ certainly leave the Lagrangian invariant, as the latter only depends on $\del b$; second, this shift symmetry is a subcase of the known $b$-field gauge symmetry, since a constant shift can be brought to the form of a $\d \Lambda$. The $GL(N)$ subgroup is also clearly a symmetry: its action \eqref{Oabbeta} on the fields is a particular example (in matrix notations) of diffeomorphisms, that are known to be a gauge symmetry of both $\L_{{\rm NSNS}}$ and $\tL_{\b}$. Let us verify this point. A diffeomorphism generically transforms the $b$-field as $b_{mn}(x')=b_{pq}(x) \frac{\del x^p}{\del {x'}^m} \frac{\del x^q}{\del {x'}^n}$. Having the $O_a$ transformation as a diffeomorphism amounts at first to satisfy the following set of differential equations
\beq
\begin{pmatrix} a & \\ & \id_{D-N} \end{pmatrix}^{\!p}_{\ m}=\frac{\del x^p}{\del {x'}^m} \ .
\eeq
This can easily be achieved since $a$ is constant. Additionally, of the coordinates obtained from this resolution, the field only depends on those not along the $N$ directions: thanks to the $\delta^p_m$, those can easily be chosen as $x'=x$. For that reason, $b_{mn}(x')=b_{pq}(x) \frac{\del x^p}{\del {x'}^m} \frac{\del x^q}{\del {x'}^n}$ can be realised by the action of $O_a$.

Finally, the $\b$-transforms \eqref{betashifts} should also be a symmetry when fields are independent of $N$ coordinates. This may look surprising from the $\L_{{\rm NSNS}}$ point of view, as it does not seem to match a known symmetry (in particular, translated on the standard supergravity fields, this transformation acts both on $b$ and $g$).\footnote{The two other subgroups of the T-duality group have been shown to correspond to subcases of gauge transformations, so one may wonder whether the same could happen for the $\b$-transforms. This is related to the footnote \ref{foot:gaugebeta}, and it looks unlikely. It may still be doable in the broader set-up of DFT, when considering $\tilde{\del}\neq 0$.} However, in view of \eqref{condnewsym2}, $\b$-transforms clearly correspond to the new symmetry of $\tL_{\b}$ discussed in section \ref{sec:newsym}: constant shifts of $\b$ along coordinate directions on which no field depends. It is then a symmetry of $\L_{{\rm NSNS}}$ up to a total derivative. We now understand that the new symmetry of section \ref{sec:newsym} can be viewed as the $\b$-transforms, a subgroup of the T-duality group.\\

We conclude this section on the symmetries of $\L_{{\rm NSNS}}$ and $\tL_{\b}$ by recalling our main idea: by considering a restriction, we enhance the symmetries of the theories, and the new symmetries can be used to build interesting geometric vacua of (the constrained) $\b$-supergravity. We considered here the subcase when fields are independent of $N$ coordinates: among various new symmetries from the T-duality group, we obtained the subgroup of $\b$-transforms, that is a manifest symmetry of $\tL_{\b}$. Those will play a crucial role in geometric vacua of $\b$-supergravity.

\subsection{To be or not to be geometric}\label{sec:nongeoglobal}

We discussed above the different symmetries of $\L_{{\rm NSNS}}$ and $\tL_{\b}$, in general but also when restricting to the presence of some isometries. We now study the effect of using these various symmetries to glue fields of these theories from one patch to the other: after proposing a precise definition of geometry and non-geometry, we discuss whether using a given symmetry leads to a geometric or non-geometric field configuration. To illustrate this discussion, we then provide an example for which we prove the non-geometry.

\subsubsection{Symmetries and (non)-geometry}\label{sec:GNG}

The original idea of non-geometry \cite{Hellerman:2002ax, Dabholkar:2002sy, Flournoy:2004vn} went as follows: a field configuration (string coordinates, supergravity fields...) is non-geometric for string theory if its fields can be defined on a set of patches (in target space), but the transformations needed to glue them from one to the other are not among the standard symmetries of a (differential) geometric configuration, meaning diffeomorphisms and gauge transformations. Still, these transformations are symmetries of string theory. As mentioned in the Introduction, it is important that these transformations correspond indeed to symmetries of a given theory \cite{Blumenhagen:2013aia}: this allows the field configuration to be described by a single theory on all patches, which is crucial for physics. Keeping this idea in mind, we extend here the notion of geometric or non-geometric field configuration to our target space theories: the transformations used to glue the fields should then be symmetries of the latter, and not only of string theory. Then, to distinguish between a geometry and a non-geometry requires to specify the symmetries used. We thus reformulate and generalize the original idea stated above into the following proposed definitions

\subsubsection*{Definitions of geometric and non-geometric field configurations}

\begin{itemize}
\item A field configuration is geometric if the fields are globally defined on the manifold considered so do not need to be glued, or if the transformations used to glue them from one patch to the other are symmetries of the theory, and the metric, dilaton and fluxes glue at most with diffeomorphisms.

\item A field configuration is non-geometric if the transformations used to glue the fields from one patch to the other are symmetries of the theory, and if the metric, dilaton or fluxes glue with something else than diffeomorphisms.
\end{itemize}
It is important to notice that the notion is theory dependent. In particular, since the metric describing the manifold may change from one theory to the other (as it is the case for us with $\L_{{\rm NSNS}}$ and $\tL_{\b}$), the notion of (non-)geometry changes accordingly. This is precisely the interest in changing theory to describe a background: it can be non-geometric for one theory, but the geometry can be restored in another theory; this is what happens for the toroidal example as we will see in details in section \ref{sec:proofnongeo}, and for the $Q$-brane as discussed below \eqref{bQ-brane}. These definitions also involve the notion of fluxes. In $\L_{{\rm NSNS}}$, respectively $\tL_{\b}$, the $H$-flux, resp. the $R$-flux, are tensors, so their transformation under diffeomorphisms is clear. But one also faces the structure constant or geometric flux, and the $Q$-flux, which are not tensors. Their transformation under diffeomorphisms can still be considered, as they correspond to building blocks of the spin connections $\omega$ and $\omega_Q$, and those evolve on a manifold. For a geometric configuration, it is important that the flux remains invariant under the other symmetries: the $H$-flux is invariant under the $b$-field gauge transformations, and the $Q$- and $R$-flux are invariant under the $\b$-transform discussed above. The latter is obvious for the $R$-flux given its definition, and for the $Q$-flux when rewritten as
\beq
Q_c{}^{ab}= \te^q{}_c \te^a{}_{m} \te^b{}_{n} \left( \del_q \b^{mn} +2 \te^d{}_{q} \b^{p[m} \del_p \te^{n]}{}_{d} \right) \label{Qfluxeee} \ .
\eeq

These definitions therefore emphasise the role of the symmetries of a theory. We identified above the symmetries of $\tL_{\b}$ and $\L_{{\rm NSNS}}$; we explained they share the same ones up to a total derivative. Those are diffeomorphisms and $b$-field/$\b$ gauge transformations. In the case where the fields are independent of $N$ coordinates (this will be implicit from now), one gets an enhancement of the symmetries to include the T-duality group $O(N,N)$. One of its subgroups, the $\b$-transforms, is of particular interest; $\tL_{\b}$ is manifestly invariant under it. Considering these various symmetries to glue the fields, let us now study whether, according to the above definitions, a field configuration is geometric (G) or non-geometric (NG) in the different theories. We give the results in table \ref{tab:GNG}.

\begin{table}
\begin{center}
\begin{tabular}{|c||c|c||c|}
\hline
Symmetry used as gluing transformation & $\L_{{\rm NSNS}}$ & $\tL_{\b}$ & Example\\
\hline
\hline
nothing or diffeo. & G & G & twisted torus \\
$b$-field gauge transfo. (and diffeo.) & G & NG (or $\times$) & $T^3 + {\rm constant}\ H$\\
$\b$-transform (and diffeo.) & NG (or $\times$) & G & toroidal example \\
$b$-field gauge transfo. and $\b$-transform (and diffeo.) & NG (or $\times$) & NG (or $\times$) & \\
Buscher transformation & NG & NG & radial inversion \\
more combinations & ? & ? &  \\
\hline
\end{tabular}
\caption{Geometric (G) or non-geometric (NG) field configuration, according to the symmetry used to glue its fields, and to the theory}\label{tab:GNG}
\end{center}
\end{table}

We denote by a $\times$ in table \ref{tab:GNG} a (tiny) possibility for a field configuration to be geometric, discussed in \cite{Andriot:2013xca}. The $b$-field gauge transformation, translated after field redefinition into a $\b$ gauge transformation, also acts on the new metric $\tg$; this is due to the non-linearity of the field redefinition. Depending on the transformation and the background, the transformation of $\tg$ could happen to correspond to a diffeomorphism \cite{Andriot:2013xca}. In that case, the field configuration would be geometric, provided the fluxes also transform properly. Such a situation is rather unlikely, but cannot be fully excluded. A similar reasoning can hold for the $\b$-transform, that would act not only on the $b$-field but also on the metric $g$, as can be seen with the field redefinition; one should determine whether this transformation could be viewed as a diffeomorphism. To study such situations properly, an analysis as the one to be performed in section \ref{sec:proofnongeo} would be necessary.

We mentioned in section \ref{sec:Td} the possibility of other elements of the T-duality group $O(N,N)$ that we have not considered. These could be built for instance by further combinations of the elements already studied here. The effect of such a generic element is not easy to guess, so we cannot conclude in full generality: this is the meaning of the last line of table \ref{tab:GNG}.\\

To conclude this study, we refer to the reasoning detailed in the Introduction, and one can see that the results of table \ref{tab:GNG} are in good agreement with it. In particular, it is worth considering a subcase that gives rise to more symmetries, and allows to go beyond the situations of the first two lines of table \ref{tab:GNG}. Considering the independence on $N$ coordinates gives the new symmetry of $\b$-transforms. The latter allows, as indicated in the third line, to get field configurations that are geometric for $\tL_{\b}$ while being non-geometric for $\L_{{\rm NSNS}}$. In that case, it is worth changing theory: this is the important outcome of this study. We have given a well-defined class of backgrounds for which $\b$-supergravity provides a better description than standard supergravity.

\subsubsection{A proof of non-geometry}\label{sec:proofnongeo}

We now illustrate the above discussion with an example of a field configuration that is geometric for $\tL_{\b}$ and non-geometric for $\L_{{\rm NSNS}}$. Being sure of the latter requires to show explicitly that some gluing transformations cannot be realised by diffeomorphisms, which is not so simple to prove. Such a proof should nevertheless be established to conclude on a non-geometry, but it is rarely worked-out in the literature. We hope here to fill this gap, at least for one example. We consider the toroidal example that was discussed in details in \cite{Andriot:2011uh, Andriot:2012vb, Andriot:2013xca}. In this field configuration, one has three directions, labelled by $m=1,2,3$. The third one is a circle, parameterized by the angle coordinate $z$. It serves as a base to a fiber where the non-geometry occurs. The fields are given as follows
\bea
& g = f_0(z) \begin{pmatrix} \frac{1}{R_1^2} & 0 & 0 \\ 0 & \frac{1}{R_2^2} & 0 \\ 0 & 0 & \frac{R^2_3}{f_0(z)} \end{pmatrix} \ , \ b =  f_0(z) \begin{pmatrix} 0 & -\frac{H z}{R_1^2 R_2^2} & 0 \\ \frac{H z}{R_1^2 R_2^2} & 0 & 0 \\ 0 & 0 & 0 \end{pmatrix}\ , \label{torexnongeo}\\
& e^{-2\p}= e^{-2\p'} R_1^2 R_2^2\ f_0^{-1}(z) \ , \ {\rm with}\ f_0(z)=\left(1 + \left(\frac{H z}{R_1 R_2} \right)^2 \right)^{-1} \ ,
\eea
where $H$ and the $R_m$ are constants, and $\p'$ is a given well-defined scalar field. Let us consider the base circle along $z$. An atlas of a circle needs at least two charts $(U_i, \varphi_i)$, $i=1,2$, where $U_i$ is an open set of points of the circle (or patch), and $\varphi_i$ maps them to a local coordinate in $\mathbb{R}$. The points of the circle can be uniquely denoted in a plane by $(\cos z, \sin z)$, and one can then take $\varphi_i^{-1}: \mathbb{R} \rightarrow U_i \ ,\ z \mapsto (\cos z, \sin z)$ (see the Example 5.2 in \cite{Nakahara}). The two coordinates $z_{1,2}$ associated to the two open sets $U_{1,2}$ are enough to cover the full circle: $z_1 \in ]-\pi , \pi[$, $z_2 \in ]0 , 2\pi[$. The maps between the coordinates $\Psi_{ij}=\varphi_i \varphi_j^{-1}$ are then defined on the (image of the) intersection of the patches: this "overlap" splits into two pieces, on which $\Psi_{12}$ is defined as follows
\beq
\Psi_{12} \begin{cases} z_2\mapsto z_1=z_2\ ,\qquad \ \ {\rm for}\  z_2 \in ]0 , \pi[ \\ z_2\mapsto z_1=z_2 - 2\pi\ ,\ {\rm for}\ z_2 \in ]\pi , 2 \pi[ \end{cases} \label{Psi12}
\eeq
$\Psi_{21}$ is its inverse, and both are $\mathcal{C}^{\infty}$. For the field configuration \eqref{torexnongeo} to be geometric, one needs at least the metric to glue with diffeomorphisms on the overlap. As the metric only depends on $z$ here, it should then satisfy
\beq
g_{mn}(z_2)=g_{pq}(z_1) \frac{\del x^p}{\del {x'}^m} \frac{\del x^q}{\del {x'}^n},\quad x^3=z_1 ,\ {x'}^3=z_2 \ ,
\eeq
on both pieces of the overlap. Let us verify this. For $m=n=3$, one can develop on both sides and gets
\beq
R_3^2=R_3^2 \left(\frac{\del z_1}{\del z_2}\right)^2 + f_0(z_1) \left( \frac{1}{R_1^2} \left(\frac{\del x^1}{\del z_2}\right)^2 + \frac{1}{R_2^2} \left(\frac{\del x^2}{\del z_2}\right)^2  \right) \ .
\eeq
The map $\Psi_{12}$ in \eqref{Psi12} gives on both pieces of the overlap $\frac{\del z_1}{\del z_2}=1$. One deduces
\beq
\frac{\del x^1}{\del z_2}=\frac{\del x^2}{\del z_2}=0 \ ,\label{x1z2}
\eeq
so that the diffeomorphism gluing is verified for $m=n=3$. For $m=1, n=3$, one gets
\beq
 0=R_3^2 \frac{\del z_1}{\del {x'}^1}\frac{\del z_1}{\del z_2} + f_0(z_1) \left( \frac{1}{R_1^2} \frac{\del x^1}{\del {x'}^1}\frac{\del x^1}{\del z_2} + \frac{1}{R_2^2} \frac{\del x^2}{\del {x'}^1} \frac{\del x^2}{\del z_2} \right) \ .
\eeq
Using that on both pieces of the overlap $\frac{\del z_1}{\del z_2}=1$ and \eqref{x1z2} holds, one deduces
\beq
\frac{\del z_1}{\del {x'}^1}=0 \ .
\eeq
Considering $m=2, n=3$, one obtains similarly $\frac{\del z_1}{\del {x'}^2}=0$. We now turn to $m=n=1$ (the case $m=n=2$ is completely identical). One gets a priori
\beq
f_0(z_2) \frac{1}{R_1^2}=R_3^2 \left(\frac{\del z_1}{\del {x'}^1}\right)^2 + f_0(z_1) \left( \frac{1}{R_1^2} \left(\frac{\del x^1}{\del {x'}^1}\right)^2 + \frac{1}{R_2^2} \left(\frac{\del x^2}{\del {x'}^1}\right)^2  \right) \ ,
\eeq
that simplifies, thanks to the above, to
\beq
\frac{f_0(z_2)}{f_0(z_1)} = \left(\frac{\del x^1}{\del {x'}^1}\right)^2 + \frac{R_1^2}{R_2^2} \left(\frac{\del x^2}{\del {x'}^1}\right)^2 \ , \label{fracf0}
\eeq
that should hold on both pieces of the overlap. There, one has by definition $z_1=\Psi_{12}(z_2)$, so the left-hand side (LHS) of \eqref{fracf0} is a function of $z_2$. However, because of \eqref{x1z2}, $x^1$ and $x^2$ do not depend on $z_2$, so neither does the right-hand side (RHS) of \eqref{fracf0}. Therefore, one must have
\beq
\frac{f_0(z_2)}{f_0(\Psi_{12}(z_2))}= {\rm constant} \ . \label{f0per}
\eeq
On the piece $z_2 \in ]0 , \pi[$, this certainly holds, but it is not the case on $z_2 \in ]\pi , 2 \pi[$, where
\beq
\frac{f_0(z_2)}{f_0(z_1)}= \frac{1 + \left(\frac{H (z_2 - 2 \pi)}{R_1 R_2} \right)^2}{1 + \left(\frac{H z_2}{R_1 R_2} \right)^2} \ .
\eeq
On $z_2 \in ]\pi , 2 \pi[$, because $\Psi_{12}(z_2)=z_2 - 2 \pi$, the condition \eqref{f0per} can be viewed as requiring $f_0$ to be periodic, up to a rescaling. In other words, the diffeomorphism gluing of the metric \eqref{torexnongeo} fails because of $f_0$, which is not periodic in $z$. The metric being diagonal, its chances of being globally defined boil down to simply being periodic, which is not the case. The $b$-field would also have required a diffeomorphism (together with a gauge transformation), that similarly fails due to $f_0$. Following the definitions of section \ref{sec:GNG}, we conclude that the field configuration is not geometric; the fact that it is non-geometric requires a little more.

This field configuration is independent of $N=2$ coordinates, corresponding to the fiber directions. As argued in section \ref{sec:Td}, the theories considered here then enjoy an enhancement of the symmetry group by the T-duality group $O(2,2)$, which is also a stringy symmetry. Gluing this field configuration by such a symmetry, knowing that it is not geometric, would make it non-geometric (from the standard supergravity point of view). It is indeed the case: more precisely, according to \eqref{Td}, one should have on both pieces of the overlap
\beq
\hhh(z_2)= O^T \hhh(z_1) O \ ,
\eeq
and we get that $O$ is a $\b$-transform. This is more easily seen using the new fields, given by
\beq
\! \tg = \begin{pmatrix} \frac{1}{R_1^2} & 0 & 0 \\ 0 & \frac{1}{R_2^2} & 0 \\ 0 & 0 & R_3^2 \end{pmatrix}\ , \ \b = \begin{pmatrix} 0 & H z & 0 \\ - H z & 0 & 0 \\ 0 & 0 & 0 \end{pmatrix}\ , \ e^{-2\tp}= e^{-2\p'} R_1^2 R_2^2 \ , \label{tgbtorex}
\eeq
and their associated generalized vielbein $\teee$. For $z_2 \in ]0 , \pi[$, $z_1=z_2$ so one can take $O=\id_6$. The non-trivial gluing is for $z_2 \in ]\pi , 2 \pi[$, where $z_1=z_2-2\pi$. The constant shift along the fiber directions between $\b(z_1)$ and $\b(z_2)$ can be compensated by the following $\b$-transform
\beq
O=\begin{pmatrix} \id_3 & \Omega \\ 0 & \id_3 \end{pmatrix} \ , \ \Omega= \begin{pmatrix} 0 & 2\pi H & 0 \\ -2 \pi H & 0 & 0 \\ 0 & 0 & 0 \end{pmatrix} \ . \label{TCtorex}
\eeq
We conclude that the field configuration \eqref{torexnongeo} is indeed non-geometric for standard supergravity (with isometries). According to the discussion of section \ref{sec:GNG}, in particular the definitions and the table \ref{tab:GNG}, we conclude as well that this field configuration, described as \eqref{tgbtorex}, is geometric for $\b$-supergravity (with isometries).

\subsection{Geometric backgrounds of $\b$-supergravity and T-duality orbits}\label{sec:geombcgdorbit}

As explained in the Introduction, backgrounds that are geometric for $\tL_{\b}$ and non-geometric for $\L_{{\rm NSNS}}$ are the most interesting ones for $\b$-supergravity. We have just established that one way to realise such backgrounds is to consider the restriction of having fields independent of $N$ coordinates, and to have the gluing transformations of the fields to be $\b$-transforms, possibly with additional diffeomorphisms (see for instance table \ref{tab:GNG}). We focus in this section on such a situation. The restriction implies that the background is on a T-duality orbit, i.e. the presence of the isometries allows to perform T-dualities on the background. We study this orbit and its consequences, first in general and then in a compact case.

\subsubsection{Always on a geometric orbit?}

We consider a background of the type just described. It is given in terms of the fields $\tg,\ \b,\ \tp$, thanks to which it is geometric (G) for $\tL_{\b}$. Through the field redefinition, it is expressed with $g,\ b,\ \p$ and is then non-geometric (NG) for $\L_{{\rm NSNS}}$.\footnote{Despite its similarity with a Buscher T-duality along all $D$ directions, let us stress that the field redefinition \eqref{fieldredef} is not such a transformation. The indices of $\tg^{-1}+\b$ are up, while those of a T-dual metric and $b$-field are down; in particular T-duality relates a $b$-field to a $b$-field, there is no notion of bivector appearing. Another way to see this is by considering the subcase $b=\b=0$, giving $g=\tg$, while a T-duality along all directions would invert the metric. This difference is crucial for the large volume limit (see a related discussion in \cite{Andriot:2013xca}). Additionally, in supergravity, a T-duality along all directions would require the fields to be constant, while the field redefinition can be performed without restriction. In DFT, such a T-duality would replace the coordinates $x^m$ by $\tilde{x}_m$, but the field redefinition does not change the coordinate dependence.} As it is independent of $N$ coordinates, one can further T-dualise along these directions. Doing so, along all $N$ directions, with Buscher T-duality on $g,\ b,\ \p$ gives the T-dual fields $g',\ b',\ \p'$, as depicted in table \ref{tab:TdN}.

\begin{table}
\begin{center}
\begin{tabular}{|c||ccc|}
\hline
Theories & $\quad \quad \tL_{\b} $ & & $\qquad \qquad \qquad \quad\ \ \L_{{\rm NSNS}} $ \\
\hline
\hline
$ \xymatrix{ \\ \mbox{T-duality frames} \\ } $  & \multicolumn{3}{c|}{ $ \xymatrix{\tg,\ \b,\ \tp\ \mbox{(G)}\ \ar@{<->}[rrr]^{\mbox{field redef.}\ } & & &\ g,\ b,\ \p \ \mbox{(NG)} \ar@{<->}[dd]|{{\small\mbox{T-d.}}\ //\ N\ \small{\mbox{dir.}}} \\ & & & \\ & & &\ g',\ b',\ \p'\ \mbox{(G)} }$ } \\
\hline
\end{tabular}
\caption{Different descriptions of a geometric background of $\tL_{\b}$}\label{tab:TdN}
\end{center}
\end{table}

Let us now show that $g',\ b',\ \p'$ provide a geometric background of $\L_{{\rm NSNS}}$. The fields $\tg,\ \b,\ \tp$ glue with a $\b$-transform and possibly a diffeomorphism $A$. These transformations can be decomposed into their blocks along the $N$ directions and the others: we introduce $A$ as in \eqref{Adiffeo} with $a$ the $N\times N$ block. Using notations of section \ref{sec:proofnongeo}, we denote by $z^p$ the $D-N$ coordinates on which the fields depend and by $y^r$ the $N$ coordinates on which they don't. Then, a generic diffeomorphism $A^m{}_n =\frac{\del x^m}{\del {x'}^n}$ becomes here
\beq
A=\begin{pmatrix}a & j \\ i & k \end{pmatrix} \ ,\qquad \begin{pmatrix} \d y \\ \d z \end{pmatrix}= \begin{pmatrix}a & j \\ i & k \end{pmatrix} \begin{pmatrix} \d y' \\ \d z' \end{pmatrix} \ . \label{Adiffeo}
\eeq
The independence of the fields on $N$ coordinates $y^r$ leads here to a constraint on the possible diffeomorphisms to be used: the $z$ and $z'$ should mix at most among themselves, i.e. should not involve any $y$ or $y'$ dependence. This implies that $\frac{\del z^p}{\del {y'}^r}=0$, i.e. $i^p{}_r=0$. As a cross-check, one should have $\frac{\del}{\del {y'}^r} k^p{}_q = \frac{\del}{\del {y'}^r} \frac{\del z^p}{\del {z'}^q}=0$. As $A$ is a diffeomorphism, this equals $\frac{\del^2 z^p}{\del z'^q\del {y'}^r}= \frac{\del}{\del {z'}^q} i^p{}_r$, that indeed vanishes for $i=0$. So $A$ is restricted as follows\footnote{The restriction on the dependence on coordinates enforces $i=0$, and this will allow us to obtain a geometric T-dual. This is a crucial point, as $i\neq0$ would have lead to a non-trivial $\b$-transform block after the T-duality, which would have implied a non-geometric T-dual. Another take on this is to consider the Maurer-Cartan one-forms that are globally defined: $\te^a(x')=\te^a(x)$. This provides the diffeomorphism matrix, as $\d {x}^n=\te^n{}_a(x) \te^a{}_m (x') \d {x'}^m$. Considering a multiple step fibration, such as the nilmanifold $n\ 3.14$, one may think that it is possible to find a vielbein leading to $i\neq 0$. But this involves a dependence on coordinates that are not well-defined, namely those corresponding to fibered directions. These, in addition, make the fields depend on the wrong coordinates after gluing. Considering a correct coordinate dependence restores $i=0$.}
\beq
A=\begin{pmatrix}a & j \\ 0 & k \end{pmatrix}\ ,\ A^{-T}=\begin{pmatrix}a^{-T} & 0 \\ -k^{-T} j^T a^{-T} & k^{-T} \end{pmatrix} \ .
\eeq
We now consider the gluing of the fields $\tg,\ \b,\ \tp$: using again notations of section \ref{sec:proofnongeo}, it is expressed with the generalized metric as
\bea
& \hhh(z_2)= O^T \hhh(z_1) O \ ,\label{patchH}\\
& O=\left(\begin{array}{cc|cc} \id_N &  & \varpi & \\  & \id_{D-N} & & 0_{D-N} \\ \hline 0_N &  & \id_N &  \\  & 0_{D-N} & & \id_{D-N} \end{array}\right) \left(\begin{array}{cc|cc} a & j & 0_N & \\  & k & & 0_{D-N} \\ \hline 0_N &  & a^{-T} &  \\  & 0_{D-N} & -k^{-T} j^T a^{-T} & k^{-T} \end{array}\right)
\eea
with $\varpi^T=-\varpi$ giving the $\b$-transform. As already mentioned, the field redefinition does not change $\hhh$, so the gluing of the fields $g,\ b,\ \p$ is expressed in the same manner. Let us now perform the Buscher T-duality along the $N$ directions. Following \eqref{ONNODD} and \eqref{Td}, we use again $\hhh$ to get the T-dual $\hhh'$ as
\beq
\hhh'=T^T \hhh T\ ,
\eeq
where $T$ is given below \eqref{Td}. By T-dualising $\hhh$ on (the image of) each patch, i.e. on both sides of \eqref{patchH}, we deduce the gluing of $\hhh'$
\beq
\hhh'(z_2)= (TOT)^T \hhh'(z_1) TOT\ ,
\eeq
where we used that $T^T=T^{-1}=T$. This gluing is therefore given by
\beq
TOT=\left(\begin{array}{cc|cc} \id_N &  & 0_N & \\  & \id_{D-N} & & 0_{D-N} \\ \hline \varpi & jk^{-1} & \id_N &  \\  -(j k^{-1})^T & 0_N & & \id_{D-N} \end{array}\right) \left(\begin{array}{cc|cc} a^{-T} &  & 0_N & \\  & k & & 0_{D-N} \\ \hline 0_N &  & a &  \\  & 0_{D-N} &  & k^{-T} \end{array}\right) \ . \label{patchH'}
\eeq
We recognise the combination of a $b$-shift and a diffeomorphism, where the former is due to the initial $\b$-transform and the off-diagonal piece $j$ of the diffeomorphism. We conclude that the fields $g',\ b',\ \p'$ form a geometric background for $\L_{{\rm NSNS}}$.

We have shown that the backgrounds that glue with $\b$-transform and diffeomorphism, i.e. geometric for $\tL_{\b}$ and non-geometric for $\L_{{\rm NSNS}}$, are T-dual to geometric ones for $\L_{{\rm NSNS}}$. So these geometric backgrounds of $\tL_{\b}$ are in a sense not new, or do not reveal new physics. One way of phrasing this is from a four-dimensional gauged supergravity point of view: these backgrounds are always on a geometric (T-duality) orbit. The converse claim may still be of interest. Consider a geometric background of four-dimensional gauged supergravity. On its T-duality orbit, there are geometric and possibly non-geometric backgrounds. If one geometric point on this orbit can be lifted to a ten-dimensional background that glues as in \eqref{patchH'}, then we know that there exists on that orbit a non-geometric one that can be lifted and described by $\b$-supergravity.\\

It is disappointing that the backgrounds of $\b$-supergravity considered above do not lead to new physics. Here is a list of ways to circumvent a similar result for other backgrounds
\begin{itemize}
\item As indicated in table \ref{tab:GNG}, there might be other T-duality elements that could be used to glue fields. They may, as for the $\b$-transform, allow geometric backgrounds for $\tL_{\b}$ and non-geometric for $\L_{{\rm NSNS}}$. Then, if a study as the above on the T-duals does not give rise to any geometric point, then the corresponding backgrounds would be fully new.

\item We only studied the NSNS sector. Considering backgrounds involving other sectors, such as RR, may alter the above conclusion.

\item One may find another restriction than the independence of coordinates, that would as well enhance the symmetries. The new symmetries could then be used again for gluing fields, possibly in the desired way. In particular, if there is no assumption on the coordinate dependence anymore, then the T-duality can a priori not be performed, preventing from the above conclusion.

\item There is a discrete symmetry of $\L_{{\rm NSNS}}$ that we have not mentioned so far: the $\mathbb{Z}_2$ transforming $b \rightarrow -b$. This also gives a sign to the $H$-flux and could therefore lead to a non-geometric field configuration, following the definitions of section \ref{sec:GNG}. This $\mathbb{Z}_2$ translates for $\tL_{\b}$ into a sign on $\b$ only. The effect on the fluxes is a sign on the $Q$-flux, but not on the $R$-flux. Then, with a vanishing $Q$-flux, such a field configuration would be geometric for $\tL_{\b}$: would that be another restriction to consider on $\b$-supergravity? Although very simple, this situation could be worth being studied more.

\item The notion of geometry used above is close to that of standard differential geometry and smooth manifolds. If singularities are present, the conclusions may be altered. Nevertheless, in the case of the $Q$-brane and $\NS 5$-brane, the previous reasonings can be applied everywhere away from the singularity, and the latter is treated in the same way for both $\tg,\ \b,\ \tp$ and $g',\ b',\ \p'$ (therefore if the singularity is acceptable on one side, it is as well on the other one).

\end{itemize}

\subsubsection{On compact purely NSNS vacua}

We discussed in \cite{Andriot:2013xca} the possibility of getting purely NSNS solutions of $\b$-supergravity, that would be of interest for compactification. Such vacua would be geometric for $\tL_{\b}$ and take the form of a given compactification ansatz. Interestingly, that ansatz was shown to be not too restrictive: the equations of motion indicated the possibility of getting non-trivial solutions. This is not the case for $\L_{{\rm NSNS}}$, for which the ansatz only leads to trivial solutions, hence the interest in getting such vacua of $\tL_{\b}$. In the above, we worked-out a well-defined class of backgrounds that are geometric for $\tL_{\b}$, and could thus serve as candidates for the vacua we are now interested in. However, we have also shown that these backgrounds are T-dual to geometric ones of $\L_{{\rm NSNS}}$, as described by the chain of relations in table \ref{tab:TdN}. Let us now study how the compactification ansatz evolves through that chain: this will constrain further the possibility of getting geometric vacua of $\tL_{\b}$ that are suited for compactification.

We recall that due to $\tL_{\b}$ and $\L_{{\rm NSNS}}$ differing only by a total derivative, and to the T-duality being a symmetry of the equations of motion, a vacua of $\tL_{\b}$ given by $\tg,\ \b,\ \tp$ leads to $g,\ b,\ \p$ and $g',\ b',\ \p'$ of table \ref{tab:TdN} being as well vacua of $\L_{{\rm NSNS}}$. Let us now look at the compactification ansatz. The metric $\tg$ has to be block diagonal in between the four-dimensional space-time and the internal six-dimensional manifold. We consider as well a separation of the corresponding coordinate dependence; in particular there is no warp factor. $\b$ has the same structure, but is in addition purely internal. This structure certainly goes through the field redefinition and the T-duality: $g'$ and $b'$ have the same block structure and coordinate dependence. Finally, our ansatz sets $\tp=\mbox{constant}$. Is that also the case of $\p'$? Let us recall that the dilaton goes through the following chain of equalities
\beq
e^{-2 \tp} \sqrt{|\tg|}=e^{-2 \p} \sqrt{|g|}=e^{-2 \p'} \sqrt{|g'|} \ .\label{dilequality}
\eeq

Having $\p'$ constant would put a severe constraint on the possibility of getting $\tg,\ \b,\ \tp$ as the type of vacua we are interested in. Indeed, one can show that a constant $\p'$ only leads to a trivial solution of $\L_{{\rm NSNS}}$, namely a flat space-time and manifold (vanishing Ricci tensor), and a vanishing $H$-flux. The corresponding background in terms of $\tg,\ \b,\ \tp$ is then most likely trivial as well: consider for instance constant $g',\ b',\ \p'$  or even a pure gauge $b'$, that do not give much freedom to get interesting $\tg,\ \b,\ \tp$. So $\p'$ should better be non-constant. Is that compatible with $\tp$ being constant? This requires the ratio
\beq
\frac{\sqrt{|\tg|}}{\sqrt{|g'|}} \label{ratio}
\eeq
to be non-constant. Note that $\tg$ and $g'$ being part of geometric backgrounds, they are globally well-defined. For $\tp$ being constant, we deduce that $\p'$ is also globally well-defined.\footnote{We also note that $g$ is part of a non-geometric background. Because of the equalities \eqref{dilequality}, if $|g|$ is ill-defined, then so is $\p$. A good supergravity limit is then lost in the non-geometric background, but $\b$-supergravity can restore it, as argued in \cite{Andriot:2013xca}. In addition, an ill-defined $\p$ is likely to be non-constant, so the compactification ansatz cannot be used for this set of fields. Then, $g,\ b,\ \p$ does not allow to conclude on the (non-)existence of solutions of $\tL_{\b}$, on the contrary here to $g',\ b',\ \p'$.} Getting it non-constant looks then like a difficult constraint.\footnote{One could also deviate from the compactification ansatz by considering warp factors and a non-constant dilaton: compact NSNS solutions with these features exist, such as wrapped $\NS$-branes, or non-K\"ahler backgrounds of heterotic string. The supergravity limit of those is nevertheless more delicate.}

The ratio \eqref{ratio} can in principle be computed in terms of one or the other set of fields, since we know how the fields are related in table \ref{tab:TdN}. A difficulty however comes from the fact that the field redefinition involves the whole fields while the T-duality only acts on certain blocks. That makes a generic computation not possible, as the inverse and the determinant of a matrix divided in blocks cannot generically be expressed in terms of those blocks. So we consider the following subcase (and basis)
\beq
\tg=\begin{pmatrix} \tg_N & \\ & \tg_{D-N} \end{pmatrix}\ ,\ \b=\begin{pmatrix} \b_N & \\ & \b_{D-N} \end{pmatrix} \ , \label{blockassumption}
\eeq
where these fields do not have off-diagonal components. One then computes $g,\ b$ and $g',\ b'$. Using some freedom of sign in the field redefinition \cite{Andriot:2011uh}, $g'$ can be simplified to
\beq
g'=\begin{pmatrix} \tg_N^{-1} & \\ & (\tg_{D-N}^{-1}+\b_{D-N})^{-1} \tg_{D-N}^{-1} (\tg_{D-N}^{-1}-\b_{D-N})^{-1} \end{pmatrix} \ .
\eeq
This result can easily be understood. The field redefinition is similar to a T-duality in all directions, although the indices are placed differently; this last point is an important distinction between the two, in particular for the large volume limit \cite{Andriot:2013xca}. This similarity still explains why the block along the $N$ directions is barely changed by the combination of the field redefinition and the T-duality, while the other block only goes through the field redefinition. Interestingly, $\b_N$ does not contribute. From this result, we deduce
\beq
\frac{\sqrt{|\tg|}}{\sqrt{|g'|}}= |\tg_N|\times |\id_{D-N} + \tg_{D-N}\b_{D-N}| \ .
\eeq
Although not impossible, having this quantity non-constant is rather unlikely, at least in usual set-ups where we look for solutions. First, $\b_{D-N}$ is likely to be constant, as it does not transform under gluing. Secondly, the metric $\tg_{D-N}$ is usually constant (for instance, that of a base circle). This makes the second factor constant. The metric $\tg_N$ can certainly be non-constant: for twisted tori, it goes through a non-trivial gluing. Its determinant is however usually constant, giving for instance a constant internal volume.\footnote{One may wonder whether a constant internal volume can be thought of as unimodularity, $f^a{}_{ab}=0$, related to the compactness of the internal manifold. One has $\del_m \ln|e|=-\te^a{}_n \del_m \te^n{}_a$, which is $f^a{}_{ab}$ up to a term in $\del_p \te^p{}_{b}$. In our context, the only non-trivial $\del_p$ are those along the $D-N$ directions. However, the inverse vielbein $\te^p{}_{b}$ along those is most likely constant, as is $\tg_{D-N}$. So $\del_m |e|=0$ (constant volume) and $f^a{}_{ab}=0$ would be equivalent.} This implies that the above ratio is constant.

We conclude that, even though we made some assumptions such as \eqref{blockassumption}, it looks unlikely to get a non-constant $\p'$. As explained above, purely NSNS solutions of $\b$-supergravity that are geometric, non-trivial, and satisfy the compactification ansatz, are thus out of reach, at least in the usual set-ups. This holds despite the apparent possibility offered by the equations of motion of $\tL_{\b}$. It would be interesting to reach the same conclusion using only those equations.

\section{Outlook}\label{sec:Ccl}

The main results of this paper have been summarized in the Introduction; let us now make a few comments beyond the scope of this work. A first set of backgrounds that has been studied here are the $\NS$-branes. We gave a detailed account on the $\NS5$-brane, the $\KK$-monopole and the $Q$-brane in section \ref{sec:Tbranes}. This description has been done at the level of supergravity. It would be interesting to go beyond and study them as stringy (or M-theory) objects. As the S-dual of the $D_5$-brane, many properties of the $\NS5$-brane are already known. In particular, $D_1$-branes should end on it. We actually expect this to hold as well for the other $\NS$-branes, because they are related in the same manner as the $D_p$-branes are: via smearing and T-duality. This could give a hint on the world-volume action of these $\NS$-branes. The case of the $\NS5$-brane is certainly studied (see e.g. \cite{Eyras:1998hn, Garousi:2011we} and references therein), but more could be learnt for the $Q$-brane. Proposals have been made in \cite{Chatzistavrakidis:2013jqa} for the latter. A mismatch with our results is however discussed in appendix \ref{ap:Hannover}. From the world-volume action, one could deduce source contributions to the equations of motion and the BI. The work done here within $\b$-supergravity should help on this point, since we obtained such contributions not only in the BI but also in the dilaton equation of motion \eqref{dilPoi} and the Einstein equation \eqref{Ex} - \eqref{Evarphi}. Interestingly, there was no such modification for the $\b$ equation of motion: this is usually expected, as long as the BI gets a source term. Finally, let us recall that the $Q$-brane is a codimension $2$ object, and is in that respect similar to the $D_7$-brane. The latter is known to have a non-perturbative description within F-theory, and one may wonder if such a construction could as well be considered for the $Q$-brane \cite{deBoer:2012ma}. The cut-off needed for its warp factor, mentioned in section \ref{sec:smearing}, could be better understood in such a context.

We also discussed in section \ref{sec:smearing} the possibility of an $R$-brane. Although the name was already used in \cite{Hassler:2013wsa}, the object proposed here is different. It would be a codimension $1$ $\NS$-brane, which is equivalent to having its warp factor given by an absolute value \eqref{fR}. The BI \eqref{delR} is a natural candidate to be corrected by such a brane, which would then source the $R$-flux. Constructing this object by performing a standard T-duality is however problematic: the lack of isometry would force us to smear the $Q$-brane fields in an unusual way. But the derivation of this warp factor and the BI \eqref{delR} still suggest the possibility for such a brane. On a similar tone, the last BI \eqref{bdelR} might be related to the existence of a codimension $0$ $\NS$-brane. But smearing the $R$-brane warp factor $f_R$, as we did for the other branes, does not bring any valuable information on the warp factor of this hypothetical object.\\

In the absence of branes, our study of BI has put forward the $Spin(D,D)\times \mathbb R^{+}$ covariant derivative and its Dirac operator $\D$. We showed that the nilpotency of the latter gives back the NSNS BI. So this object is an important tool to characterise vacua; understanding its cohomology should for instance be helpful. The formalism of Generalized Geometry or DFT would certainly help to study this operator. The specific Generalized Geometry with $Spin(D,D) \times \mathbb{R}^+$ structure group worked-out in \cite{Strickland-Constable:2013xta} could also be related. In addition, this object $\D$ should appear and characterise supersymmetric vacua, in the context of $SU(3)\times SU(3)$ structures. In the future work \cite{Future}, we expect to obtain it in $\b$-supergravity Killing spinor equations, similarly to \cite{Grana:2005sn}, and consequently in the superpotential (a discussion and references on the latter can be found in \cite{Andriot:2011uh}). The $\D$ given in \eqref{Dbetaintro} should then provide a characterisation of internal manifolds analogous to the standard twisted Generalized Calabi-Yau \cite{Hitchin:2004ut, Gualtieri:2003dx}. Its cohomology could thus again play a role, this time in dimensional reductions on those manifolds, or maybe on the generalized parallelizable spaces of \cite{Lee:2014mla}.\\

In the last part of the paper, we studied the symmetries of standard and $\b$-supergravity, and how those could be used to construct geometric backgrounds. In the presence of isometries, the symmetries were shown to be enhanced by the T-duality group. One of its elements, the $\b$-transforms, turned out to be a manifest symmetry of $\b$-supergravity, and played an important role in our analysis. Using those as gluing transformations would always lead to geometric backgrounds of $\b$-supergravity. The restriction of having isometries and the use of $\b$-transforms could then help in constructing the generalized cotangent bundle $E_{T^*}$, introduced in \cite{Grana:2008yw, Andriot:2013xca}. This counterpart of the generalized tangent bundle $E_T$ was argued in \cite{Andriot:2013xca} to be the correct Generalized Geometry bundle for the generalized frames built with $\teee(\b)$ \eqref{genvielb}. It would be interesting to have one concrete construction of $E_{T^*}$. This point could be related to the behaviour of the Courant bracket under $\b$-transforms, provided the isometries: this could be worth being studied as well.

Our analysis lead us to determine a class of geometric backgrounds of $\b$-supergravity, while clarifying some related notions. These vacua were however shown to be on a geometric T-duality orbit, preventing them from leading to new physics. Similar results were obtained in \cite{Dibitetto:2012rk} when considering reductions from DFT to some supergravities in seven dimensions or higher. Although we rather have in mind here physics of four-dimensional supergravities, these results might be related. We proposed in section \ref{sec:geombcgdorbit} various possibilities to circumvent this result, at the level of ten-dimensional supergravity. It was suggested in \cite{Dibitetto:2012rk} that truly new vacua and new physics would rather be accessible beyond that level, and similar proposals have been made in \cite{Condeescu:2013yma, Chatzistavrakidis:2013wra, Blumenhagen:2013zpa, Hassler:2014sba}. The extension of our formalism to the Ramond-Ramond sector or to include the gauge fluxes of heterotic string, as discussed in \cite{Andriot:2013xca}, would in any case bring a more complete picture of the properties of these backgrounds with non-geometric fluxes.

Even if we do not get new physics from $\b$-supergravity, as in the case studied here, this reformulation of standard supergravity may offer a better description of some backgrounds. It is for instance the case of the $Q$-brane, that is T-dual to the smeared $\NS5$-brane: its brane picture is much clearer in terms of $\b$-supergravity fields, and the BI are then nicely formulated with non-geometric fluxes. We expect to find other examples of (non-compact) backgrounds better described by $\b$-supergravity in the AdS/CFT context, where $\b$-transforms already play a role.

\vspace{0.4in}

\subsection*{Acknowledgments}

We would like to thank R. Blumenhagen, A. Deser and F. Rennecke for useful discussions related to section \ref{sec:geomvac}. A. B. thanks the Max-Planck-Institut f\"ur Gravitationsphysik, AEI Potsdam, and D. A. thanks the Institute of Advanced Study, HKUST, for their warm hospitality during the completion of this project. The work of D. A. is part of the Einstein Research Project "Gravitation and High Energy Physics", which is funded by the Einstein Foundation Berlin.

\newpage

\begin{appendix}

\section{Conventions}\label{ap:conv}

We give in this appendix various conventions used throughout the paper. The space-time is $D$-dimensional. The flat (tangent space) indices are $a \dots l$ and the curved ones are $m \dots z$. $|\tg|$ denotes the absolute value of the determinant of the metric $\tg$, and $\R(\tg)$ denotes its Ricci scalar, for a Levi-Civita connection. The squares introduced are defined as
\bea
& (\del \p)^2 \equiv g^{mn} \del_m \p\ \del_n \p\ , \ H^2 \equiv \frac{1}{3!} H_{mnp} H_{qrs} g^{mq} g^{nr} g^{ps} \ , \ R^2 \equiv \frac{1}{3!} R^{mnp} R^{qrs} \tg_{mq} \tg_{nr} \tg_{ps} \ ,\label{squares} \\
& (\del \tp)^2 \equiv \tg^{mn} \del_m \tp\ \del_n \tp\ , \ (\b^{mp}\del_p \tp - \T^m)^2 \equiv \tg_{mn} (\b^{mp}\del_p \tp - \T^m) (\b^{nq}\del_q \tp - \T^n) \ .\nn
\eea
Going to flat indices, we use the vielbein $\te^a{}_m$ and its inverse $\te^n{}_b$, associated to the metric $\tg_{mn}= \te^a{}_m \te^b{}_n \eta_{ab}$, with $\eta_{ab}$ the components of the flat metric $\eta_D$. Tensors with flat indices are obtained after multiplication by the appropriate (inverse) vielbein(s), e.g. $\b^{ab}=\te^a{}_m \te^b{}_n \b^{mn}$, and we also denote $\del_a= \te^m{}_a \del_m$. The structure constant or geometric flux $f^{a}{}_{bc}$ is defined from the vielbeins as
\beq
f^{a}{}_{bc} = 2 \te^a{}_m \del_{[b} \te^m{}_{c]} = - 2 \te^m{}_{[c} \del_{b]} \te^a{}_{m}\ ,\ 2 \del_{[a} \del_{b]}=f^c{}_{ab} \del_c \label{fabc} \ .
\eeq
The spin connection coefficient, given for Levi-Civita connection by \eqref{defof}, satisfies
\beq
\eta^{dc} \o^a_{bc} = - \eta^{ac} \o^d_{bc} \ , \ f^{a}{}_{bc} = 2 \o^a_{[bc]} \ ,\ f^a{}_{ab}=\o^a_{ab} \ .\label{prop}\\
\eeq
A $p$-form $A$ is given by
\beq
A=\frac{1}{p!}A_{m_1 \dots m_p}  \d x^{m_1} \w \dots \w \d x^{m_p} = \frac{1}{p!}A_{a_1 \dots a_p}  \te^{a_1} \w \dots \w \te^{a_p} \ .
\eeq
We deduce for a $p$-form $A$ and a $q$-form $B$ the coefficient
\beq
(A\w B)_{\mu_1...\mu_{p+q}}=\frac{(p+q)!}{p!q!}A_{[\mu_1...\mu_p} B_{\mu_{p+1}...\mu_{p+q}]} \ .
\eeq
The contraction of a vector $V=V^m \del_m=V^a \del_a$ on $A$ is defined by
\beq
V \vee A= \frac{1}{(p-1)!}V^{m_1} A_{m_1 \dots m_p} \d x^{m_2} \w \dots \w \d x^{m_p} \ . \label{Contraction}
\eeq
It is also denoted by $\iota_a=\te^m{}_a \iota_m$, that satisfies the following commutation relations
\beq
V\vee A= V^a \iota_a A \ ,\quad \{\te^{a},\iota_{b}\}=\delta^{a}_{b}\ , \quad \{\iota_{a},\iota_{b}\}=0 \ ,
\eeq
and a contraction on scalar vanishes. In the case of multiple contractions, such as $Q_c{}^{ab} \iota_a \iota_b$, one should be careful with their order, that may generate signs when acting on a form. Finally, we introduce the totally antisymmetric quantity $\epsilon$, given by $\epsilon_{m_1 \dots m_n}=+1/-1$ for $(m_1 \dots m_n)$ being an even/odd permutation of $(1 \dots n)$, and $0$ otherwise. The one with flat indices $\epsilon_{a_1 \dots a_n}$ has the same value, i.e. $\epsilon$ is not a tensor. This can be seen by preserving the volume form. We also consider (constant) matrices $\gamma^a$, satisfying the Clifford algebra
\beq
\{ \gamma^a , \gamma^b \}= 2 \eta^{ab} \ , \ [ \gamma^a , \gamma^b ]= 2 \gamma^{ab} \ {\rm with} \ \gamma^{a_1 a_2 \dots a_p} \equiv \gamma^{[a_1} \gamma^{a_2} \dots \gamma^{a_p]} \  , \label{g1}
\eeq
and further useful properties listed in the appendix of \cite{Andriot:2013xca}.

\section{Derivation of the equations of motion in flat indices}\label{ap:eom}

In this appendix, we give details on the rewriting of the equations of motion \eqref{dileom}, \eqref{Einstein} and \eqref{beom} in flat indices, following section \ref{sec:eom}. This is achieved with two methods: first a direct approach, and secondly using the Generalized Geometry formalism. As a side remark, let us mention that it would be interesting to apply a Palatini formalism to the $\b$-supergravity objects to rederive these equations. They should also be obtainable from the DFT ones of \cite{Geissbuhler:2013uka}.

\subsection{Direct approach}\label{ap:eomdirect}

As explained in section \ref{sec:eom}, the $\b$ equation of motion requires more work than the other two; we only focus on this one here. We start by multiplying the equation in curved indices \eqref{beom} by the appropriate vielbeins to get it in flat indices. We then separate the terms in $\del \tp$ and $\T$ from the others, as they may vanish upon standard assumptions when looking for solutions \cite{Andriot:2013xca}. We obtain
\bea
- & \frac{1}{2} \eta_{ab} \eta_{cd} \eta_{ef} \cN^a R^{bdf} + 2 \eta_{f[e} \R_{c]d} \b^{fd} + \frac{1}{2} \eta_{cd} \eta_{ef} \eta^{ab} \N_a \N_b \b^{fd} - \N_a \left( \eta_{f[e} \N_{c]} \b^{fa} \right) \label{beomflat1} \\
= &\ \eta_{ab} \eta_{cd} \eta_{ef} R^{bdf} \left(\b^{ag} \del_g \tp - \T^a \right) + \eta^{ab} \eta_{cd} \eta_{ef} \N_b \b^{fd}\ \del_a \tp + 4 \b^{ab} \eta_{a[c} \N_{e]} \del_b \tp + 2 \eta_{a[c} \N_{e]} \b^{ab}\ \del_b \tp \ . \nn
\eea
We now focus on the LHS of \eqref{beomflat1}. A key ingredient is $\N \b$: it can be written in terms of fluxes as
\beq
\N_b \b^{fd} = Q_b{}^{fd} + \b^{h[f} f^{d]}{}_{bh} + 2 \eta^{h[d} \b^{f]g} f^{i}{}_{h(b} \eta_{g)i} \ . \label{Nbeta}
\eeq
Using this expression and the definitions of the fluxes, a tedious computation gives a lengthy expression for $\N_a \N_b \b^{fd}$. From the latter, we get two terms of \eqref{beomflat1}. We first deduce an expression for $\N_a \N_c \b^{fa}$, and obtain further
\bea
\eta_{fe} \N_a \N_c \b^{fa} = & \ \eta_{fe} \del_a Q_c{}^{fa} + \eta_{fe} \b^{h[f} \del_a f^{a]}{}_{ch} \label{term2} \\
& + \frac{1}{2} \eta_{fe} \eta^{ha} \b^{fg} \left( \eta_{gi} \del_a f^{i}{}_{hc} + \eta_{ci} \del_a f^{i}{}_{hg} \right) + \frac{1}{2} \b^{ga} \left( \eta_{gi} \del_a f^{i}{}_{ec} + \eta_{ci} \del_a f^{i}{}_{eg} \right) \nn \\
& + \eta_{fe} f^{a}{}_{ah} Q_c{}^{fh} + \frac{1}{2} \eta_{fe} f^{f}{}_{ch} Q_a{}^{ah} + \frac{1}{2} \eta_{fe} f^{f}{}_{ah} Q_c{}^{ha} \nn\\
& + \frac{1}{2} Q_a{}^{ag} \left( \eta_{ci} f^{i}{}_{ge} + \eta_{gi} f^{i}{}_{ce} \right) + \frac{1}{2} Q_c{}^{ga} \left( \eta_{ai} f^{i}{}_{eg} + \eta_{gi} f^{i}{}_{ea} \right) \nn\\
& + \frac{1}{2} \eta_{fe} \left( \frac{1}{2} f^{a}{}_{hc} f^{f}{}_{ag} \b^{gh} + f^{a}{}_{hg} f^{f}{}_{ca} \b^{gh} + f^{a}{}_{ah} f^{h}{}_{cg} \b^{gf} + f^{a}{}_{hc} f^{h}{}_{ag} \b^{gf} \right) \nn\\
& + \frac{1}{2} \eta_{lc} \left( \frac{1}{2} f^{l}{}_{ka} f^{k}{}_{ej} \b^{aj} + f^{l}{}_{eg} f^{g}{}_{aj} \b^{aj} \right) + \frac{1}{2} \eta_{gi} f^{i}{}_{ec} f^{g}{}_{aj} \b^{aj} \nn\\
& + \frac{1}{4} f^{g}{}_{ac} \left( \eta_{ij} f^{i}{}_{eg} \b^{aj} + \eta_{ig} f^{i}{}_{ej} \b^{aj} + \eta_{gj} f^{a}{}_{ek} \b^{jk} \right) \nn\\
& + \frac{1}{2} \eta_{fe} f^{g}{}_{ac} \eta^{ha} \b^{fj} \left(  \eta_{gi} f^{i}{}_{jh} +  \eta_{gj} f^{k}{}_{kh} \right) + \frac{1}{4} \eta_{fe} f^{i}{}_{hc} f^{f}{}_{aj} \eta_{gi} \eta^{hj} \b^{ga} \nn\\
& + \frac{1}{4} \eta_{lc} f^{l}{}_{ka} f^{i}{}_{eg} \eta_{ij} \eta^{gk} \b^{aj} + \frac{1}{4} f^{i}{}_{hc} f^{l}{}_{ae} \eta_{gl} \eta_{ij} \eta^{ha} \b^{gj} \nn\\
&  + \frac{1}{2} \eta_{lc} \eta_{fe} \left( f^{l}{}_{hg} f^{a}{}_{aj} \eta^{gj} \b^{hf} + \frac{1}{2} f^{f}{}_{hg} f^{l}{}_{ak} \eta^{gk} \b^{ha} \right) + \frac{1}{4} \eta_{lc} \eta_{fe} \eta_{ij} \eta^{ah} \eta^{gk} f^{l}{}_{ak} f^{i}{}_{hg} \b^{fj} \ . \nn
\eea
Secondly, we contract $\N_a \N_b \b^{fd}$ with a metric to get
\bea
\eta^{ab} \N_a \N_b \b^{fd} = &\ \eta^{ab} \del_a Q_b{}^{fd} + \eta^{ab} \b^{h[f} \del_a f^{d]}{}_{bh} + \eta^{h[d} \b^{f]g} \left( \eta^{ab} \eta_{gi} \del_a f^{i}{}_{hb} + \del_a f^{a}{}_{hg} \right)  \label{term1} \\
& + Q_b{}^{fd} \eta^{gk} f^{a}{}_{ak} + 2 Q_a{}^{g[d} \eta^{f]h} \left( \eta^{ab} \eta_{gi} f^{i}{}_{hb} + f^{a}{}_{hg} \right) \nn\\
& + \frac{1}{2} \eta^{ab} \b^{gh} f^{[f}{}_{ag} f^{d]}{}_{bh} + f^{a}{}_{hg} \left( \eta^{h[d} f^{f]}{}_{ja} \b^{jg} + f^{g}{}_{ak} \eta^{k[f} \b^{d]h} \right) \nn\\
& + f^{a}{}_{ak} \left( \eta^{gk} \b^{h[f} f^{d]}{}_{gh} + \eta^{h[d} \b^{f]j} ( \eta^{gk} \eta_{ij} f^{i}{}_{hg} + f^{k}{}_{hj} ) \right) \nn\\
& + \eta^{ab} \eta_{gi} f^{g}{}_{ja} f^{i}{}_{hb} \eta^{h[d} \b^{f]j} + \eta^{ab} \eta_{gi} \b^{gj} f^{i}{}_{hb} \eta^{h[d} f^{f]}{}_{aj} + \frac{1}{2} \eta^{ab} \eta^{hg} \eta_{ij} f^{i}{}_{hb} \b^{j[d} f^{f]}{}_{ag} \nn\\
& + \eta^{h[d} \eta^{f]k} \b^{gj} \left( \frac{1}{2} \eta_{ab} f^{a}{}_{kg} f^{b}{}_{hj} + \eta_{jl} f^{a}{}_{hg} f^{l}{}_{ak} \right) + \frac{1}{2} \eta^{ab}  \eta^{hd} \eta^{fk} \eta_{gl} \eta_{ij} \b^{jg} f^{l}{}_{ka} f^{i}{}_{bh} \ .\nn
\eea
We finally sum the two terms of \eqref{beomflat1} just obtained, together with a third one involving $\R_{cd}$ that we get using \eqref{Ricciflat}. Many simplifications occur to eventually give
\bea
2 & \eta_{f[e} \R_{c]d} \b^{fd} + \frac{1}{2} \eta_{cd} \eta_{ef} \eta^{ab} \N_a \N_b \b^{fd} - \N_a \left( \eta_{f[e} \N_{c]} \b^{fa} \right) \label{term3} \\
= & \ \frac{1}{2} \eta_{ef} \eta_{cd} \eta^{ab} \del_a Q_b{}^{fd} + \del_a ( \eta_{f[e} Q_{c]}{}^{af}) \nn\\
& + 2 \b^{hf} \del_a f^{a}{}_{h[c} \eta_{e]f} - \b^{ha} \del_a f^{f}{}_{h[c} \eta_{e]f} + \frac{1}{2} \eta_{gi} \b^{ga} \del_a f^{i}{}_{ce} - 2 \b^{fd} \eta_{f[e} \del_{c]} f^{a}{}_{ad} \nn\\
& + Q_a{}^{gf} f^{a}{}_{g[c} \eta_{e]f} + \frac{1}{2} f^{f}{}_{ha} Q_{[c}{}^{ha} \eta_{e]f} + f^{a}{}_{ah} Q_{[c}{}^{hf} \eta_{e]f} +  Q_a{}^{ah} f^{f}{}_{h[c} \eta_{e]f} + \frac{1}{2} Q_a{}^{ag} f^{i}{}_{ec} \eta_{gi} \nn\\
& +  \frac{1}{2} \eta_{ef} \eta_{cd} \eta^{gk} Q_g{}^{fd} f^{a}{}_{ak} + \eta_{gi} \eta^{ab} Q_a{}^{dg} f^{i}{}_{b[e} \eta_{c]d} \nn\\
& + 2 \b^{jf} f^{a}{}_{ak} f^{k}{}_{j[c} \eta_{e]f} + \b^{gh} f^{a}{}_{hg} f^{f}{}_{a[c} \eta_{e]f} + \frac{1}{2} \b^{aj} f^{i}{}_{ce} f^{g}{}_{aj} \eta_{gi} \ . \nn
\eea
We now rewrite this expression in a more convenient manner. To do so, one can first show the following identity using \eqref{delf}
\beq
2 \b^{hf} \del_a f^{a}{}_{h[c} \eta_{e]f} - 2 \b^{fd} \eta_{f[e} \del_{c]} f^{a}{}_{ad} + 2 \b^{jf} f^{a}{}_{ak} f^{k}{}_{j[c} \eta_{e]f} = 2 \b^{df} \del_d f^{a}{}_{a[c} \eta_{e]f} \label{trick1} \ .
\eeq
Secondly, thanks to definitions, including the one of $\T^a$ given in \eqref{Ricciflat}, one can derive
\bea
- 2 \eta_{f[e} \N_{c]} \T^f = &\ 2 \eta_{f[e} \del_{c]} Q_{a}{}^{af} -  \b^{hi} \eta_{f[e} \del_{c]} f^{f}{}_{hi} \label{trick2} \\
& - \eta_{f[e} Q_{c]}{}^{hi} f^{f}{}_{hi} + \eta_{gd} Q_{a}{}^{ag} f^{d}{}_{ec} + 2 \b^{jh} f^{f}{}_{hi} f^{i}{}_{j[c} \eta_{e]f} - \frac{1}{2} \b^{hi} f^{g}{}_{hi} f^{d}{}_{ec} \eta_{gd} \ . \nn
\eea
Thirdly, one can show that
\beq
-\b^{ha} \del_a  f^{f}{}_{hc} = \b^{ha} \del_a  f^{f}{}_{hc} - 3 \b^{ha} \del_{[a}  f^{f}{}_{hc]} + \b^{ha} \del_{c}  f^{f}{}_{ah} \label{trick3} \ ,
\eeq
where the RHS can be further rewritten with \eqref{delf}. Then, using
\beq
\b^{ha} \del_a  f^{f}{}_{hc} = -2 \b^{a[h} \del_a  f^{f]}{}_{hc} - \b^{af} \del_a f^{h}{}_{hc} \ ,\label{trick35}
\eeq
together with \eqref{delQ} on the RHS of \eqref{trick35}, one gets an expression for $\b^{ha} \del_a  f^{f}{}_{hc}$. The latter should be inserted in the RHS of \eqref{trick3}. The resulting expression, antisymmetrized with $\eta_{ef}$, can be rewritten using \eqref{trick2} into
\bea
-\b^{ha} \del_a  f^{f}{}_{h[c} \eta_{e]f} =& \ - 2 \eta_{f[e} \N_{c]} \T^f + \b^{fd} \del_{d} f^{a}{}_{a[c} \eta_{e]f} - \eta_{f[e} \del_{c]} Q_{a}{}^{af} - \del_{a} Q_{[c}{}^{af} \eta_{e]f} \label{trick4} \\
& + \b^{ah} f^{g}{}_{ah} f^{f}{}_{g[c} \eta_{e]f} + \frac{1}{2} \b^{hi} f^{g}{}_{hi} f^{d}{}_{ec} \eta_{gd} \nn\\
& - f^{a}{}_{ag} Q_{[c}{}^{gf} \eta_{e]f} - Q_{a}{}^{ag} \left( f^{f}{}_{g[c} \eta_{e]f} +  f^{d}{}_{ec} \eta_{gd} \right) \ .\nn
\eea
Using \eqref{trick1} and \eqref{trick4}, we rewrite \eqref{term3} as follows
\bea
2 & \eta_{f[e} \R_{c]d} \b^{fd} + \frac{1}{2} \eta_{cd} \eta_{ef} \eta^{ab} \N_a \N_b \b^{fd} - \N_a \left( \eta_{f[e} \N_{c]} \b^{fa} \right) \label{terms} \\
= & \ - 2 \eta_{f[e} \N_{c]} \T^f + \frac{1}{2} \eta_{gi} \b^{ga} \del_a f^{i}{}_{ce} +  \b^{df} \del_{d} f^{a}{}_{a[c} \eta_{e]f} - \eta_{f[e} \del_{c]} Q_{a}{}^{af} + \frac{1}{2} \eta_{ef} \eta_{cd} \eta^{ab} \del_a Q_b{}^{fd} \nn\\
& + Q_a{}^{gf} f^{a}{}_{g[c} \eta_{e]f} + \frac{1}{2} f^{f}{}_{ha} Q_{[c}{}^{ha} \eta_{e]f} -\frac{1}{2} Q_{a}{}^{ag} f^{i}{}_{ec} \eta_{gi} \nn\\
& + \frac{1}{2} \eta_{ef} \eta_{cd} \eta^{gk} Q_g{}^{fd} f^{a}{}_{ak} + \eta_{gi} \eta^{ab} Q_a{}^{dg} f^{i}{}_{b[e} \eta_{c]d} \nn \ .
\eea
From this \eqref{terms}, we finally rewrite the $\b$ equation of motion from \eqref{beomflat1} to
\bea
- & \frac{1}{2} \eta_{ab} \eta_{cd} \eta_{ef} \cN^a R^{bdf} + Q_a{}^{gf} f^{a}{}_{g[c} \eta_{e]f} + \frac{1}{2} f^{f}{}_{ha} Q_{[c}{}^{ha} \eta_{e]f} -\frac{1}{2} Q_{a}{}^{ag} f^{i}{}_{ec} \eta_{gi}  \label{beomflat4} \\
& + \frac{1}{2} \eta_{ef} \eta_{cd} \eta^{gk} Q_g{}^{fd} f^{a}{}_{ak} + \eta_{gi} \eta^{ab} Q_a{}^{dg} f^{i}{}_{b[e} \eta_{c]d} \nn\\
= &\ 2 \eta_{f[e} \N_{c]} \T^f - \frac{1}{2} \eta_{gi} \b^{ga} \del_a f^{i}{}_{ce} - \b^{df} \del_{d} f^{a}{}_{a[c} \eta_{e]f} + \eta_{f[e} \del_{c]} Q_{a}{}^{af} - \frac{1}{2} \eta_{ef} \eta_{cd} \eta^{ab} \del_a Q_b{}^{fd} \nn\\
& + \eta_{ab} \eta_{cd} \eta_{ef} R^{bdf} \left(\b^{ag} \del_g \tp - \T^a \right) + \eta^{ab} \eta_{cd} \eta_{ef} \N_b \b^{fd}\ \del_a \tp + 4 \b^{ab} \eta_{a[c} \N_{e]} \del_b \tp + 2 \eta_{a[c} \N_{e]} \b^{ab}\ \del_b \tp \ , \nn
\eea
as given in \eqref{beomflat2}.

\subsection{Using the Generalized Geometry formalism}\label{ap:GGeom}

We explain in section \ref{sec:eom} the main procedure to derive the equations of motion in flat indices from the Generalized Geometry formalism. Here, we give some details on the computation of the generalized Ricci tensor \eqref{defR_ab}. We start from its expression \eqref{GenRicci1}. We observe that all derivatives acting on the spinor $\eps^+$ should vanish, since the generalized Ricci tensor only acts on the spinor via a multiplication by a $\gamma$-matrix. One can therefore verify that
\bea
&\Big(\g^a \del_a \del_b - \g^a \ov{\eta_{bg}} \del_a \b^{\ov{ge}} \del_{\ov{e}} - \g^a \ov{\eta_{bg}} \b^{\ov{ge}} \del_a\del_{\ov{e}} + \g^a \gamma^{gh} Y_{\ov{b}gh}\del_a \\
&\ + \gamma^a \eta_{ad} \b^{dc} \del_c \del_{\ov{b}} - \gamma^a \eta_{ad}\ov{\eta_{bg}} \b^{dc} \del_c\b^{\ov{ge}} \del_{\ov{e}} - \gamma^a \eta_{ad}\ov{\eta_{bg}} \b^{dc} \del_c\b^{\ov{ge}} \del_c\del_{\ov{e}}+ \gamma^a \gamma^{gh} Y_{\ov{b}gh}\eta_{ad} \b^{dc}  \del_c \nn \\
&\ +X_{acd} \gamma^{acd}\del_{\ov{b}} - X_{acd} \gamma^{acd}\ov{\eta_{bg}} \b^{\ov{ge}} \del_{\ov{e}} + \gamma^a X_a  \del_{\ov{b}} - \gamma^a X_a\ov{\eta_{bg}} \b^{\ov{ge}} \del_{\ov{e}} \nn \\
&\ -\g^a\o_{a\ov{b}}^{\ov{c}}\del_{\ov{c}} +\g^a\o_{a\ov{b}}^{\ov{c}}\ov{\eta_{cg}} \b^{\ov{ge}} \del_{\ov{e}}+ \g^a\eta_{ad}{\o_Q}^{d\ov{c}}_{\ov{b}}\del_{\ov{c}} - \g^a\eta_{ad}{\o_Q}^{d\ov{c}}_{\ov{b}}\ov{\eta_{cg}} \b^{\ov{ge}} \del_{\ov{e}} \nn \\
&\ -\frac{1}{2}\g^a \eta_{ad}\ov{\eta_{bf}}R^{d\ov{fc}} \del_{\ov{c}} +\frac{1}{2}\g^a \eta_{ad}\ov{\eta_{bf}}R^{d\ov{fc}} \ov{\eta_{cg}} \b^{\ov{ge}} \del_{\ov{e}} \nn \\
&\ -\gamma^a \del_{\ov{b}} \del_a - \gamma^a \eta_{ad} \del_{\ov{b}} \b^{dc} \del_c - \gamma^a \eta_{ad}  \b^{dc}\del_{\ov{b}} \del_c -\gamma^{acd} X_{acd} \del_{\ov{b}} -\gamma^a X_a  \del_{\ov{b}} \nn \\
&\ +\gamma^a \ov{\eta_{bg}} \b^{\ov{ge}} \del_{\ov{e}}\del_a + \gamma^a \eta_{ad} \ov{\eta_{bg}} \b^{\ov{ge}} \del_{\ov{e}} \b^{dc} \del_c + \gamma^a \eta_{ad} \ov{\eta_{bg}} \b^{\ov{ge}}  \b^{dc}\del_{\ov{e}} \del_c + \gamma^{acd}X_{acd}\ov{\eta_{bg}} \b^{\ov{ge}} \del_{\ov{e}}  + \gamma^aX_a \ov{\eta_{bg}} \b^{\ov{ge}} \del_{\ov{e}} \nn \\
&\ -\gamma^{gh}\gamma^a Y_{\ov{b}gh}\del_a - \gamma^{gh}\gamma^a Y_{\ov{b}gh}\eta_{ad} \b^{dc} \del_c \Big)\eps^+=0 \nn\ .
\eea
We are then left with $\gamma$-matrices acting on $\eps^+$. Using several identities on $\gamma$-matrices listed in the appendix of \cite{Andriot:2013xca}, we obtain
\bea
\frac{1}{2}R_{a\ov{b}}\g^a\eps^+=&\left((\g^{agh}+2\eta^{a[g}\g^{h]}) \del_a Y_{\ov{b}gh} + (\g^{agh}+2\eta^{a[g}\g^{h]}) \eta_{ad} \b^{dc} \del_c Y_{\ov{b}gh}\right. \\
&\ + [\gamma^{acd},\gamma^{gh}]X_{acd} Y_{\ov{b}gh}  + [\gamma^a,\gamma^{gh}]X_a Y_{\ov{b}gh} \nn  \\
&\ -(\g^{agh}+2\eta^{a[g}\g^{h]})\o_{a\ov{b}}^{\ov{c}}Y_{\ov{c}gh} + (\g^{agh}+2\eta^{a[g}\g^{h]})\eta_{ad}{\o_Q}^{d\ov{c}}_{\ov{b}}Y_{\ov{c}gh} \nn  \\
&\ -\frac{1}{2}(\g^{agh}+2\eta^{a[g}\g^{h]})\eta_{ad}\ov{\eta_{bf}}R^{d\ov{fc}}Y_{\ov{c}gh} \nn  \\
&\ -\left. \gamma^{acd} \del_{\ov{b}} X_{acd} - \gamma^a\del_{\ov{b}}X_a + \gamma^{acd}\ov{\eta_{bg}} \b^{\ov{ge}} \del_{\ov{e}}X_{acd} + \gamma^a\ov{\eta_{bg}} \b^{\ov{ge}} \del_{\ov{e}}X_a \right) \eps^+  \nn
\eea
Similarly to the calculation of the scalar $S$ in \cite{Andriot:2013xca}, we should then distinguish the different orders in $\gamma$-matrices. Here, we only consider the lowest order in $\gamma^a$, and assume that all higher orders vanish: this would be analogous to the computation of $S$, where the BI \eqref{delf2} - \eqref{bdelR2} played an important role; we expect the same here. In addition, the lowest order will be enough to obtain the equations of motion. Then at first order in $\gamma^a$, $\frac{1}{2}R_{a\ov{b}}\g^a$ gives
\bea
&\left(\frac{1}{2} \R_{\ov{b}a}-\frac{1}{2}  \eta_{ae}\ov{\eta_{bg}} \cR^{\ov{g}e}+\frac{1}{8} \eta_{ae}\ov{\eta_{bg}}\eta_{if}\ov{\eta_{cd}} R^{i\ov{gc}}  R^{\ov{d}fe}-\frac{1}{4} \eta_{ae}\ov{\eta_{bg}}e^{2\tp}  \N_d(e^{-2\tp} R^{\ov{g}de})\right.\\
&\ +\N_{\ov{b}} \N_a\tp -  \eta_{ae}\ov{\eta_{bg}} \cN^{\ov{g}}(\cN^{e}\tp)-\eta_{ae}\ov{\eta_{bg}} \cN^{\ov{g}} \T^e \nn \\
&\ +\frac{1}{4} \ov{\eta_{bg}}\del_d  Q_a{}^{\ov{g}d}+\frac{1}{4} \eta_{ae}\ov{\eta_{bg}}\del_d  Q_{f}{}^{e\ov{g}}\eta^{df}+\frac{1}{4} \eta_{ae}\ov{\eta_{bg}}\del_d  Q_{\ov{f}}{}^{ed}\ov{\eta^{gf}}-\frac{1}{2} \eta_{ae}\del_{\ov{b}} Q_d{}^{de} \nn \\
&\ -\frac{1}{4}  \eta_{ae} \b^{gc} \del_c f^e{}_{\ov{b}g}-\frac{1}{4}  \b^{gc} \del_c  f^{e}{}_{a\ov{b}}\eta_{ge}-\frac{1}{4}  \b^{gc} \del_c  f^{\ov{e}}{}_{ag}\ov{\eta_{be}}+\frac{1}{2}  \ov{\eta_{bg}} \b^{\ov{gc}} \del_{\ov{c}} f^d{}_{da} \nn \\
&\ +\frac{1}{4}\ov{\eta_{bg}}f^d{}_{dc} Q_a{}^{\ov{g}c}+\frac{1}{4}\eta_{ae}f^d{}_{dc}Q_{\ov{b}}{}^{ec}+\frac{1}{4}\ov{\eta_{bg}}\eta_{ae}\eta^{ch}f^d{}_{dc}Q_h{}^{e\ov{g}} \nn \\
&\ -\frac{1}{4}\eta_{ae} Q_d{}^{dc}f^e{}_{\ov{b}c}- \frac{1}{4}\ov{\eta_{bh}} Q_d{}^{dc}f^{\ov{h}}{}_{ac}- \frac{1}{4}\eta_{ch} Q_d{}^{dc}f^{h}{}_{a\ov{b}}  \nn \\
&\ +\frac{1}{8}\ov{\eta_{bg}}f^{\ov{g}}{}_{c\ov{d}}Q_a{}^{\ov{d}c}+\frac{1}{8}\eta_{ch} f^{h}{}_{\ov{bd}}Q_a{}^{\ov{d}c}+\frac{1}{8}\ov{\eta_{dh}} f^{\ov{h}}{}_{\ov{b}c}Q_a{}^{\ov{d}c} \nn  \\
&\ +\frac{1}{8}\eta_{ae}\ov{\eta_{bg}}\eta^{cf}f^{\ov{g}}{}_{c\ov{d}}Q_{f}{}^{e\ov{d}} +\frac{1}{8}\eta_{ae}f^{h}{}_{\ov{gd}}Q_{h}{}^{e\ov{d}}+\frac{1}{8}\eta_{ae}\ov{\eta_{dh}} \eta^{ci}f^{\ov{h}}{}_{\ov{b}c}Q_{i}{}^{e\ov{d}} \nn  \\
&\ +\frac{1}{8}\eta_{ae}\ov{\eta_{bg}}\ov{\eta^{df}}f^{\ov{g}}{}_{c\ov{d}}Q_{\ov{f}}{}^{ec}+\frac{1}{8}\eta_{ae}\eta_{ch} \ov{\eta^{di}}f^{h}{}_{\ov{bd}}Q_{\ov{i}}{}^{ec}+\frac{1}{8}\eta_{ae} f^{\ov{h}}{}_{\ov{b}c} Q_{\ov{h}}{}^{ec} \nn \\
&\ -\frac{1}{8} \eta_{ae}f^e{}_{\ov{c}d}Q_{\ov{b}}{}^{d\ov{c}}-\frac{1}{8} \eta_{ae}\ov{\eta_{bg}}\eta^{dh}f^e{}_{\ov{c}d}Q_{h}{}^{\ov{gc}}-\frac{1}{8} \eta_{ae}\ov{\eta_{bg}}\ov{\eta^{ch}}f^e{}_{\ov{c}d}Q_{\ov{h}}{}^{\ov{g}d} \nn  \\
&\ -\frac{1}{8} \eta_{de} f^e{}_{a\ov{c}}Q_{\ov{b}}{}^{d\ov{c}}-\frac{1}{8} \ov{\eta_{bg}}f^e{}_{a\ov{c}}Q_{e}{}^{\ov{gc}} -\frac{1}{8} \ov{\eta_{bg}}\ov{\eta^{ch}}\eta_{de}f^e{}_{a\ov{c}}Q_{\ov{h}}{}^{\ov{g}d}  \nn \\
&\ -\frac{1}{8} \ov{\eta_{ce}}f^{\ov{e}}{}_{ad}Q_{\ov{b}}{}^{d\ov{c}}-\frac{1}{8} \ov{\eta_{bg}}\eta^{dh}\ov{\eta_{ce}}f^{\ov{e}}{}_{ad}Q_{h}{}^{\ov{gc}}-\frac{1}{8} \ov{\eta_{bg}}f^{\ov{e}}{}_{ad}Q_{\ov{e}}{}^{\ov{g}d} \nn  \\
&\ -\eta_{ae}\N_{\ov{b}}(\cN^{e}\tp)-\eta_{ae}\N_{\ov{b}}\T^e+  \ov{\eta_{bg}} \cN^{\ov{g}} \N_a\tp \nn \\
&\ -\left. \frac{1}{2}\eta_{ae}\ov{\eta_{bg}}\eta_{fc}   R^{\ov{g}fe}\T^c+\frac{1}{4} \eta_{ae}\ov{\eta_{bg}} \eta_{df}  e^{2\tp}\cN^{d}(e^{-2\tp} R^{\ov{g}fe})\right) \g^{a}\nn\ .
\eea
By considering aligned vielbeins, the previous expression reduces to
\bea
&\left(\frac{1}{2} \R_{ba}-\frac{1}{2}  \eta_{ae}\eta_{bg} \cR^{ge}+\frac{1}{8} \eta_{ae}\eta_{bg}\eta_{if}\eta_{cd} R^{igc}  R^{dfe}-\frac{1}{4} \eta_{ae}\eta_{bg}e^{2\tp}  \N_d(e^{-2\tp} R^{gde})\right.\\
&\ +\N_{b} \N_a\tp -  \eta_{ae}\eta_{bg} \cN^{g}(\cN^{e}\tp)-\eta_{ae}\eta_{bg} \cN^{g} \T^e \nn \\
&\ +\frac{1}{2} \del_d  Q_{(a}{}^{gd}\eta_{b)g}+\frac{1}{4} \eta_{ae}\eta_{bg}\eta^{df}\del_d  Q_{f}{}^{eg}-\frac{1}{2} \eta_{ae}\del_{b} Q_d{}^{de}  \nn \\
&\ -\frac{1}{4}  \b^{gc} \del_c  f^{e}{}_{ab}\eta_{ge}+\frac{1}{2}  \b^{gc} \del_c  f^{e}{}_{g(a}\eta_{b)e}+\frac{1}{2}  \eta_{bg} \b^{gc} \del_{c} f^d{}_{da} \nn \\
&\ +\frac{1}{2}f^d{}_{dc} Q_{(a}{}^{gc}\eta_{b)g}+\frac{1}{4}\eta_{bg}\eta_{ae}\eta^{ch}f^d{}_{dc}Q_h{}^{eg}+ \frac{1}{2} Q_d{}^{dc}f^e{}_{c(a}\eta_{b)e}- \frac{1}{4}\eta_{ch} Q_d{}^{dc}f^{h}{}_{ab} \nn \\
&\ +\frac{1}{4}f^{g}{}_{cd}Q_{[a}{}^{dc}\eta_{b]g}+\frac{1}{2}\eta_{e[a}f^h{}_{b]d}Q_{i}{}^{ec}\eta_{ch} \eta^{di}+\frac{1}{2}\eta_{e[a} f^{h}{}_{b]c} Q_{h}{}^{ec} \nn  \\
&\ -\eta_{ae}\N_{b}(\cN^{e}\tp)-\eta_{ae}\N_{b}\T^e+  \eta_{bg} \cN^{g} \N_a\tp \nn \\
&\ -\left. \frac{1}{2}\eta_{ae}\eta_{bg}\eta_{fc} R^{gfe}\T^c+\frac{1}{4} \eta_{ae}\eta_{bg} \eta_{df} e^{2\tp}\cN^{d}(e^{-2\tp} R^{gfe})\right) \g^{a} \nn\ .
\eea
We can further simplify the above using the following identities. First, one can show
\beq
\eta_{g(a} \cN^{g} \N_{b)}\tp-\eta_{g(a}\N_{b)}(\cN^{g}\tp)=0 \ ,\quad  - \eta_{e[a}\eta_{b]g} \cN^{g}(\cN^{e}\tp)=\frac{1}{2} \eta_{e[a}\eta_{b]g}R^{ged}\N_d \tp\ ,
\eeq
where the second one cancels the term coming from $-\frac{1}{4} \eta_{ae}\eta_{bg}e^{2\tp}  \N_d(e^{-2\tp} R^{gde})$. In addition, three terms antisymmetric in $(a,b)$ at second order in $\beta$ vanish thanks to the following identity using \eqref{delQ} and \eqref{delR}\footnote{One also has the identity $2\cR^{[ab]}=-\N_c R^{cab}$ \cite{Andriot:2012an}, related to \eqref{delR}.}
\beq
-\frac{1}{2}  \eta_{e[a}\eta_{b]g} \cR^{ge}-\eta_{e[a}\eta_{b]g} \cN^{g} \T^e-\frac{1}{4} \eta_{ae}\eta_{bg} \N_d R^{gde}=0\ ,\\
\eeq
and the seven terms symmetric in $(a,b)$ at linear order in $\beta$ cancel using \eqref{delf} and \eqref{delQ}
\bea
&\frac{1}{2} \del_d  Q_{(a}{}^{gd}\eta_{b)g}-\frac{1}{2} \eta_{e(a}\del_{b)} Q_d{}^{de}+\frac{1}{2}  \b^{gc} \del_c  f^{e}{}_{g(a}\eta_{b)e}+\frac{1}{2}   \b^{gc} \del_{c} f^d{}_{d(a}\eta_{b)g}\\
-&\eta_{e(a}\N_{b)}\T^e+ \frac{1}{2}f^d{}_{dc} Q_{(a}{}^{gc}\eta_{b)g}+ \frac{1}{2} Q_d{}^{dc}f^e{}_{c(a}\eta_{b)e}=0\ .\nn
\eea
Using all those, we are finally left with the following expression for $\frac{1}{2}R_{ab}\g^a$ at first order in $\gamma$-matrices, that we give also in \eqref{Rab}
\bea
&\Big(\frac{1}{2} \R_{ba}-\frac{1}{2}  \eta_{e(a}\eta_{b)g} \cR^{ge}+\frac{1}{8} \eta_{ae}\eta_{bg}\eta_{if}\eta_{cd} R^{igc}  R^{dfe} \label{Rabgamma1}\\
&\ +\N_{b} \N_a\tp -  \eta_{e(a}\eta_{b)g} \cN^{g}(\cN^{e}\tp)-\eta_{e(a}\eta_{b)g} \cN^{g} \T^e \nn \\
&\ +\frac{1}{4} \eta_{ae}\eta_{bg}\eta^{df}\del_d  Q_{f}{}^{eg}-\frac{1}{2} \eta_{e[a}\del_{b]} Q_d{}^{de} -\frac{1}{4}  \b^{gc} \del_c  f^{e}{}_{ab}\eta_{ge}+\frac{1}{2}   \b^{gc} \del_{c} f^d{}_{d[a}\eta_{b]g} \nn \\
&\ +\frac{1}{4}\eta_{bg}\eta_{ae}\eta^{ch}f^d{}_{dc}Q_h{}^{eg}- \frac{1}{4}\eta_{ch} Q_d{}^{dc}f^{h}{}_{ab}\nn \\
&\ +\frac{1}{4}f^{g}{}_{cd}Q_{[a}{}^{dc}\eta_{b]g}+\frac{1}{2}\eta_{e[a}f^h{}_{b]d}Q_{i}{}^{ec}\eta_{ch} \eta^{di}+\frac{1}{2}\eta_{e[a} f^{h}{}_{b]c} Q_{h}{}^{ec} \nn  \\
&\ -\eta_{e[a}\N_{b]}(\cN^{e}\tp)-\eta_{e[a}\N_{b]}\T^e+  \eta_{g[b} \cN^{g} \N_{a]}\tp \nn \\
&\ - \frac{1}{2}\eta_{ae}\eta_{bg}\eta_{fc} R^{gfe}\T^c+\frac{1}{4} \eta_{ae}\eta_{bg} \eta_{df} e^{2\tp}\cN^{d}(e^{-2\tp} R^{gfe})\Big) \g^{a}\nn\ .
\eea

\subsection{Relation to the subcase with simplifying assumption}\label{ap:relbetadel}

A simplifying assumption was considered in \cite{Andriot:2011uh}, given by the conditions $\b^{mn} \del_n \cdot =0$, where the dot stands for any field, and $\del_p \b^{np}=0$. This provided a simple Lagrangian, corresponding to a subcase of $\b$-supergravity: one can reduce $\tL_{\b}$ to the former upon the assumption. Let us study here the simplification of the equations of motion. First, the assumption leads to $R^{abc}=0$ and $\T^a=0$. In addition, the $Q$-flux gets reduced as in \eqref{Qsimplif}, implying that $Q_a{}^{ab}=0$ and $Q_c{}^{ha}f^b{}_{ha}=0$. The dilaton equation of motion \eqref{dileom} and the Einstein equation \eqref{Einstein}, rewritten in flat indices, boil down to
\bea
 & \frac{1}{4} \left(\R(\tg) + \cR(\tg) \right) - (\del \tp)^2 + \N^2 \tp =0 \ ,\label{dileomsimplif1}\\
 & \R_{ab} - \eta_{c(a} \eta_{b)d} \cR^{cd} + 2 \N_a \N_b \tp =0 \label{Einsteinsimplif1} \ ,
\eea
where $\cR$ and $\cR^{ab}$ can be further simplified using \eqref{cRRqRf} and \eqref{cRtensor}. The $\b$ equation of motion in flat indices \eqref{beomflat2} becomes
\bea
& Q_a{}^{gf} f^{a}{}_{g[c} \eta_{e]f} + \frac{1}{2} \eta_{ef} \eta_{cd} \eta^{gk} Q_g{}^{fd} f^{a}{}_{ak} + \eta_{gi} \eta^{ab} Q_a{}^{dg} f^{i}{}_{b[e} \eta_{c]d} \label{beomflatsimpl1} \\
& = - \frac{1}{2} \eta_{ef} \eta_{cd} \eta^{ab} \del_a Q_b{}^{fd} + \eta^{ab} \eta_{cd} \eta_{ef} \N_b \b^{fd}\ \del_a \tp + 2 \b^{ab} \eta_{a[c} \N_{e]}\del_b \tp  \ , \nn
\eea
where the last term does not vanish due to the connection terms. Using for the penultimate term \eqref{Nbeta} and for the last term the different definitions, one can show that all explicit dependence on $\b$ vanishes with the assumption, leaving the $\b$ equation of motion as
\bea
& \eta_{ef} \eta_{cd} \eta^{gk} Q_g{}^{fd} f^{a}{}_{ak} + 2 \eta_{gi} \eta^{ab} Q_a{}^{dg} f^{i}{}_{b[e} \eta_{c]d} + e^{2\tp} \eta_{ef} \eta_{cd} \eta^{ab} \del_a (e^{-2\tp} Q_b{}^{fd}) \label{beomflatsimpl2} \\
& + 2 Q_a{}^{gf} f^{a}{}_{g[c} \eta_{e]f} =0 \ .\nn
\eea
The last term can be simplified further by the assumption towards $2 Q_a{}^{gf}\ \te^a{}_m \eta_{f[c} \del_{e]} \te^m{}_g$. It is interesting to compare this equation \eqref{beomflatsimpl2} to the one obtained in \cite{Andriot:2011uh}:
\beq
\del_m (e^{-2\tp} \sqrt{|\tg|}\ \tg^{mn} \tg_{pq} \tg_{rs} \del_n \b^{qs} )=0 \ .\label{beomflatsimplold}
\eeq
This comparison was initiated in curved indices in \cite{Andriot:2011uh}. Here, we turn \eqref{beomflatsimplold} into flat indices and get, using the assumption,
\bea
& \eta_{ef} \eta_{cd} \eta^{gk} Q_g{}^{fd} f^{a}{}_{ak} + 2 \eta_{gi} \eta^{ab} Q_a{}^{dg} f^{i}{}_{b[e} \eta_{c]d} + e^{2\tp} \eta_{ef} \eta_{cd} \eta^{ab} \del_a (e^{-2\tp} Q_b{}^{fd}) \label{beomflatsimpl3} \\
& + 2 Q_a{}^{gf} \eta_{gd} \eta^{ab}\ \te^d{}_m \eta_{f[e} \del_{c]} \te^m{}_b  =0 \ .\nn
\eea
We see that \eqref{beomflatsimpl2} and \eqref{beomflatsimpl3} do not match: they differ by their second rows, i.e. their last term. This fact can be understood as follows: applying the simplifying assumption to the Lagrangian and deriving the $\b$ equation of motion do not commute. This can be seen for instance on a Lagrangian term like $\b^{mn} \del_n \tg^{pq} \del_q \tg_{mp}$, that would contribute to \eqref{beomflatsimpl2} but not to \eqref{beomflatsimpl3}. This problem does not affect the other equations of motion (one can verify directly the matching) because the assumption does not involve the other fields. So to conclude, the correct $\b$ equation of motion for field configurations satisfying the simplifying assumption of \cite{Andriot:2011uh} is \eqref{beomflatsimpl2} and not \eqref{beomflatsimplold}. Note though that for the toroidal example and the $Q$-brane, the two differing terms vanish.

\section{On sourceless NSNS Bianchi identities}\label{ap:BI}

\subsection{Relations to other Bianchi identities in the literature}\label{ap:BIlit}

Our Bianchi identities (BI) \eqref{delf2} - \eqref{bdelR2} provide a generalization to non-constant fluxes of the BI \eqref{BIstw1} - \eqref{BIstw5}, for $H=0$. As mentioned in the Introduction and in section \ref{sec:BIsourcelesslit}, such generalizations have already been proposed in two other approaches. We show in this appendix that the BI obtained there can be reduced and matched with the simpler expressions given by our \eqref{delf2} - \eqref{bdelR2}.

In \cite{Blumenhagen:2012pc} are introduced some straight and some curly fluxes. They are identical once one sets the $H$-flux to vanish, and then match the definition of our fluxes, up to a minus sign on the $R$-flux. Four BI are derived there, as described in section \ref{sec:BIsourcelesslit}, and are given in our conventions by
\bea
0=&\ \del_{[a}f^{e}{}_{bf]}-f^{e}{}_{d[a}f^{d}{}_{bf]}\ ,\label{BIBlum1}\\
0=&\ \b^{dg}\del_g f^{e}{}_{af} + 2 \del_{[a}Q_{f]}{}^{de} - Q_{g}{}^{de}f^{g}{}_{af} + 4 Q_{[a}{}^{g[d}f^{e]}{}_{f]g} \label{BIBlum2}\\
&\ + \b^{eg} \left(2 \del_{[a} f^{d}{}_{f]g} - 3 f^{d}{}_{h[g}f^{h}{}_{af]} \right)\ ,\nn\\
0=&\ -\del_a R^{ghi} +2 \b^{d[g}\del_d Q_{a}{}^{h]i} + 3 Q_{a}{}^{d[g}Q_{d}{}^{hi]} - 3 R^{d[gh}f^{i]}{}_{ad} \label{BIBlum3}\\
&\ + \b^{id} \left( 2 \b^{e[g} \del_e f^{h]}{}_{ad} - \del_a Q_d{}^{gh} + Q_e{}^{gh} f^e{}_{ad} - 4 Q_{[a}{}^{e[g} f^{h]}{}_{d]e}  \right) \ ,\nn\\
0=&\ \b^{g[a}\del_g R^{bc]d}+ 2 R^{g[da}Q_{g}{}^{bc]} + \b^{ed} \left( -\b^{f[a} \del_f Q_e{}^{bc]} - f^{[a}{}_{fe} R^{bc]f} + Q_f{}^{[ab} Q_e{}^{c]f} \right)\ .\label{BIBlum4}
\eea
The set of conditions \eqref{BIBlum1} - \eqref{BIBlum4} turns out to match our \eqref{delf2} - \eqref{bdelR2}. This can be verified using the identities
\bea
& 2 \del_{[a} f^d{}_{f]g} = 3 \del_{[a} f^d{}_{fg]}  - \del_{g} f^d{}_{af}  \ , \label{idtrick1}\\
& 2 \b^{d[g} \del_{d} Q_a{}^{h]i} = 3 \b^{d[g} \del_{d} Q_a{}^{hi]}  - \b^{di} \del_{d} Q_a{}^{gh}   \ , \label{idtrick2}\\
& 3 \b^{g[a} \del_{g} R^{bc]d} = 4 \b^{g[a} \del_{g} R^{bcd]} + \b^{gd} \del_{g} R^{abc}   \ . \label{idtrick3}
\eea
To start with, \eqref{BIBlum1} matches \eqref{delf2}. Using the latter and \eqref{idtrick1}, one shows that \eqref{BIBlum2} matches \eqref{delQ2}. Then, using the latter and \eqref{idtrick2}, one shows that \eqref{BIBlum3} matches \eqref{delR2}. Eventually, using the latter and \eqref{idtrick3}, one verifies that \eqref{BIBlum4} matches \eqref{bdelR2}.

At the level of Double Field Theory (DFT) were obtained in \cite{Geissbuhler:2013uka} some generalized BI. One of them, given by a quantity denoted $\mathcal{Z}_{ABCD}$, was further decomposed into its various $O(D,D)$ components to get a set of DFT conditions. If we set again $H=0$ and use the strong constraint $\tilde{\del}^m=0$, we can show that these conditions match precisely \eqref{delf2} - \eqref{bdelR2}. Indeed, the notations there then become $\mathcal{D}_a=\del_a \ ,\ \mathcal{D}^a=\b^{ab} \del_b\ ,\ \tau_{bc}{}^a=f^a{}_{bc}$, and the fluxes are the same as ours, up to a minus sign on the $R$-flux; this allows to verify the matching. As a confirmation, the conditions of \cite{Geissbuhler:2013uka} were mentioned to reproduce those of \cite{Blumenhagen:2012pc}, namely \eqref{BIBlum1}-\eqref{BIBlum4}, that we have just shown to match our BI \eqref{delf2} - \eqref{bdelR2}.

\subsection{Derivation of BI from the $Spin(D,D)\times \mathbb R^{+}$ covariant derivative}\label{ap:spinder}

In section \ref{sec:Spinder}, we introduced a $Spin(D,D)\times \mathbb R^{+}$ derivative and its associated Dirac operator in \eqref{Diracopgen}. Before studying its nilpotency condition \eqref{nilpocond}, let us first give some details on how to compute a piece of it, namely $\D_2$. This piece is given by
\beq
\D_2=\frac{1}{4} \Omega_{ABC}\Ga^{ABC}=\frac{1}{4} \Omega_{[ABC]}\Ga^{A}\Ga^{B}\Ga^{C} \ ,
\eeq
where the index ${}_B$ is lowered by an $O(D,D)$ metric. To compute this antisymmetry, we use
\beq
\Omega_{ABC} \Ga^B \equiv \Omega_A{}^D{}_C\ \eta_{DB} \Ga^B = \frac{1}{2} \left(\Omega_A{}^b{}_C \Ga_b + \Omega_{A b C} \Ga^b \right) \ .
\eeq
One then gets for instance
\beq
(\Omega_{ABC} - \Omega_{ACB}) \Ga^A \Ga^B \Ga^C=  \Ga^A (\Omega_{Abc}  \Ga^b \Ga^c+ \Omega_A{}^{bc}  \Ga_b \Ga_c + \Omega_A{}^b{}_c  \Ga_b \Ga^c - \Omega_A{}^c{}_b  \Ga^b \Ga_c) \ ,
\eeq
using the antisymmetry properties of the connection coefficient \cite{Andriot:2013xca}. The six terms from $\Omega_{[ABC]}$ can be grouped two by two to use the above formula, and further combinations give
\bea
\D_2 &=\frac{8}{24}\Big( 3 \Omega_{[abc]}\te^a\!\w \te^b\!\w \te^c\!\w \\
&\phantom{=\frac{8}{24}} +2 \Omega_{[a}{}^{b}{}_{c]}\te^{a}\!\w\iota_b\, \te^c\!\w+2 \Omega_{[b}{}^{c}{}_{a]}  \te^a\!\w \te^b\!\w \iota_c +2 \Omega_{[c}{}^{a}{}_{b]}  \iota_a\, \te^b\!\w \te^c\!\w \nn\\
&\phantom{=\frac{8}{24}} + 2 \Omega^{[a}{}_{b}{}^{c]}\iota_a\, \te^b\!\w \iota_c + 2 \Omega^{[b}{}_{c}{}^{a]} \iota_a\, \iota_b\, \te^c\!\w + 2 \Omega^{[c}{}_{a}{}^{b]} \te^a\!\w \iota_b\, \iota_c \nn\\
&\phantom{=\frac{8}{24}} + 3 \Omega^{[abc]}\iota_a\,\iota_b\,\iota_c\Big) \ , \nn
\eea
where we also set some connection coefficients to zero following \cite{Andriot:2013xca}, and the $\G$-matrices have been rewritten with the Clifford map of section \ref{sec:Spinder}. Using the commutation properties of forms and contractions, and the value of the connection coefficients derived in \cite{Andriot:2013xca}, one obtains eventually the two $\D_2$ given in section \ref{sec:Spinder}.

We now turn to the derivation of the BI using the nilpotency condition \eqref{nilpocond} on the Dirac operator $\D$ \eqref{Diracopgen}. We focus only on the $\b$-supergravity case, and use the expressions for the three parts $\D_1$, $\D_2$ and $\D_3$ given in section \ref{sec:Spinder}. We start with $\D_2$, that we showed to be related to the derivative $\D_{\sharp}$ of \cite{Ihl:2007ah}. As mentioned in \eqref{Dsharpsquare}, the vanishing square of this last derivative is known to reproduce the Bianchi identities for constant fluxes, together with an additional constraint. So this piece should be a good starting point. That square, acting on a $p$-form $A$, was computed explicitly in \cite{Ihl:2007ah} and can be translated here as follows (we use conventions of appendix \ref{ap:conv})
\bea
\frac{1}{4}\D_2^2 A=\D_{\sharp}^2A=& +\frac{1}{4}f^{g}{}_{gd} f^{d}{}_{ab}\te^a \w \te^b \w A\\
&+ \frac{1}{2} f^{d}{}_{ga}f^{g}{}_{bc} \te^a \w \te^b \w \te^c \w \iota_d A\nn\\
&+\frac{1}{4} f^{g}{}_{gd}Q_{a}{}^{da}A\nn\\
&-\frac{1}{2}(f^{b}{}_{cd} Q_{a}{}^{cd}+f^{c}{}_{cd} Q_{a}{}^{db}+f^{b}{}_{da} Q_{c}{}^{cd}) \te^a \w \iota_b A\nn\\
&+\frac{1}{4}\Big(4f^{c}{}_{ga}Q_{b}{}^{gd}+f^{g}{}_{ab}Q_{g}{}^{cd}\Big)\te^a \w \te^b \w \iota_c \iota_d A \nn\\
&-\frac{1}{2}\Big(f^{a}{}_{cd} R^{cdb}+\frac{1}{2}f^{c}{}_{cd} R^{dab}+\frac{1}{2} Q_{c}{}^{cd}Q_{d}{}^{ab}\Big)\iota_a \iota_b A\nn\\
&- \frac{1}{2} \Big(f^{d}{}_{ga} R^{gbc}+Q_{g}{}^{bc} Q_{a}{}^{gd}\Big)\te^a \w \iota_b \iota_c \iota_d A\nn\\
&-\frac{1}{4}Q_{g}{}^{ab} R^{gcd}\iota_a \iota_b \iota_c \iota_d  A \nn\ .
\eea
Let us now add to $\D_2$ the derivative part $\D_1$
\bea
\frac{1}{4}\Big(\D_1^2+\D_1\D_2+\D_2\D_1\Big) A =& -\frac{1}{2}\del_a f^{d}{}_{db}\te^a \w \te^b \w A - \frac{1}{2}\del_a f^{d}{}_{bc} \te^a \w \te^b \w \te^c \w \iota_d A \\
&+\frac{1}{2}\Big(\b^{ac}f^{d}{}_{ca}\del_d -\b^{de}f^{g}{}_{gd}\del_e+Q_{d}{}^{db}\del_b-\b^{de}\del_e f^{g}{}_{gd}\Big)A\nn\\
&+ \Big(-\b^{de}\del_e f^{b}{}_{da}+\frac{1}{2}(\del_a Q_{d}{}^{db}+\b^{be}\del_e f^{d}{}_{da})\Big) \te^a \w \iota_b A\nn\\
&-\frac{1}{2}\Big(\del_a Q_{b}{}^{cd}-\b^{gc}\del_g f^{d}{}_{ab}\Big)\te^a \w \te^b \w \iota_c \iota_d A \nn\\
&+\frac{1}{6}\Big(-3\b^{dc}\del_c Q_{d}{}^{ab}+3\b^{ac}\del_c Q_{d}{}^{db}\Big)\iota_a \iota_b A \nn\\
&+ \frac{1}{6} \Big(\del_a R^{bcd}-3\b^{eb}\del_e Q_{a}{}^{cd}\Big)\te^a \w \iota_b \iota_c \iota_d A\nn\\
&-\frac{1}{6}\b^{ga}\del_g R^{bcd}\iota_a \iota_b \iota_c \iota_d  A \nn\ .
\eea
Bringing indices in the right order and writing out antisymmetries, we obtain a set of identities by adding the above to $\frac{1}{4}\D_2^2$. Among those are already present our four BI \eqref{delf2} - \eqref{bdelR2}. However the additional identities are independent and non-trivial; they contain in particular derivatives acting on $A$. To get rid of those, the missing part $\D_3$ of the Dirac operator is then necessary. Note that this last part contains terms that include the dilaton. So the additional terms to the square are
\bea
&\frac{1}{4}\Big(\D_1 \D_3 + \D_3\D_1 + \D_2\D_3+ \D_3\D_2+\D_3^2\Big) A\\
&= \Big(-\frac{1}{4}f^{g}{}_{gd} f^{d}{}_{ab}+\frac{1}{2}f^{c}{}_{ab}\del_c\tp+\frac{1}{2}\del_a f^{d}{}_{db}-\del_a\del_b\tp\Big)\te^a \w \te^b \w A \nn\\
&+\Big(\frac{1}{4}Q_{d}{}^{da}f^{g}{}_{ga}-\frac{1}{2}f^{d}{}_{da}(\b^{ab}\del_b \tp -\T^a)-\frac{1}{2}Q_{d}{}^{da}\del_a\tp+\del_a\tp(\b^{ab}\del_b \tp -\T^a) +\frac{1}{2}Q_{d}{}^{da}\del_a \nn\\
& \phantom{+\Big(} + \T^a \del_a +\frac{1}{2}\b^{ac}\del_c f^{d}{}_{da} + \frac{1}{2}\b^{ac}f^{d}{}_{da}\del_c - \b^{ac}\del_c\del_a \tp + \frac{1}{2}f^{g}{}_{gd}(\b^{dc}\del_c \tp -\T^d)-\frac{1}{2}Q_{d}{}^{da}\del_a\tp\Big) A\nn\\
&+\Big(\frac{1}{2}\del_a Q_{d}{}^{db}- \del_a (\b^{bc}\del_c \tp -\T^b)-\frac{1}{2}\b^{bc}\del_c f^{d}{}_{da}-\b^{bc}\del_c\del_a\tp\nn\\
&\phantom{+\Big(} +f^{b}{}_{da}(\b^{dc}\del_c \tp -\T^d)+ Q_{a}{}^{bc}\del_c\tp +\frac{1}{2}f^{b}{}_{ad} Q_{g}{}^{gd}-\frac{1}{2}f^{g}{}_{gc} Q_{a}{}^{bc}\Big) \te^a \w \iota_b A\nn\\
&+\frac{1}{2}\Big(\b^{ac}\del_c Q_{d}{}^{db}-2\b^{ac}\del_c (\b^{bd}\del_d \tp -\T^b)\nn\\
&\phantom{+2\Big(} +\frac{1}{2}f^{g}{}_{gd} R^{abd}-R^{abd}\del_d\tp-\frac{1}{2} Q_{d}{}^{ab}Q_{g}{}^{gd}+Q_{d}{}^{ab}(\b^{dc}\del_c \tp -\T^d)\Big)\iota_a \iota_b A\nn\ .
\eea

All these contributions add-up to the following identities
\bea
\frac{1}{2}\del_{[a} f^{d}{}_{b]d}+\frac{1}{4}f^{g}{}_{gd} f^{d}{}_{ab}-\frac{1}{4}f^{g}{}_{gd} f^{d}{}_{ab}+\frac{1}{2}f^{c}{}_{ab}\del_c\tp-\frac{1}{2}\del_{[a} f^{d}{}_{b]d}-\del_{[a}\del_{b]}\tp&=0\label{fcontr2}\\
- \frac{1}{2}\del_{[a} f^{d}{}_{bc]}+ \frac{1}{2} f^{d}{}_{g[a}f^{g}{}_{bc]}&=0 \label{BIdelf2}\\
\frac{1}{2}(\b^{ac}f^{d}{}_{ca}\del_d -\b^{de}f^{g}{}_{gd}\del_e+Q_{d}{}^{db}\del_b-\b^{de}\del_e f^{g}{}_{gd})+\frac{1}{4} f^{g}{}_{gd}Q_{a}{}^{da}+\frac{1}{4}Q_{d}{}^{da}f^{g}{}_{ga}& \nn\\
-\frac{1}{2}f^{d}{}_{da}(\b^{ab}\del_b \tp -\T^a)-\frac{1}{2}Q_{d}{}^{da}\del_a\tp+\del_a\tp(\b^{ab}\del_b \tp -\T^a)+\frac{1}{2}Q_{d}{}^{da}\del_a + \T^a \del_a &\nn \\
+\frac{1}{2}\b^{ac}\del_c f^{d}{}_{da} + \frac{1}{2}\b^{ac}f^{d}{}_{da}\del_c + \b^{ac}\del_c\del_a \tp+ \frac{1}{2}f^{g}{}_{gd}(\b^{dc}\del_c \tp -\T^d)-\frac{1}{2}Q_{d}{}^{da}\del_a\tp&=0 \label{fullcontr2}\\
-\b^{de}\del_e f^{b}{}_{da}+\frac{1}{2}(\del_a Q_{d}{}^{db}+\b^{be}\del_e f^{d}{}_{da})-\frac{1}{2}(f^{b}{}_{cd} Q_{a}{}^{cd}+f^{c}{}_{cd} Q_{a}{}^{db}+f^{b}{}_{da} Q_{c}{}^{cd})&\nn \\
+\frac{1}{2}\del_a Q_{d}{}^{db}- \del_a (\b^{bc}\del_c \tp -\T^b)-\frac{1}{2}\b^{bc}\del_c f^{d}{}_{da}-\b^{bc}\del_c\del_a\tp&\nn \\
+f^{b}{}_{da}(\b^{dc}\del_c \tp -\T^d)+ Q_{a}{}^{bc}\del_c\tp +\frac{1}{2}f^{b}{}_{ad} Q_{g}{}^{gd}-\frac{1}{2}f^{g}{}_{gc} Q_{a}{}^{bc}&=0\label{wedgecontr2}\\
-\frac{1}{2}(\del_{[a} Q_{c]}{}^{de}-\b^{g[d}\del_g f^{e]}{}_{ac})+\frac{1}{4}(-4f^{[d}{}_{g[a}Q_{c]}{}^{e]g}+f^{g}{}_{ac}Q_{g}{}^{de})&=0 \label{BIdelQ2}\\
\frac{1}{6}(-3\b^{dc}\del_c Q_{d}{}^{ab}+3\b^{c[a}\del_c Q_{d}{}^{b]d})-\frac{1}{2}(f^{[a}{}_{cd} R^{b]cd}+\frac{1}{2}f^{c}{}_{cd} R^{dab}+\frac{1}{2} Q_{c}{}^{cd}Q_{d}{}^{ab})& \nn\\
+\frac{1}{2}(\b^{ac}\del_c Q_{d}{}^{db}-2\b^{ac}\del_c (\b^{bd}\del_d \tp -\T^b)&\nn \\
+\frac{1}{2}f^{g}{}_{gd} R^{abd}-R^{abd}\del_d\tp-\frac{1}{2} Q_{d}{}^{ab}Q_{g}{}^{gd}+Q_{d}{}^{ab}(\b^{dc}\del_c \tp -\T^d))&=0 \label{twocontr2}\\
 \frac{1}{6} (\del_a R^{bcd}-3\b^{e[b}\del_e Q_{a}{}^{cd]})- \frac{1}{2} ( -R^{g[bc}f^{d}{}_{a]g}+Q_{a}{}^{g[d}Q_{g}{}^{bc]} )&=0 \label{BIdelR2}\\
-\frac{1}{6}\b^{g[a}\del_g R^{bcd]}-\frac{1}{4}Q_{g}{}^{[ab} R^{cd]g}&=0\ . \label{BIbdelR2}
\eea
Using in particular the expression of $\T^a$ in terms of the other fluxes, \eqref{fullcontr2}, \eqref{wedgecontr2} and \eqref{twocontr2} can be simplified respectively to
\bea
-\frac{1}{2}Q_{d}{}^{da}f^{g}{}_{ga}&=0\label{fullcontr3}\\
-\frac{3}{2}\b^{de}\del_{[e} f^{b}{}_{da]}+\frac{3}{2}\b^{de}f^{b}{}_{h[a}f^{h}{}_{ed]}&=0\label{betadelf2}\\
-\frac{1}{2}\b^{dc}\del_c Q_{d}{}^{ab}-\frac{1}{2}\b^{cd}\b^{g[a}\del_g f^{b]}{}_{cd} -\b^{dc}Q_{c}{}^{g[a}f^{b]}{}_{dg}+\frac{1}{4}\b^{dc}Q_{g}{}^{ab}f^{g}{}_{cd}&=0\label{betadelQ2}\ .
\eea
In addition, \eqref{fcontr2} simply vanishes. We are then left with seven identities, namely \eqref{BIdelf2}, \eqref{fullcontr3}, \eqref{betadelf2}, \eqref{BIdelQ2}, \eqref{betadelQ2}, \eqref{BIdelR2} and \eqref{BIbdelR2}, that we respectively give in \eqref{BIdelf} - \eqref{BIbdelR}. As we show there, only five of those are independent and give our four BI \eqref{delf2} - \eqref{bdelR2} together with the expected scalar condition.

\section{The $Q$-brane background and the related Bianchi identity}\label{ap:Qbrane}

\subsection{The $Q$-brane is a vacuum of $\b$-supergravity}\label{ap:Qbranevac}

The $\NS 5$-brane and the $\KK$-monopole are known vacua of standard supergravity. We verify explicitly in this appendix that the $Q$-brane, given in sections \ref{sec:NSbranesol} and \ref{sec:TdBI}, satisfies the equations of motion of $\b$-supergravity. We recall that this makes the $Q$-brane a vacuum of standard supergravity as well. As discussed in section \ref{sec:betasugra} and appendix \ref{ap:relbetadel}, for a field configuration satisfying $\b^{mn} \del_n \cdot =0$ and $\del_p \b^{np}=0$, $\b$-supergravity gets simplified to the theory worked out in \cite{Andriot:2011uh}. These two conditions turn out to be verified by the $Q$-brane, even at the singularity. Using this property, the $Q$-brane was verified in \cite{Hassler:2013wsa} to solve the simple equations of motion of \cite{Andriot:2011uh}. We show however in appendix \ref{ap:relbetadel} that the $\b$ equation of motion of \cite{Andriot:2011uh} is a priori not correct. In addition, the warp factor was considered in \cite{Hassler:2013wsa} to be harmonic, which only holds away from the singularity. Here we will get some new information at the singularity. So we start with the full equations of motion of $\b$-supergravity, obtained in this paper in flat indices. Using the two above conditions, the three equations of motion have been simplified towards \eqref{dileomsimplif1}, \eqref{Einsteinsimplif1}, and \eqref{beomflatsimpl2}.

For the $Q$-brane, given the non-zero components of the fluxes, each term of the $\b$ equation of motion \eqref{beomflatsimpl2} simply vanishes because of the indices contractions: it is trivially satisfied. So let us turn to the dilaton equation of motion \eqref{dileomsimplif1}. One computes
\bea
& \R=-\frac{5}{2} f^{-3} (\del_{\rho} f)^2 + f^{-2} \Delta_2 f \ ,\ \cR= -\frac{1}{2} f^{-3} (\del_{\rho} f)^2 \ ,\\
& (\del \tp)^2 = \frac{1}{4} f^{-3} (\del_{\rho} f)^2 \ ,\ \N^2 \tp= f^{-3} (\del_{\rho} f)^2 -\frac{1}{2} f^{-2} \Delta_2 f \ .
\eea
Note that in these expressions and the following ones, the LHS is given in flat indices, whereas the RHS involves derivatives in curved indices. One way to compute $\N^2 \tp$ is to use
\beq
\eta^{ab} \N_a V_b = \eta^{ab} \del_a V_b + \eta^{cd} f^b{}_{bc} V_d \ .
\eeq
This leads to
\beq
\frac{1}{4} \left(\R(\tg) + \cR(\tg) \right) - (\del \tp)^2 + \N^2 \tp =-\frac{1}{4} f^{-2} \Delta_2 f \label{dilPoi}\ .
\eeq
So away from the singularity, \eqref{dileomsimplif1} is satisfied, since $\Delta_2 f=0$ for $\rho >0$. At the singularity, we get a $\delta$, which is expected. Indeed, one should in principle add a source action to the bulk action, and the former would contribute to the equations of motion by a $\delta$ within the energy-momentum tensor. This is what we obtain here.

Finally, we focus on the simplified Einstein equation \eqref{Einsteinsimplif1}. The only non-zero components of the Ricci tensor in flat indices are
\bea
\R_{xx} &=\R_{yy} =-f^{-3}(\del_\rho  f)^{2}+\frac{1}{2}f^{-2}\Delta f  \\
\R_{\rho\rho} &=-\frac{3}{2}f^{-3}(\del_\rho  f)^{2}+\frac{1}{2}f^{-2}\del_\rho^2 f-\frac{1}{2}f^{-2}\rho^{-1}\del_\rho f\\
\R_{\varphi\varphi} &=f^{-3}(\del_\rho f)^{2}-\frac{1}{2}f^{-2}\del_\rho^2 f+\frac{1}{2}f^{-2}\rho^{-1}\del_\rho f \ .
\eea
The other curvature tensor takes the form
\bea
\cR^{ab}=\beta^{cd} \del_d {\omega_{Q}}_c^{ab}-\beta^{ad} \del_d {\omega_{Q}}_c^{cb} +{\omega_{Q}}_c^{ab}{\omega_{Q}}_d^{dc} -{\omega_{Q}}_d^{ca}{\omega_{Q}}_c^{db} -\frac{1}{2}R^{adc}f^{b}{}_{dc} \simeq -{\omega_{Q}}_d^{ca}{\omega_{Q}}_c^{db}\ , \label{cRtensor}
\eea
where the last equality is obtained thanks to the aforementioned simplifications verified by the $Q$-brane. The non-zero components are
\bea
\cR^{xx} &=\cR^{yy}=-\frac{1}{2}(Q_{\varphi}{}^{yx})^2=-\frac{1}{2}f^{-3}(\del_\rho f)^{2}\\
\cR^{\varphi\varphi} &=\frac{1}{2}f^{-3}(\del_\rho f)^{2}\ ,\quad \cR^{\rho\rho}=0\ .
\eea
In addition the dilaton terms in flat indices yield
\bea
\N_x \N_x \tp&=-\o^\rho_{xx}f^{-\frac{1}{2}}\del_\rho \tp= \frac{1}{4}f^{-3}(\del_\rho f)^2\\
\N_y \N_y \tp&=-\o^\rho_{yy}f^{-\frac{1}{2}}\del_\rho \tp= \frac{1}{4}f^{-3}(\del_\rho f)^2\\
\N_\rho \N_\rho \tp&=f^{-\frac{1}{2}}\del_\rho (f^{-\frac{1}{2}}\del_\rho \tp)= -\frac{1}{2}f^{-\frac{1}{2}}\del_\rho(f^{-\frac{3}{2}}\del_\rho f)=\frac{3}{4}f^{-3}(\del_\rho f)^2 -\frac{1}{2}f^{-2}\del_\rho^2 f\\
\N_\varphi \N_\varphi \tp&=-\o^\rho_{\varphi\varphi}f^{-\frac{1}{2}}\del_\rho \tp=-\frac{1}{4}f^{-3}(\del_\rho f)^2-\frac{1}{2}f^{-2}\rho^{-1}\del_\rho f\ ,
\eea
from which we eventually deduce
\bea
\R_{xx}-\cR^{xx}+2\N_x \N_x \tp&= \frac{1}{2}f^{-2}\Delta f \label{Ex} \\
\R_{yy}-\cR^{yy}+2\N_y \N_y \tp&=\frac{1}{2}f^{-2}\Delta f \label{Ey} \\
\R_{\rho\rho}-\cR^{\rho\rho}+2\N_\rho \N_\rho \tp &= - \frac{1}{2}f^{-2}\Delta f \label{Erho} \\
\R_{\varphi\varphi}-\cR^{\varphi\varphi} +2\N_\varphi \N_\varphi \tp &=- \frac{1}{2}f^{-2}\Delta f \ .\label{Evarphi}
\eea
As explained for the dilaton equation of motion \eqref{dilPoi}, the above equations vanish away from the singularity as \eqref{Einsteinsimplif1}, and receive at the singularity an energy-momentum tensor contribution in the form of a $\delta$, due to the $Q$-brane action to be added.

\subsection{The Bianchi identity with $Q$-brane source term}\label{ap:Hannover}

We comment here on a BI with a $Q$-brane source term obtained in (5.24) of \cite{Chatzistavrakidis:2013jqa}, and compare it to our proposal \eqref{BIsourcedQbrane}. It is given by
\beq
\d \left( \del_m \b^{np}\ \tg_{nu} \tg_{pv} \d x^m \wedge \d u \wedge \d v \right) = \mbox{constant}\ {\rm vol}_4\ \delta^{(4)} \ ,\label{BIHannover}
\eeq
where the RHS contains a constant times a four-dimensional volume form, and the LHS involves two specific directions $u$ and $v$. This BI looks similar to the one for the $H$-flux, in presence of an $\NS5$-brane, since it is a four-form and the source is localised in four dimensions by the $\delta^{(4)}$. This last point looks however unexpected, since the $Q$-brane is only a codimension $2$ object. One can still wonder whether, upon smearing two dimensions, \eqref{BIHannover} reduces to our proposal \eqref{BIsourcedQbrane} that contains a $\delta^{(2)}$. The two BI are given in rather different fashions, so to ease the comparison, let us rewrite \eqref{BIHannover}, partially evaluated on the $Q$-brane solution given in section \ref{sec:Tbranes}.

In this background, the metric is diagonal and $\b$ has only one non-trivial component. Therefore we can replace $u$ and $v$ by generic directions: on the $Q$-brane solution, the two expressions have the same value up to a factor $2$. Using \eqref{Qsimplif}, we then rewrite \eqref{BIHannover} on this background as
\bea
& *_4 \d \left( Q_a{}^{bc}\ \eta_{bd} \eta_{cf} \te^a \wedge \te^d \wedge \te^f \right) = \mbox{constant}'\ \delta^{(4)} \ ,\\
\Leftrightarrow\ & \epsilon^{gadf} \left(\del_g Q_a{}^{bc} \eta_{bd} \eta_{cf} - \frac{1}{2} f^h{}_{ga} Q_h{}^{bc} \eta_{bd} \eta_{cf} + Q_g{}^{bc} f^h{}_{ad} \eta_{bh} \eta_{cf}  \right) = \mbox{constant}''\ \delta^{(4)} \ ,
\eea
the indices of $\epsilon$ being lifted with $\eta$. With the non-zero fluxes of the $Q$-brane solution, we get
\beq
2\epsilon^{\rho \varphi xy}\left( f^{-\frac{1}{2}} \del_{\rho}Q_{\varphi}{}^{xy}-  Q_{\varphi}{}^{xy} (f^{\varphi}{}_{\rho \varphi} + f^{y}{}_{\rho y} + f^{x}{}_{\rho x} ) \right) = \mbox{constant}''\ \delta^{(4)} \ .
\eeq
This expression is close to ours for $S^{xy}_{\rho \varphi}$ in \eqref{computationSQbr}, but is still different: the signs in front of $f^{y}{}_{\rho y}, f^{x}{}_{\rho x}$ differ. Another way to see this mismatch is through the related term $Q_g{}^{bc} f^h{}_{ad} \eta_{bh} \eta_{cf}$ that is generically different from the one in our BI \eqref{delQ2}, although it is again only a matter of sign when evaluated on the solution. We believe that smearing would not change this sign.

So the two proposals \eqref{BIHannover} and \eqref{BIsourcedQbrane} differ, at least when evaluated on the $Q$-brane solution, which would have been a minimal requirement. As consequence, we doubt that \eqref{BIHannover} could reduce to the two-dimensional Poisson equation, even when smeared. We actually believe that an explicit tensorial expression for a BI with a $Q$-brane source term is not given by a four-form, but rather involves contractions, e.g. $\cN^a\cdot \iota_a$, as indicated by \eqref{tensorialformBI}.

\section{Proofs about symmetries}\label{ap:proofs}

In this appendix, we prove various statements that appeared in our study of symmetries in section \ref{sec:sym}.

\subsection{Proof of the equivalence \eqref{condnewsym}}

Having isometries generated by Killing vectors translates into Killing equations on each of our fields. Those are given in terms of the Lie derivative ${\cal L}_{V_{\iota}}$. For constant Killing vectors, it boils down to the conditions
\beq
\forall \iota \in \{1 \dots N\}, p, q,\ V_{\iota}^m\del_m \tg_{pq}=0 \ ,\  V_{\iota}^m\del_m \b^{pq}=0 \ , \  V_{\iota}^m\del_m \tp=0 \ . \label{proofcond1}
\eeq

Let us first prove the implication $\Rightarrow$. The $N$ Killing vectors are constant and independent. So they form a basis of an $N$-dimensional vector space. Using constant rotations, one can thus bring them to a form where $V_{\iota}^m = \delta_{\iota}^m\ v_{(\iota)}$ (no sum on $\iota$), $v_{(\iota)}\neq 0$. As the rotations are constant, they can be performed on the coordinates as well, and on the $\del_m$. So without changing notation, we now consider to have such Killing vectors. The conditions \eqref{proofcond1} now become
\beq
\forall \iota \in \{1 \dots N\}, p, q,\ \del_{\iota} \tg_{pq}=0 \ ,\  \del_{\iota} \b^{pq}=0 \ , \  \del_{\iota} \tp=0 \ . \label{proofcond2}
\eeq
As the vectors are constant and independent, $N$ cannot be bigger than the dimension of the space-time. Let us now consider any constant antisymmetric bivector of coefficient $\varpi^{pq}$ that is non-zero only along these $N$ directions, i.e. $\forall p \in \{1 \dots N\},\ \exists q\ /\ \varpi^{pq} \neq 0$ and $\forall p \notin \{1 \dots N\},\ \varpi^{pq}=0$. Thanks to the antisymmetry of $\varpi^{pq}$, this means that only the diagonal block along $(1\dots N) \times (1\dots N)$ is non-zero. Note that this requires $N>1$, as assumed. Because of this block structure, one has $\varpi^{pr} \del_r=\sum_{\iota=1}^{N} \varpi^{p \iota} \del_{\iota}$. This operator applied on any of the three fields vanishes, thanks to \eqref{proofcond2}. In addition, it also vanishes on any of their derivatives, by commuting the derivatives. So we eventually obtain $\varpi^{pr} \del_r \cdot =0$.\\

Let us now prove the reverse implication $\Leftarrow$. We start with a constant antisymmetric bivector $\varpi^{pq}$ non-zero along a diagonal $N \times N$ block. Up to relabeling the directions, having this block translates into $\forall p \in \{1 \dots N\},\ \exists q\ /\ \varpi^{pq} \neq 0$ and $\forall p \notin \{1 \dots N\},\ \varpi^{pq}=0$. Let us now assume that $N$ is even. We then consider a particular $\varpi^{pq}$ such that the block only has one non-zero entry on each line, i.e. $\forall p \in \{1 \dots N\},\ \exists ! p_0\ /\ \varpi^{pp_0} \neq 0$. Thanks to the antisymmetry, this means that each column of the block also has only one non-zero entry. So it is clear that $\{ p_0 \}$ spans $\{1 \dots N\}$. Let us provide an example of such a block of $\varpi$ (viewed as a matrix), to show that it can exist\footnote{Such $\varpi$ are only possible for an even $N$, that we assumed; indeed, for $N$ being odd, the determinant of the block would be zero (a property of antisymmetric matrices), which would prevent to get from it (alone) $N$ independent vectors, as we will see.}
{\small
\beq
\begin{pmatrix} 0 & 1 & & & \\ -1 & 0 & & & \\  & & \ddots & & \\ & & & 0 & 1 \\ & & & -1 & 0 \end{pmatrix} \ .
\eeq}
In addition, one has by assumption $\forall p,\ \varpi^{pr} \del_r \cdot =0$. The peculiar structure of the block just considered then implies that $\forall p \in \{1 \dots N\},\ \varpi^{p p_0} \del_{p_0} \cdot = 0$ (without sum on $p_0$). We then define $N$ vectors $V_{\iota}$, $\iota \in \{1 \dots N\}$, of components $V_{\iota}^m = \delta_{\iota}^m\ v_{(\iota)}$ (no sum on $\iota$) with $v_{(p_0)}\equiv\varpi^{p p_0}\neq 0$. Given these components, the $N$ vectors are constant and independent. One can verify that they satisfy $\forall \iota \in \{1 \dots N\}, \ V_{\iota}^m \del_m \cdot =0$. So they satisfy the condition \eqref{proofcond1}, and they are Killing vectors.

Let us now look at the case where $N$ is odd. As $N>1$, we deduce $N\geq3$. We then consider a $\varpi$ having a non-zero diagonal $N \times N$ block that splits into two diagonal blocks of size $(N-3) \times (N-3)$ and $3 \times 3$. The first block is of even size; from that one we can construct as above $N-3$ constant and independent Killing vectors, along directions that do not mix with the remaining $3$. We will now construct a similar set of $3$ vectors along these last directions, and overall, the $N$ Killing vectors will then be independent. To construct two of the three missing Killing vectors, one can consider a block of the form
{\small
\beq
\begin{pmatrix} 0 & 1 & 0 \\ -1 & 0 & 0 \\ 0 & 0 & 0 \end{pmatrix} \ ,
\eeq}
possibly with coefficients different than $1$. Either by proceeding as above on the $2 \times 2$ non-zero sub-block, or by diagonalising this block, one can get two more constant and independent Killing vectors. However, with this $\varpi$, we cannot get a Killing vector along the last direction; we need to consider a different $\varpi$. We only change the $3 \times 3$ block towards
{\small
\beq
\begin{pmatrix} 0 & 0 & 0 \\ 0 & 0 & 1 \\ 0 & -1 & 0 \end{pmatrix} \ ,
\eeq}
and proceed similarly. By linear combinations, we can then get one new constant Killing vector along the last direction, which is independent from all others.

\subsection{T-duality is a symmetry for the NSNS sector}

We show here the invariance of $\L_{{\rm NSNS}}$, up to a total derivative, under the T-duality transformation $O(N,N)$ given in \eqref{Td}, when the fields are independent of $N$ coordinates. To do so, we recall two approaches in the literature.
\begin{itemize}
\item Maharana-Schwarz \cite{Maharana:1992my} and the compactification along the isometries

We consider that the NSNS fields are independent of $N$ coordinates, in a $D$-dimensional space-time. One can then develop the Lagrangian $\L_{{\rm NSNS}}$ by separating the components of the fields that are along these $N$ directions and those that are not. The latter do not transform under the $O(N,N)$, while the former do. One can then look at how the various terms in the Lagrangian transform. This was precisely done in \cite{Maharana:1992my}: the resulting rewritten Lagrangian was shown to be $O(N,N)$ invariant.

The corresponding action can also be viewed as the compactified one. Because of the independence on $N$ coordinates, the corresponding volume factor can be factorized out (it is set to $1$ in \cite{Maharana:1992my}), leaving the action to be $D-N$ dimensional. It is actually a well-known fact that the reduced action has this $O(N,N)$ symmetry. It is however only a matter of volume factor to make it a $D$-dimensional action, and it then still has the symmetry.

\item Double Field Theory

The Double Field Theory (DFT) Lagrangian can be formulated as follows \cite{Hohm:2010pp}
\bea
\L_{{\rm DFT}} = e^{-2d} \Big(& \frac{1}{8} \hhh^{MN}\del_M \hhh^{PQ} \del_N \hhh_{PQ} - \frac{1}{2} \hhh^{MN} \del_N \hhh^{PQ} \del_Q \hhh_{MP} \\
 & - 2 \del_M d \del_N \hhh^{MN} + 4 \hhh^{MN} \del_M d \del_N d \Big)\ .\nn
\eea
The fields $\hhh$ and $d$ can be defined in terms of $g,\ b,\ \p$ as in section \ref{sec:betasugra} ($\hhh^{MN}$ is the component of $\hhh^{-1}$). However, they depend here on $2D$ coordinates $X^M=(\tilde{x}_m, x^m)$; the latter also define the derivative $\del_M$ accordingly. An interesting property of this Lagrangian is that it reproduces the standard NSNS Lagrangian up to a total derivative if one enforces the strong constraint, that we take here to be $\tilde{\del}=0$
\beq
\L_{{\rm DFT}}|_{\tilde{\del}=0} = \L_{{\rm NSNS}} + \del(\dots) \ . \label{DFTNSNS}
\eeq
Another property of this Lagrangian is its invariance under constant $O(D,D)$ transformations. Those are given by the same action as in \eqref{Td} for a generic $O \in O(D,D)$, together with a transformation of the coordinates and of the derivatives
\beq
X'= O^{-1} X \ ,\ \del' = O\ \del \ . \label{OXdel}
\eeq
Because of the contraction of indices and the invariance of $d$, it is straightforward to see that these constant $O(D,D)$ transformations are a symmetry of the Lagrangian, i.e. $\L_{{\rm DFT}}$ is invariant under them.

Let us now consider an independence on $N$ standard coordinates $x^m$, together with the strong constraint $\tilde{\del}=0$. This implies that the only non-trivial derivatives are the $\del_p$, where $x^p$ is not one of the $N$ coordinates. Similarly, the fields in $\L_{{\rm DFT}}$ then only depend on such $x^p$. Let us now consider $O_N$, one of the $O(N,N)$ transformations discussed in \eqref{ONNODD} and \eqref{Td}. Because of its $O(D,D)$ invariance, $\L_{{\rm DFT}}$ is invariant under this $O(N,N)$ subgroup. Let us now look at the action of such an $O_N$ on the derivatives and coordinates \eqref{OXdel}: on the $x^p$ that are the coordinates on which the Lagrangian depends, the action is trivial (it is the $\id_{D-N}$). The same holds for the derivatives $\del_p$. Therefore, when the fields are independent of $N$ coordinates $x^m$ and the strong constraint $\tilde{\del}=0$ is enforced, the effective transformation on the coordinates and derivatives in the Lagrangian under $O_N$ is
\beq
X'=X \ ,\ \del' = \del \ ,
\eeq
i.e. they do not transform. The action of this $O(N,N)$ subgroup then boils down to that of the T-duality group: indeed, the latter does not change the coordinates nor the derivatives, but only acts on $\hhh$ and $d$ as in \eqref{Td}. As mentioned above, this $O(N,N)$ leaves $\L_{{\rm DFT}}$ invariant. Therefore, thanks to \eqref{DFTNSNS}, we deduce that $\L_{{\rm NSNS}}$ is invariant under the T-duality group transformations, up to a total derivative, when fields are independent of $N$ coordinates.

\end{itemize}

\end{appendix}


\newpage

\providecommand{\href}[2]{#2}\begingroup\raggedright

\endgroup

\end{document}